\documentclass[aps,footinbib,superscriptaddress,groupedaddres,showpacs,twocolumn,prb,longbibliography,bibnotes,showkeys]{revtex4-1}

\usepackage{changepage}
\usepackage{fancyhdr}
\usepackage{float}
\usepackage{floatflt}
\usepackage[colorlinks=true,citecolor=blue,linkcolor=magenta]{hyperref}
\usepackage[normalem]{ulem}
\usepackage{epstopdf}
\usepackage{amsmath,amssymb}
\usepackage{dsfont}
\usepackage{braket}
\usepackage{graphicx}
\usepackage{color}
\usepackage{stackengine}
\usepackage{verbatim}

\newcommand{\Nf}{\mathcal{N}_F}
\newcommand{\Fv}{\mathcal{F}_{vol}}
\newcommand{\Fl}{\mathcal{F}_{lin}}
\newcommand{\E}{\mathcal{E}}

\def \titlename{ Diagnostics of nonergodic extended states and many body localization proximity effect through real-space and Fock-space excitations}
\def \authornames{Nilanjan Roy$^1$, Jagannath Sutradhar$^{1,2,3}$, and Sumilan Banerjee}
\def \affiliations{Centre for Condensed Matter Theory, Department of Physics,
Indian Institute of Science, Bangalore 560012, India\\
$^{2}$Department of Physics, Bar Ilan University, Ramat Gan 5290002, Israel
\\
$^{3}$Department of Physics, Faculty of Natural Sciences, Ariel  University,  Ariel  40700,  Israel 
}

\begin{document}
\title{\titlename}
\author{\authornames}
\affiliation{\affiliations}
\date{\today}
\begin{abstract}
We provide real-space and Fock-space (FS) characterizations of ergodic, nonergodic extended (NEE) and many-body localized (MBL) phases in an interacting quasiperiodic system, namely generalized Aubry-Andr\'e-Harper model, which possesses a mobility edge in the non-interacting limit. We show that a mobility edge in the single-particle (SP) excitations survives even in the  presence of interaction in the NEE phase. In contrast, all single-particle excitations get localized in the MBL phase due to the MBL proximity effect. We give complementary insights into the distinction of the NEE states from the ergodic and MBL states by computing local FS self-energies and decay length associated, respectively, with the local and the non-local FS propagators. Based on a finite-size scaling analysis of the typical local self-energy across the NEE to ergodic transition, we show that MBL and NEE states exhibit qualitatively similar multifractal character.  However, we find that the NEE and MBL states can be distinguished in terms of the distribution of local self-energy and the decay of the non-local propagator in the FS, whereas the typical local FS self-energy cannot tell them apart.

\end{abstract}

\maketitle

\section{Introduction}
Understanding thermalization or ergodicity, and its breakdown in isolated quantum systems has been one of the central problems of recent times in many-body quantum physics. While a typical interacting quantum system thermalizes adhering to the eigenstate thermalization hypothesis (ETH)~\cite{srednicki1994chaos,deutsch1991quantum}, a nonergodic behavior may arise in the presence of strong disorder leading to many-body localization (MBL)~\cite{baa,gornyi2005interacting}.  MBL and its phenomenology in one dimension have been studied quite extensively both in theory~\cite{oganesyan2007localization,kjall2014many,vznidarivc2008many,pal2010many,serbyn2015criterion,imbrie2016many,morningstar2019renormalization,goremykina2019analytically} and experiments~\cite{schreiber2015observation,rispoli2019quantum,lukin2019probing}. These works provide strong  evidences in favor of the existence of MBL phase in one dimension. To this end, while the universal properties of MBL to thermal transition \cite{luitz2015many,dumitrescu2019kosterlitz,kiefer2020evidence,vsuntajs2020quantum,abanin2021distinguishing,Sierant2020,Laflorencie2020,Suntajs2020} and the regime of stability \cite{Polkovnikov,Huse,Chandran,Chandran2,Sels2022,Sierant2022} of MBL phase remain under active debate, MBL is arguably the only well-established generic example of the non-equilibrium states of quantum matter that violates ETH~\cite{abanin2019colloquium,alet2018many,nandkishore2015many2,abanin2017recent}. 

\begin{table*}[htpb]
\centering
\includegraphics[width=\textwidth]{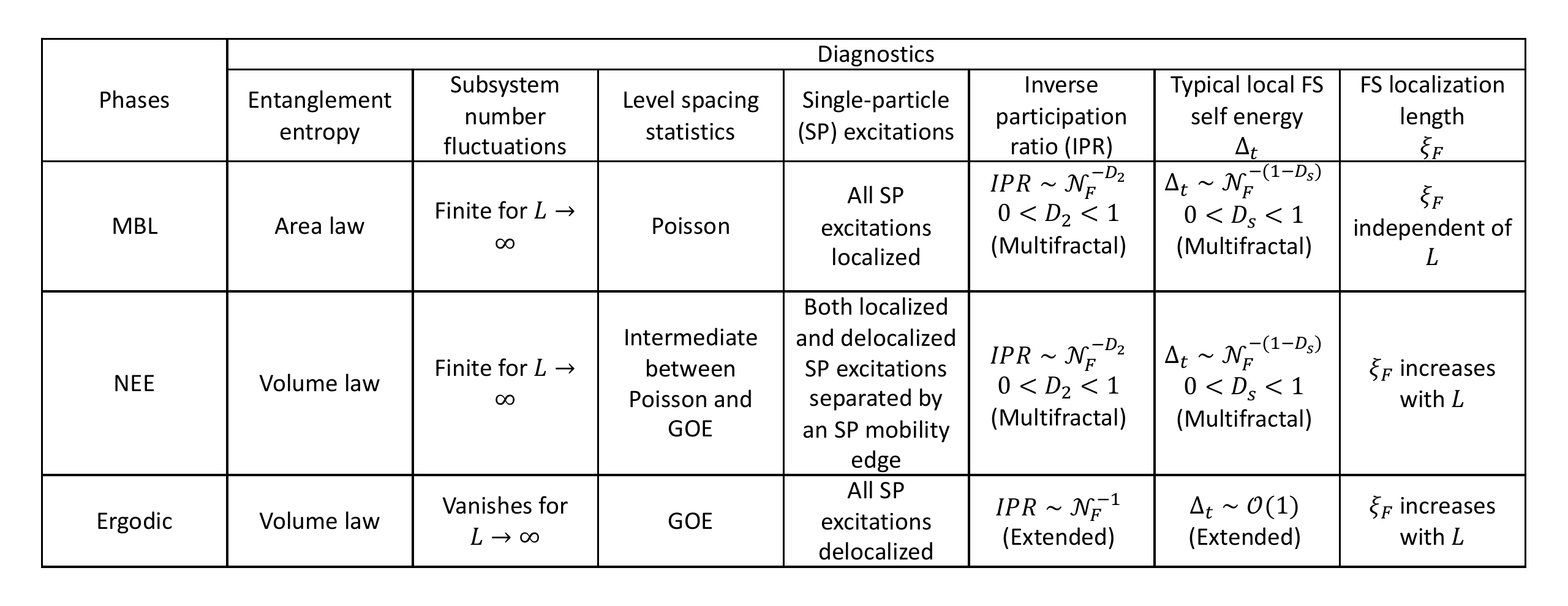}
\caption{{\bf Classification of MBL, NEE and ergodic phases in the GAAH model based on various real-space and Fock-space diagnostics:} Here $L$ and $\mathcal{N}_F$ are the number of sites on the real-space and Fock-space lattices, respectively.}
\label{tab:Table}
\end{table*}

It is thus an interesting question whether there are other type of non-ergodic many-body phases \emph{intermediate} between ergodic and MBL. There have been several proposals in different situations, and associated debates~\cite{Biroli2012,DeLuca2014,Altshuler2016,altshuler2016multifractal,Tikhonov2016,Tikhonov2016a,garcia2017scaling,Tikhonov2019,Kravtsov2015,Pino2016,Herrera2017,Micklitz2019,Prasad2021,Das2022,Tang2022,li2015many,modak2015many,Li2016,Deng2017,Modak2018,Nag2017,ghosh2020transport,Deng2019} for realizing such states, typically dubbed as non-ergodic extended (NEE) states, as a distinct phase.
Strong evidence \cite{li2015many,modak2015many,Li2016,Deng2017,Modak2018,Nag2017,ghosh2020transport} of NEE states have been found in systems with a particular type of quasiperiodic disorder, namely the generalized Aubry-Andr\'e-Harper (GAAH) model ~\cite{ganeshan2015nearest,modak2015many}, in the certain regime of many-body energy density and quasiperiodic potential strength.
In this system, the NEE states, in contrast to the ergodic and MBL states, are characterized by volume-law entanglement entropy and finite eigenstate-to-eigenstate fluctuations of local observable~\cite{li2015many,ghosh2020transport}. The non-interacting GAAH model hosts a single-particle (single-particle) mobility edge, and thus, naively, all the many-body eigenstates are expected to be ergodic due to coupling between localized and delocalized single-particle (single-particle) states through interaction. The existence of the MBL state in the GAAH model is rationalized \cite{modak2015many} in terms of ``MBL proximity effect"~\cite{nandkishore2015many,Hyatt2017,Marino2018}. Through this mechanism, a strongly localized system can localize a weakly \emph{ergodic bath} when coupled with each other in the absence of any symmetry or topological protections \cite{nandkishore2014marginal,Potter2016,Banerjee2016,Bhatt2021} of the delocalized states in the bath. In the GAAH model, the localized and delocalized single-particle states, existing in the different parts of the energy spectrum, constitute the localized system and the ergodic bath, respectively.

In this work, we unravel direct signatures of the MBL proximity effect and the fate of the \emph{single-particle mobility edge} in the MBL and NEE states of the interacting GAAH model through real-space single-particle excitations. We further characterize the non-ergodic and ergodic states by considering the interacting problem on the real-space lattice as an effective \emph{non-interacting} problem on the Fock-space (FS) graph or lattice, albeit with correlated disorder~\cite{Altland2017,Logan,logan2019many,Roy2020,Roy2020a,Ghosh.2019}. From this perspective, 
the MBL states are themselves multifractal or \emph{non-ergodic extended} in nature~\cite{roy2021fock,luitz2015many,mace2019multifractal,de2021rare}, i.e. they are extended over $\sim \Nf^D$ ($0<D<1$) FS sites, a vanishing fraction of the total $\Nf\sim \exp{(L)}$ FS sites corresponding to a real-space lattice with $L$ sites; here $D$ is a fractal dimension. In this scenario, how are the NEE states of GAAH model different from the MBL states then? We show that, though the NEE and MBL states are both multifractal in nature, they can be distinguished based on the existence or absence of a mobility edge in real-space single-particle excitations and an FS \emph{localization length} extracted from the non-local propagation of an effective \emph{excitation} on the FS lattice.

We characterize the single-particle excitation in real space by computing via exact diagonalization (ED) the typical value $\rho_t(\omega)$ of local density of states at excitation energy $\omega$. To quantify localization properties in Fock space, we obtain local and non-local FS propagators of an excitation on the FS lattice using a recursive Green's function method~\cite{sutradhar2022scaling}. We extract the statistical properties, e.g., the typical values $\Delta_t$ and $\xi_F$, of the imaginary part of the local Feenberg self-energy \cite{logan2019many} and an FS length scale, respectively, from the local and non-local FS propagator; $\xi_F$ captures the decay length of the non-local propagator on the FS graph. 

We demonstrate that the above two diagnostics along with the multifractality, provide much more clear-cut distinctions of MBL, NEE and  ergodic phases for the GAAH model, compared to the probes used in earlier studies~\cite{modak2015many,ghosh2020transport}, at least, within finite-size numerics. Thus, our work serves two main goals, (a) classification of the MBL, NEE and ergodic phases in terms of real- and Fock-space properties, and (b) characterization of NEE-ergodic phase transition through FS propagator in the GAAH model. We study the NEE-ergodic transition as a function of many-body energy density $\mathcal{E}$ through a finite-size scaling of $\Delta_t$. In this work, we do not study the MBL-NEE transition with $\mathcal{E}$ in much detail since the MBL phase only appears over a limited region in the many-body spectrum, near the lower edge close to the ground state. As a result, it is hard to carry out a controlled finite-size scaling analysis for the MBL-NEE transition with $\mathcal{E}$. For marking the critical value of $\mathcal{E}$ for the MBL-NEE transition we use the estimate from earlier studies~\cite{ghosh2020transport}, which are consistent with our results. As far as the classification of the phases is concerned, we tune both the energy density and quasiperiodic strength to access robust MBL, NEE, and ergodic states.
Our main results are the following:\\
1.~By tuning either the many-body energy density and/or quasiperiodic potential strength, we explicitly demonstrate the MBL proximity effect on the single-particle excitations from the system size dependence of $\rho_t(\omega)$. Remarkably, we find a single-particle mobility edge, even in the presence of interaction, separating localized and delocalized excitations in the NEE phase. This is in contrast to the ergodic and MBL phases, where all the excitations get delocalized and localized, respectively.\\
2. We show that $\Delta_t$ remains finite in the ergodic phase and vanishes as $\Delta_t\sim\Nf^{-(1-D_s)}$ with the fractal dimension $0<D_s<1$ in the NEE and MBL phases, reflecting multifractal nature of both kinds of nonergodic states on the FS lattice.\\
3. We show that the FS length scale $\xi_F$ varies with system size in the NEE phase, like in the ergodic phase but unlike the MBL phase, where $\xi_F$ becomes system size independent.

The remainder of the paper is organized as follows.
In Sec.~\ref{sec2}, we describe the model and the parameters. In Sec.~\ref{sec3}, we calculate the standard diagnostics such as the entanglement entropy, level-spacing ratio, etc. to distinguish the three phases: ergodic, NEE and MBL. We discuss the characterization of the phases in terms of single-particle excitations in real space for noninteracting and interacting systems in Sec.~\ref{sec4a}  and Sec.~\ref{sec4b}, respectively. Then in Sec.~\ref{sec5}, we define the FS propagator and discuss the FS lattice structure. We study the nonergodic-ergodic phase transition in FS in Sec.~\ref{sec6}, followed by a finite-size scaling analysis for the same transition in Sec.~\ref{sec7}. We also provide a comparative analysis of the inverse participation ratio and self-energy for our system in Sec.~\ref{sec9}. 
In Sec.~\ref{sec8}, we analyze the distribution of the FS Feenberg self-energy. In Sec.~\ref{sec10}, we define FS localization length and distinguish different phases using it. Finally, we conclude in Sec.~\ref{sec11}.

\section{Model}\label{sec2}
We consider interacting spinless fermions in a 1D quasiperiodic potential described by the following Hamiltonian,
\begin{align} 
H&=-t\sum_{i=1}^{L-1}[c_{i}^{\dagger}c_{i+1} + h.c.] +  \sum_{i=1}^{L} h_i n_i + V\sum\limits_{i=1}^{L-1} n_i n_{i+1},
\label{ham}
\end{align}
with open boundary conditions. Here $c_i$ is the fermion annihilation operator at site $i$ and $n_i=c^\dagger_ic_i$.
The nearest-neighbor hopping and the nearest-neighbor interaction strengths are $t=1$ and $V=1$, respectively. The quasiperiodic potential is $h_i=h\cos(2\pi \chi i + \phi)/(1-\alpha \cos(2\pi \chi i + \phi))$, where $\chi=(\sqrt{5}-1)/2$, $\alpha=-0.8$, and $h$ is the strength of the potential with a global phase $\phi\in (0,2\pi]$.
 The non-interacting model $(V=0)$ has a single-particle mobility edge (SPME) at an energy $\epsilon_c=\text{sgn}(h)(2|t|-|h|)/\alpha$~\cite{ganeshan2015nearest} i.e., all states with energy $\epsilon<\epsilon_c$ are localized and those with energy $\epsilon>\epsilon_c$ are delocalized. For the interacting system $(V=1)$, we consider quarter filling, i.e., $N=L/4$ fermions on $L$ sites. The choice of the filling is motivated by the earlier studies \cite{ghosh2020transport}, where MBL, NEE, and ergodic phases were identified at the quarter filling. In our ED calculations, we use $L=8,12,16,20$, and for the recursive Green's function calculations, we access larger systems $L=16,20,24$, consistent with the filling. As mentioned earlier, we study the phases for the GAAH model as a function of quasiperiodic potential strength $h$ and the energy density (per site) $\mathcal{E}=E/L$, where $E$ is the many-body energy. To compute a quantity, e.g., entanglement entropy, for eigenstates at an energy density $\mathcal{E}$, we bin the energy eigenvalues and compute the average over around $5$ eigenstates in a given bin whereas for energy level-statistics we average over around $10-100$ eigenstates in a given bin for smaller to larger system sizes. We also average over the phase $\phi$ appearing in the quasiperiodic potential, taking $5000,2000,500,100$ samples in the ED for $L=8,12,16,20$, respectively and $2000,1000,400$ samples in the recursive calculation for $L=16,20,24$, respectively.

\section{Distinguishing MBL, NEE and ergodic phases with standard diagnostics}\label{sec3}
\begin{figure}[h!]
\centering
\includegraphics[width=0.49\columnwidth,height=3.6cm]{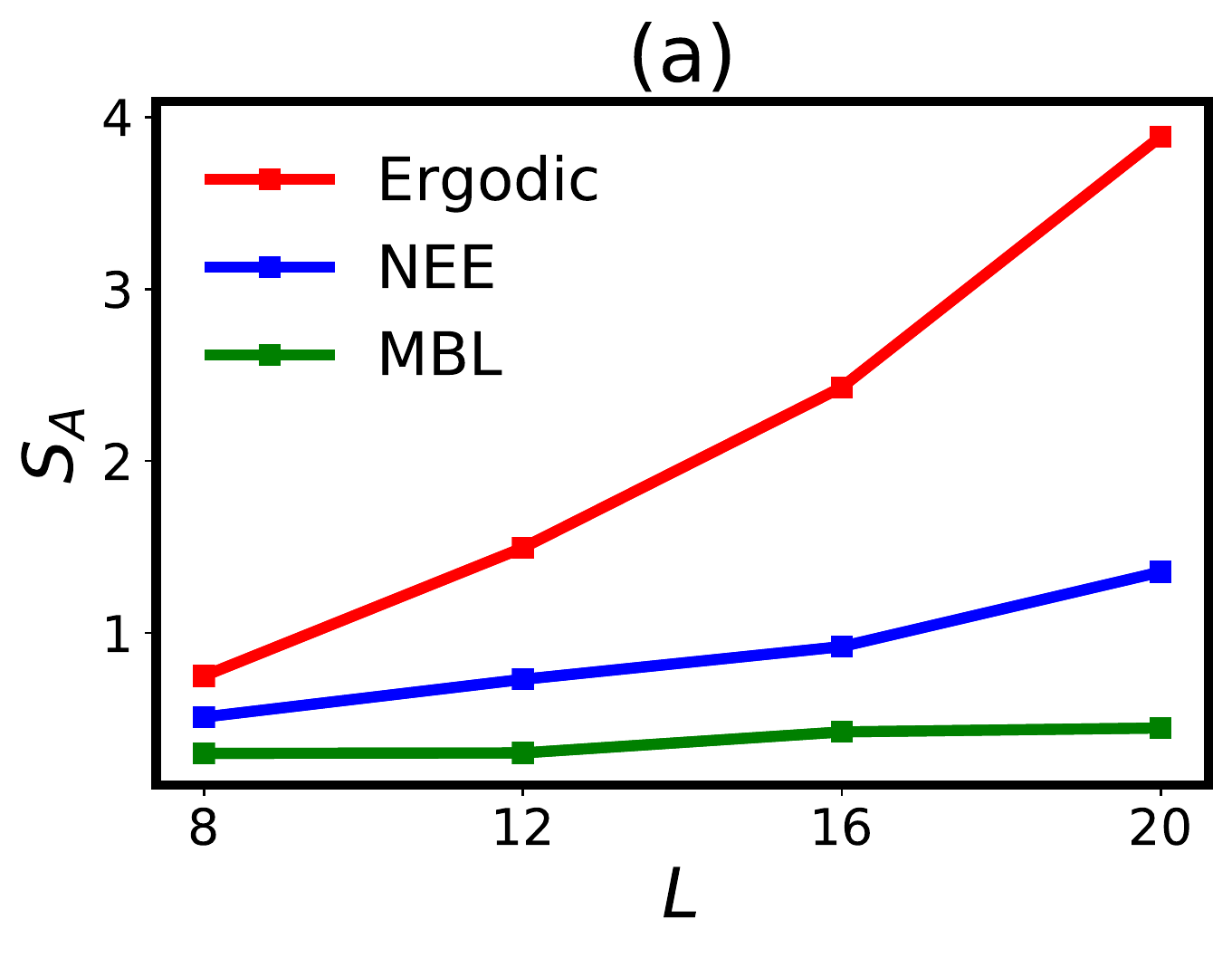}
\includegraphics[width=0.49\columnwidth,height=3.6cm]{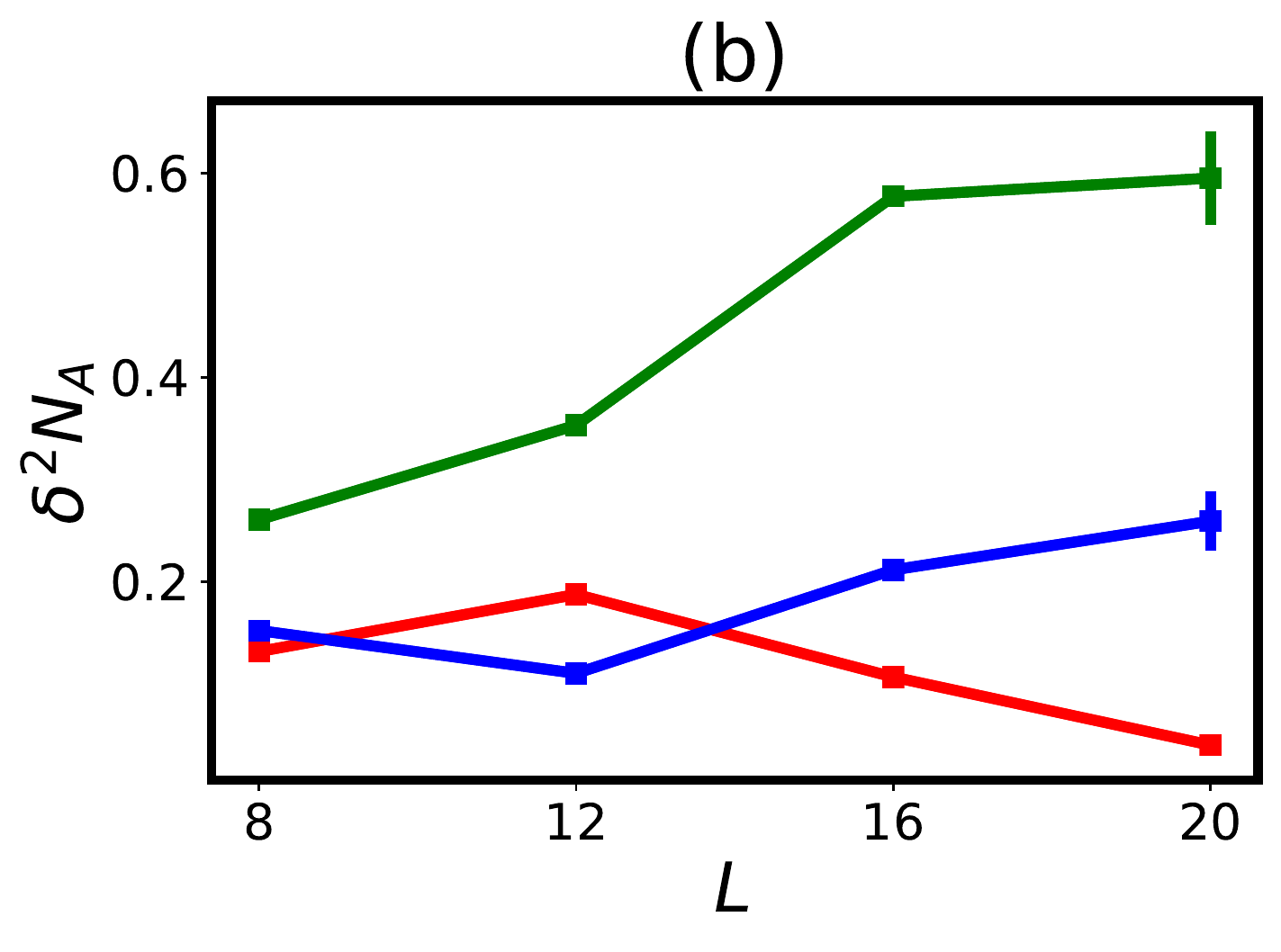}
\includegraphics[width=0.49\columnwidth,height=3.6cm]{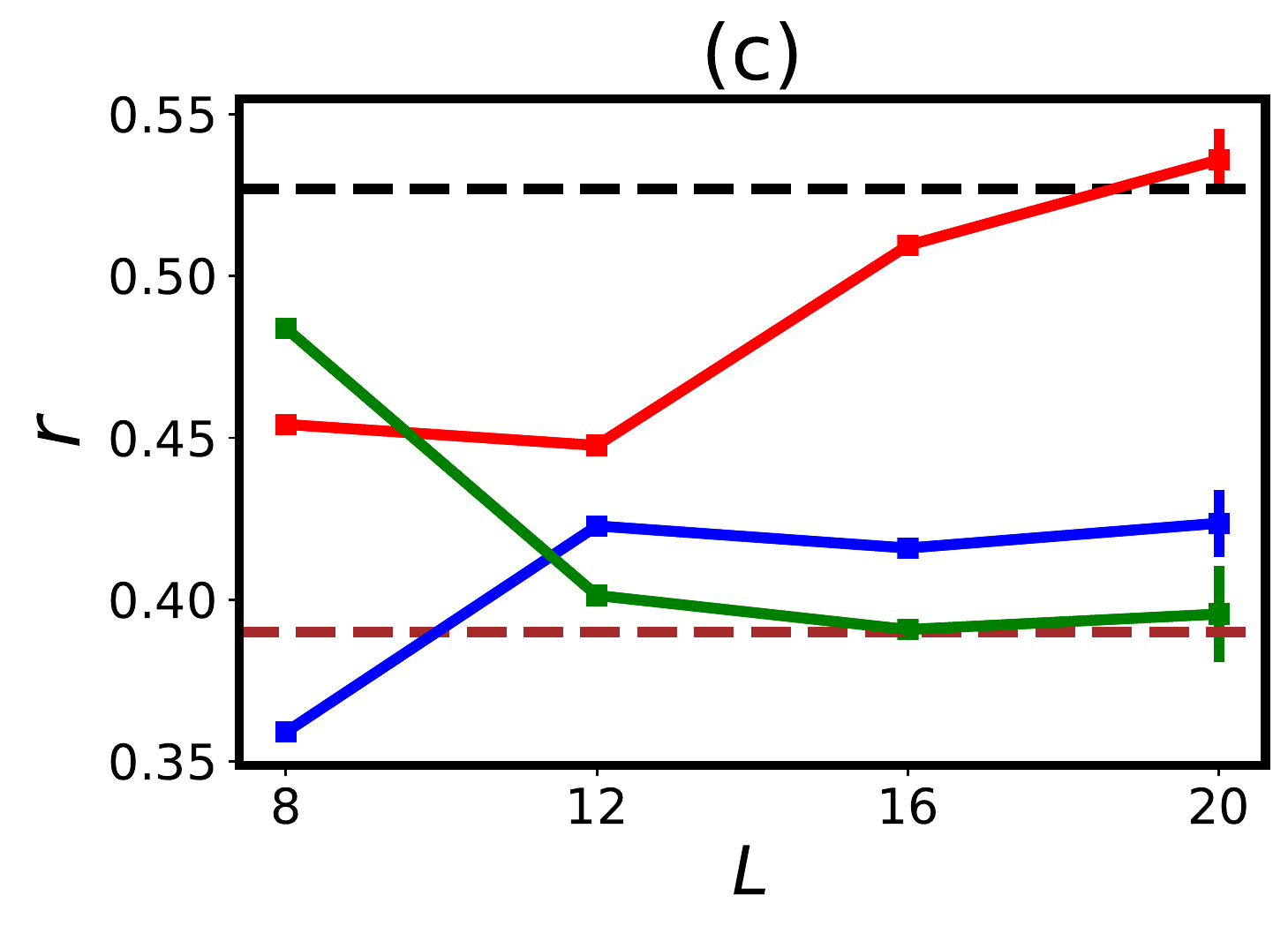}
\includegraphics[width=0.49\columnwidth,height=3.6cm]{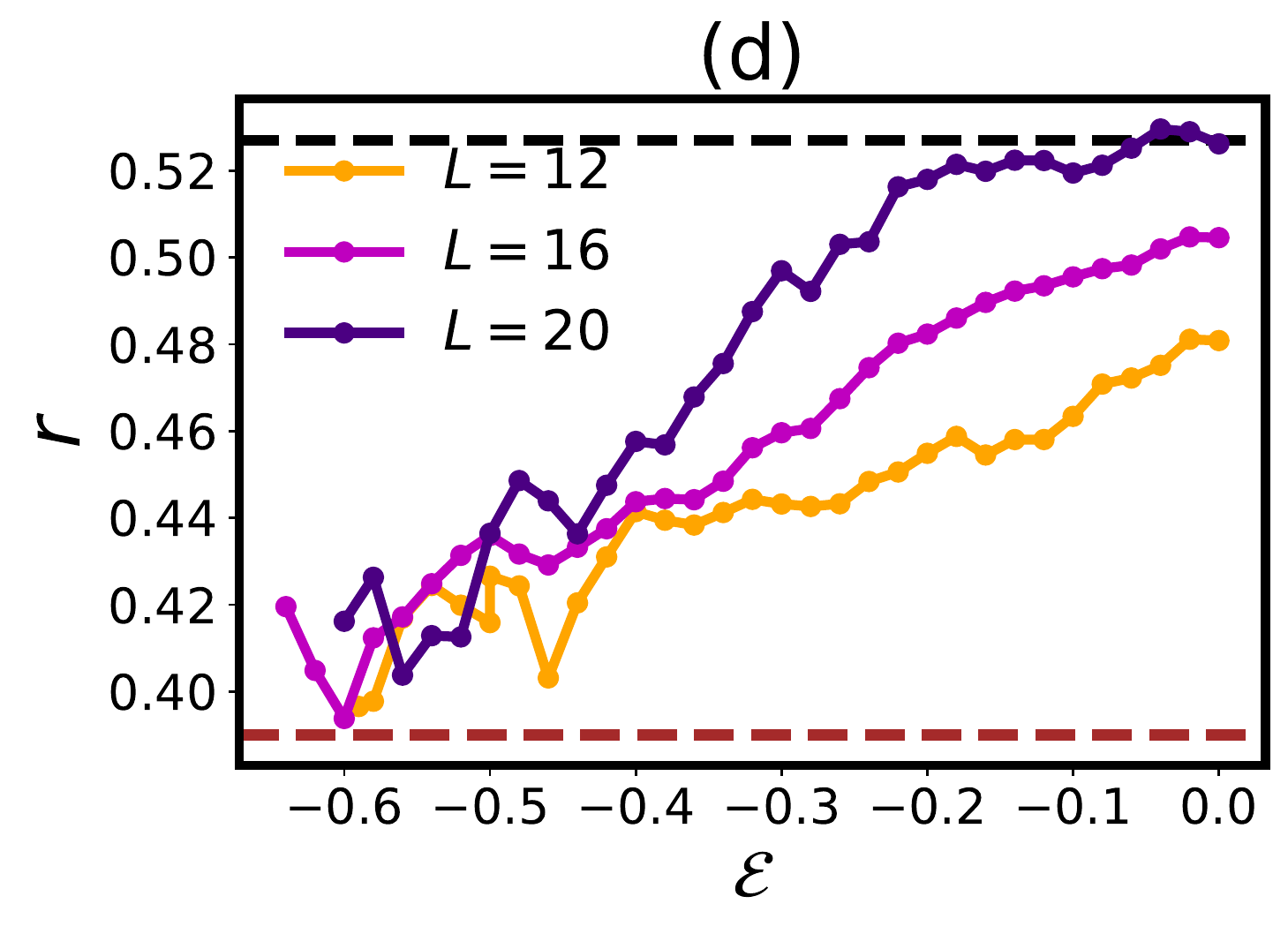}
\caption{{\bf Standard diagnostics for MBL, NEE and ergodic phases:} (a) The half-chain ($A$) entanglement entropy $S_A$, (b) subsytem particle number fluctuations (variance) $\delta^2N_A$, and (c) energy level-spacing ratio $r$  as function of $L$ in the three phases, ergodic ($h=0.6,\E=0$), NEE ($h=0.6,\E=-0.49$), and MBL ($h=1.8,\E=-0.49$). (d) $r$ as a function of $\mathcal{E}$ for increasing $L$ for $h=0.6$.}
\label{old_measures}
\end{figure}

Previous studies~\cite{ghosh2020transport,li2015many} have identified MBL, NEE and ergodic phases in the many-body energy spectrum of the interacting model of Eq.\eqref{ham} as a function of energy density $\mathcal{E}$ and $h$. 
We reconfirm this in Fig.~\ref{old_measures}(a-c) by computing three well-known quantities, the bipartite entanglement entropy $S_A$ of the subsystem $A$ consisting of the half of the chain, the variance of particle number $\delta^2N_A$ in $A$, and the energy level-spacing ratio $r$. We calculate these quantities via ED by choosing three combinations of $h$ and $\mathcal{E}$ such that we have robust ergodic ($h=0.6,\E=0$), NEE ($h=0.6,\E=-0.49$), and MBL ($h=1.8,\E=-0.49$) states, i.e., we are deep within the phases. We also look at another combination, $h=0.6,\E=-0.66$, which should correspond to the MBL phase based on previous studies \cite{ghosh2020transport}. However, as we discuss later, we find that states for this parameter do not show very clear-cut MBL behaviors; they appear MBL-like in some diagnostics and NEE-like in others. Below we briefly describe the classification of the phases based on these diagnostics. A summary can be found in Table \ref{tab:Table}. 

{\it Half-chain entanglement entropy $S_A$.--} The entanglement entropy is obtained as $S_A=-\mathrm{Tr}(\rho_A\ln\rho_A)$ from the reduced density matrix $\rho_A=\mathrm{Tr}_B(\rho)$ for the pure-state density matrix, $\rho=\ket{\Psi_\E}\bra{\Psi_\E}$. Here $|\Psi_\mathcal{E}\rangle$ is a many-body eigenstate  at an energy density $\mathcal{E}$. As shown in Fig.\ref{old_measures}(a), $S_A$ increases with $L$ in ergodic and NEE phases, implying a volume-law entanglement $\sim L$. On the contrary, $S_A$ remains almost independent of system size in the MBL phase, i.e., exhibits an area law $S_A\sim L^0$, as expected \cite{luitz2015many}. Thus, though the system-size dependence of bipartite entanglement can tell MBL and the extended states apart, NEE and ergodic phases cannot be distinguished easily based on this diagnostic. Ergodic eigenstates have a thermal volume-law entanglement, i.e., $S_A\simeq s_{th}(\mathcal{E})L/2$, with $s_{th}(\mathcal{E})$ thermal entropy per site at energy density $\mathcal{E}$, for large $L$. NEE states, on the other hand, are expected to exhibit \cite{Modak2018,wang2021many} a sub-thermal volume-law entanglement entropy, i.e., the coefficient of linear $L$ dependence less than $s_{th}(\mathcal{E})$. However, this distinction might be hard to verify for the limited system sizes accessed in ED \cite{Modak2018}.

{\it Subsystem particle number variance.--} $\delta^2N_A$ measures fluctuations of total number of particles $\hat{N}_A=\sum_{i=1}^{L/2}n_i$ in the subsystem $A$ compared to the average  number $N_A=\sum_{i=1}^{L/2} \bra{\Psi_\E}n_i\ket{\Psi_\E}$ at an energy density $\mathcal{E}$. As shown in Fig.~\ref{old_measures}(b), $\delta^2N_A$ decreases with $L$ in the ergodic phase, as expected from ETH~\cite{abanin2019colloquium,alet2018many,nandkishore2015many2,abanin2017recent}, while it increases and then tends to saturate with $L$ in MBL and the NEE phases~\cite{ghosh2020transport}. As a result, this quantity can differentiate the ergodic states from non-ergodic states.

{\it Level-spacing ratio.--} The level-spacing ratio \cite{oganesyan2007localization,Atas2013} $r_i=\min(s_i,s_{i+1})/\max(s_i,s_{i+1})$ is obtained from $s_i=E_{i+1}-E_i$ with $E_i$'s being the many-body energy eigenvalues arranged in ascending order. We compute the arithmetic mean of $r_i$ to obtain the average level-spacing ratio $r(\mathcal{E})$ at energy density $\mathcal{E}$. The ergodic phase can be identified with the gaussian-orthogonal ensemble (GOE) value $r\simeq 0.528$ and the MBL phase with the Poissonian value $r\simeq 0.386$. In Fig.~\ref{old_measures}(c), $r$ approaches GOE and Poissonian values with increasing $L$ for the ergodic and MBL phases, respectively, whereas $r$ tends to an intermediate value for the NEE phase. We also discuss the level-spacing distribution in the three phases in Appendix~\ref{app1}.


Since $r$ is expected to change discontinuously across the MBL-to-ergodic transition, $r$ has been used \cite{oganesyan2007localization,luitz2015many,abanin2021distinguishing,Suntajs2020,Aramthottil2021} as an ``order parameter" to detect the MBL transition, e.g., through finite-size scaling analysis of $r$ in the models with the random and quasiperiodic disorder. However, as we show in Fig.\ref{old_measures}(d) for $h=0.6$, $r$ is not a good diagnostic of the MBL-NEE and NEE-MBL transitions in the quasiperiodic model [Eq.\eqref{ham}] for the system sizes accessed in ED. $r$ fluctuates \cite{ghosh2020transport} a lot as a function of $\mathcal{E}$ and $L$ in the putative non-ergodic phases, even after averaging over a large number of values of $\phi$. As a result, we cannot do a reasonable finite-size scaling analysis of $r$ for the transitions in the GAAH model. We show later that the FS diagnostics vary smoothly across the NEE-MBL transition and thus enable us to do more controlled finite-size scaling. We also show in Appendix~\ref{app1} that the level spacing statistics do not exhibit proper Poisson statistics in the putative MBL phase for $h=0.6$, presumably because the corresponding states are too close to the ground state in energy. As a result, distinguishing states at finite-energy density (relative to the ground state)  and obtaining good statistics for them by energy binning becomes challenging. 
Hence, to attain a clear distinction of the phases, we look at MBL states for ($h=1.8,\E=-0.49$), where the level statistics convincingly converge to Poisson distribution, as shown in Appendix~\ref{app1}. A comparison between the level statistics for ($h=1.8,\E=-0.49$) and ($h=0.6,\E=-0.66$) can be found in Fig.~\ref{levelstat}(c,d).

In the next section, we provide the anatomy of the above phases in terms of single-particle excitations. 


\section{Single-particle excitations and MBL proximity effect}\label{sec4}
In this section, we characterize single-particle excitations in the MBL, NEE, and ergodic phases via eigenstate single-particle Green's function and the associated local density of states (LDOS).  The single-particle Green's function in the $n$-th many-body eigenstate $\ket{\Psi_n}$ with energy $E_n$ of the $N$-particle system is obtained from $\mathcal{G}_n(i,j,t)=-i\theta(t)\bra{\Psi_n}\{c_i(t),c^\dagger_j(0)\}\ket{\Psi_n}$ for sites $i$ and $j$. The Fourier transform of the onsite element $\mathcal{G}_n(i,i,t)=\mathcal{G}_n(i,t)$ is   
\begin{align} \label{eq:SPGreenFn}
\mathcal{G}_n(i,\omega)&=\sum\limits_{m}\left[\frac{|\bra{\Psi^+_{m}}c_i^\dagger \ket{\Psi_n}|^2}{\omega  + i\eta - E_{m} + E_n}+ \frac{|\bra{\Psi^-_{m}}c_i \ket{\Psi_n}|^2}{\omega + i\eta + E_{m} - E_n}\right]. 
\end{align}
$\ket{\Psi^+_{m}}$ and $\ket{\Psi^-_{m}}$ are the $m$-th eigenstate with energy $E_m$ of the system with $N+1$ and $N-1$ particles, respectively. For the interacting system ($V\neq 0$), the broadening parameter $\eta$ is taken to be the typical value or the geometric mean of the many-body level spacing ($\sim e^{-L}$) at energy $E_n$ (see Appendix \ref{app3} for details). 

The single-particle excitation at energy $\omega$ is characterized by the local density of states (LDOS) $\rho_n(i,\omega) = -(1/\pi) \mathrm{Im}[\mathcal{G}_n(i,\omega)]$. In particular, we obtain the typical LDOS $\rho_t(\omega)$, the geometric mean, from $\ln\rho_t(\omega)=\langle \ln \rho_n(i,\omega)\rangle$ and the average LDOS as $\rho_a(\omega)=\langle \rho_n(i,\omega)\rangle$, where $\langle... \rangle$ denotes an arithmetic average over the lattice sites and $\phi$. In the localized phase, for both non-interacting ($V=0$) and interacting ($V\neq 0$) systems, the local single-particle excitations originate from a finite number of discrete poles of the Green's function $\mathcal{G}_n(i,\omega)$, effectively corresponding to a finite system having the size of the localization length. Thus the poles of $\mathcal{G}_n(i,\omega)$ lead to discrete peaks in $\omega$, having zero measure in the LDOS even in the thermodynamic limit.
As a result, the typical value $\rho_t(\omega)$ decreases with system sizes and $\rho_t(\omega)\to 0$ in the thermodynamic limit. In contrast, the poles of $\mathcal{G}_n(i,\omega)$ form a continuum in the delocalized phase for $L\to\infty$ and $\rho_t(\omega)$ approaches a non-zero value with increasing system size~\cite{ganeshan2015nearest} for $\omega$ lying within the single-particle bands of states. Thus, the typical LDOS $\rho_t(\omega)$ acts as a probabilistic order parameter~\cite{dobrosavljevic1997mean,dobrosavljevic2003typical,jana2021local} for localization of an excitation at energy $\omega$. 
In contrast, the arithmetic mean $\rho_a(\omega)$ averages LDOS over all the sites and approaches non-zero value with increasing system size both in the delocalized and localized phases. 


\subsection{Noninteracting system: $V=0$}
\label{sec4a}
\begin{figure}[h]
\centering
\stackon{\includegraphics[width=0.4925\columnwidth,height=3.6cm]{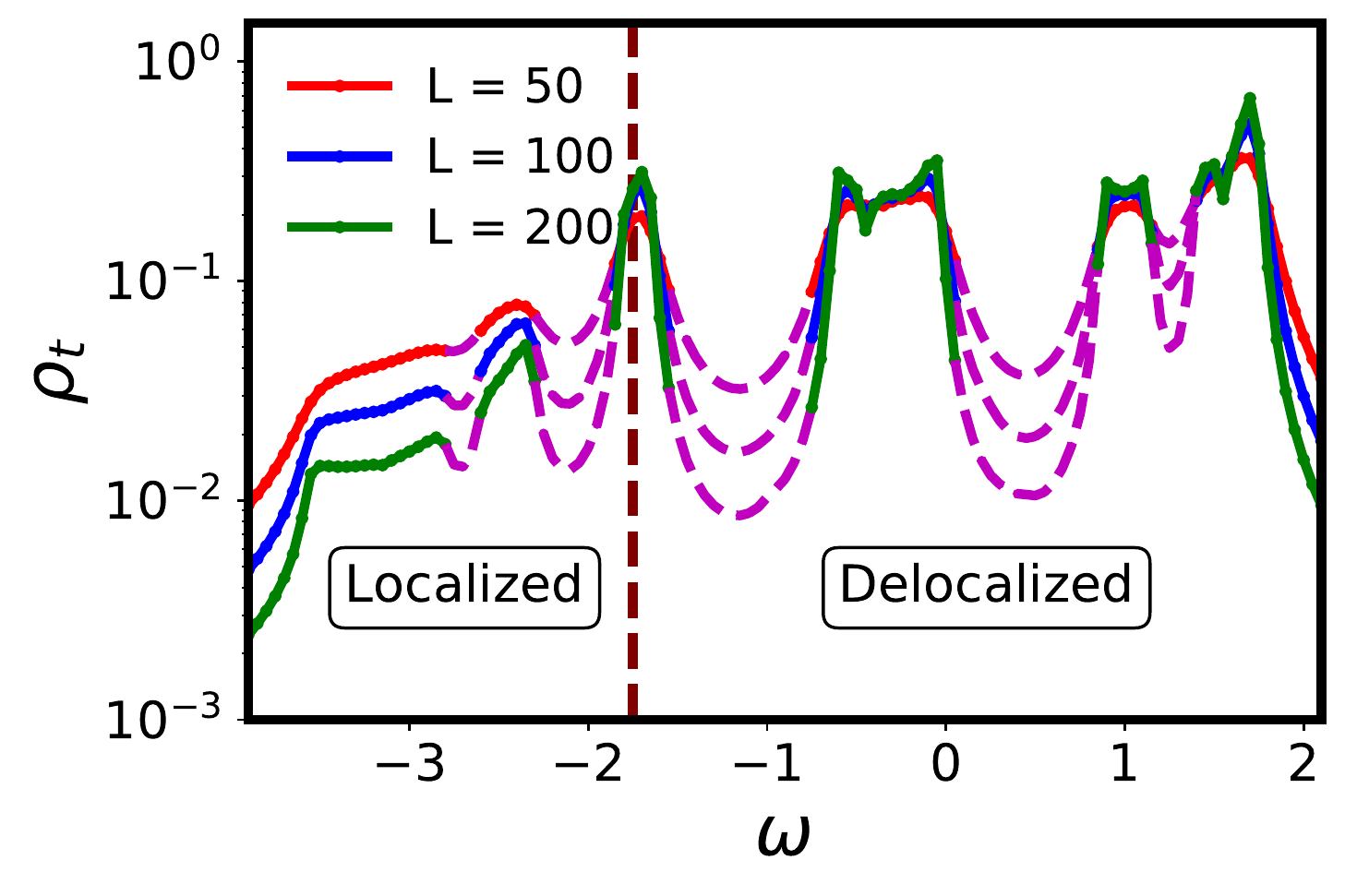}}{(a)}
\stackon{\includegraphics[width=0.4925\columnwidth,height=3.6cm]{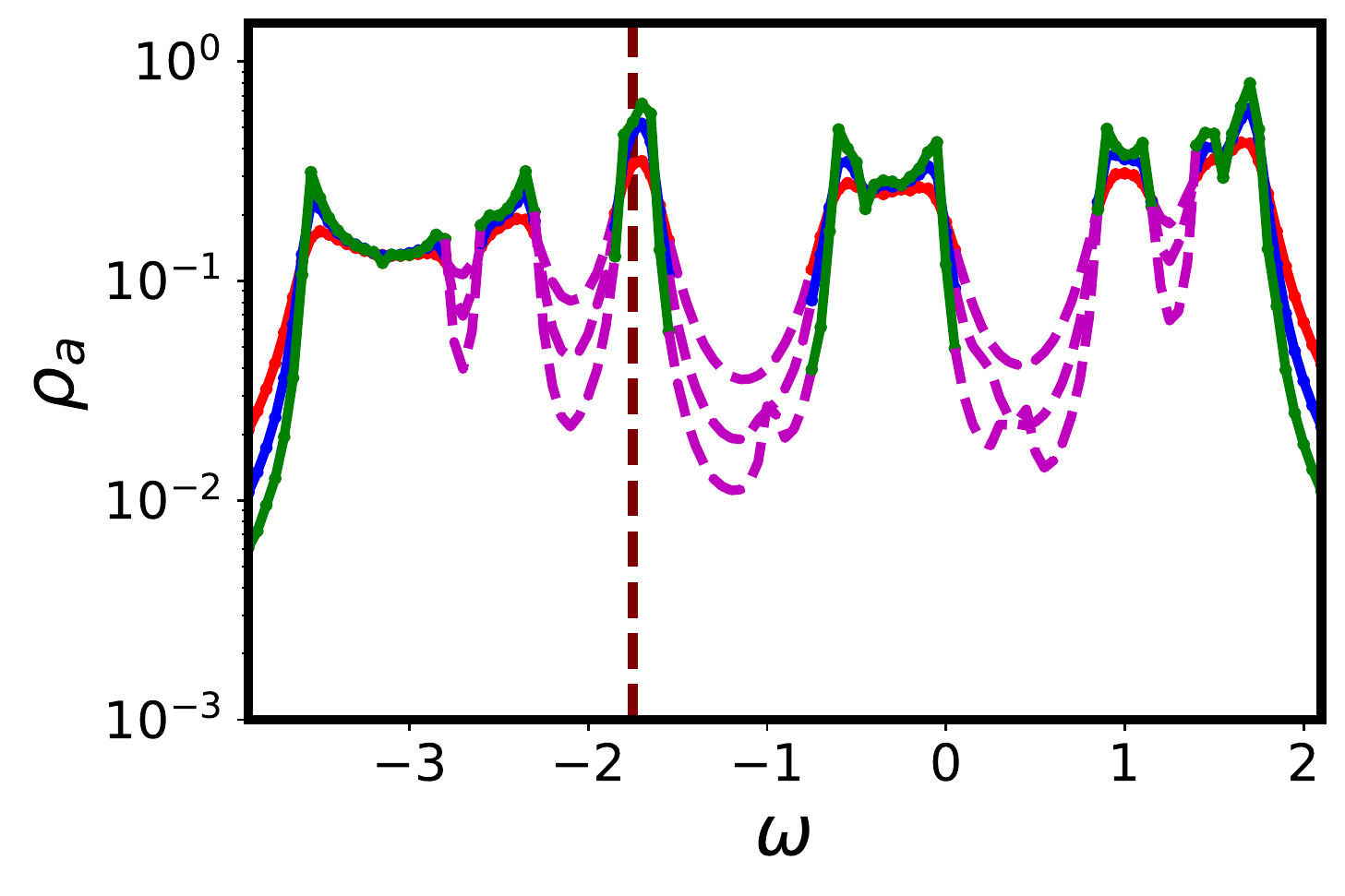}}{(b)}
\caption{{\bf LDOS for noninteracting $(V=0)$ GAAH model:} (a) $\rho_t(\omega)$ vs. $\omega$ for increasing system sizes $L$. (b) $\rho_a(\omega)$ vs. $\omega$ for increasing $L$. The vertical dashed line shows the position of the single particle mobility edge for $\alpha=-0.8$ and $h=0.6$. The number of disorder realizations over $\phi$ is at least $100$ for all the plots.}
\label{nonint}
\end{figure}
\begin{figure*}[htpb]
\centering
\includegraphics[width=0.49\columnwidth,height=3.75cm]{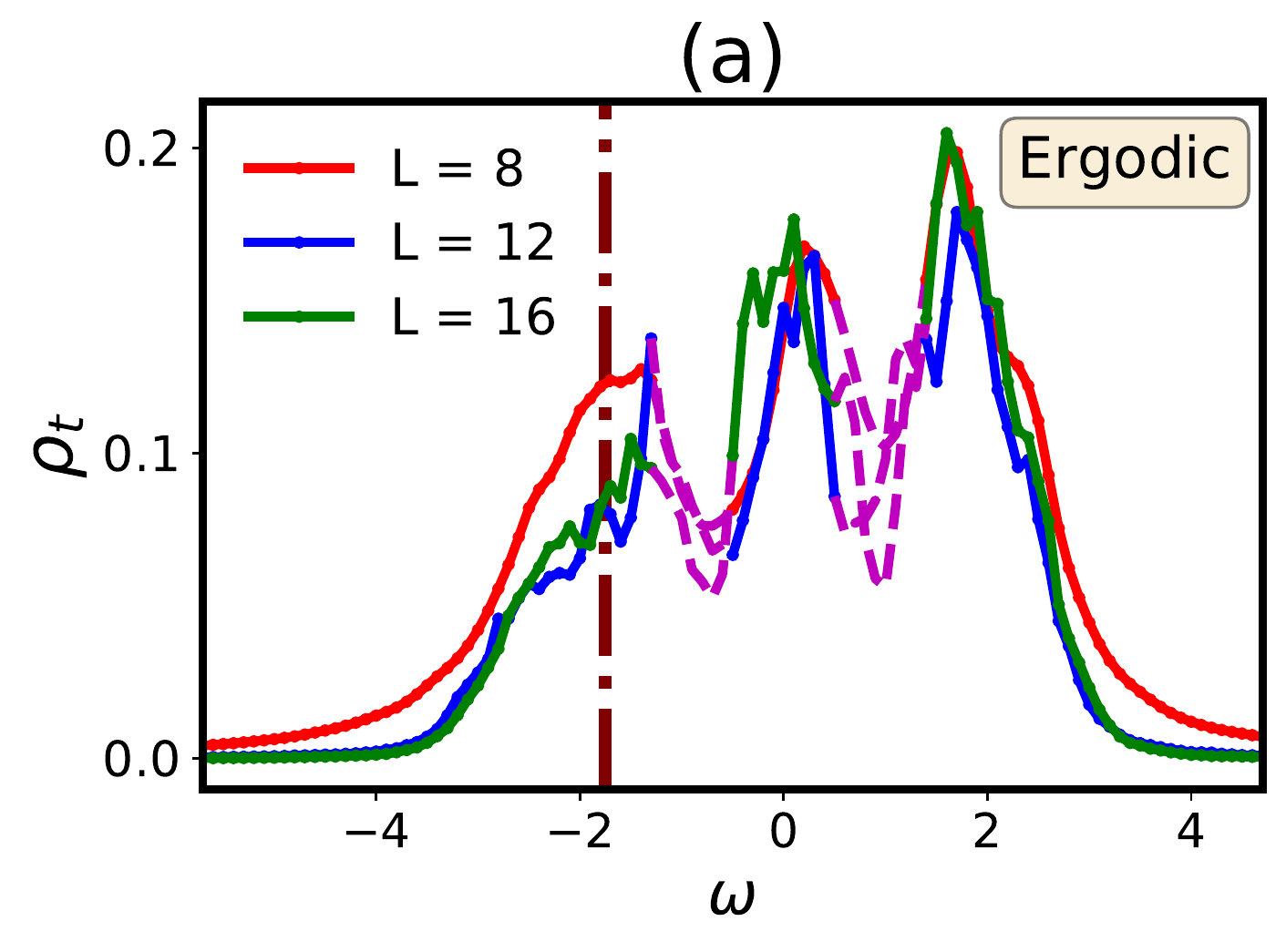}
\includegraphics[width=0.49\columnwidth,height=3.75cm]{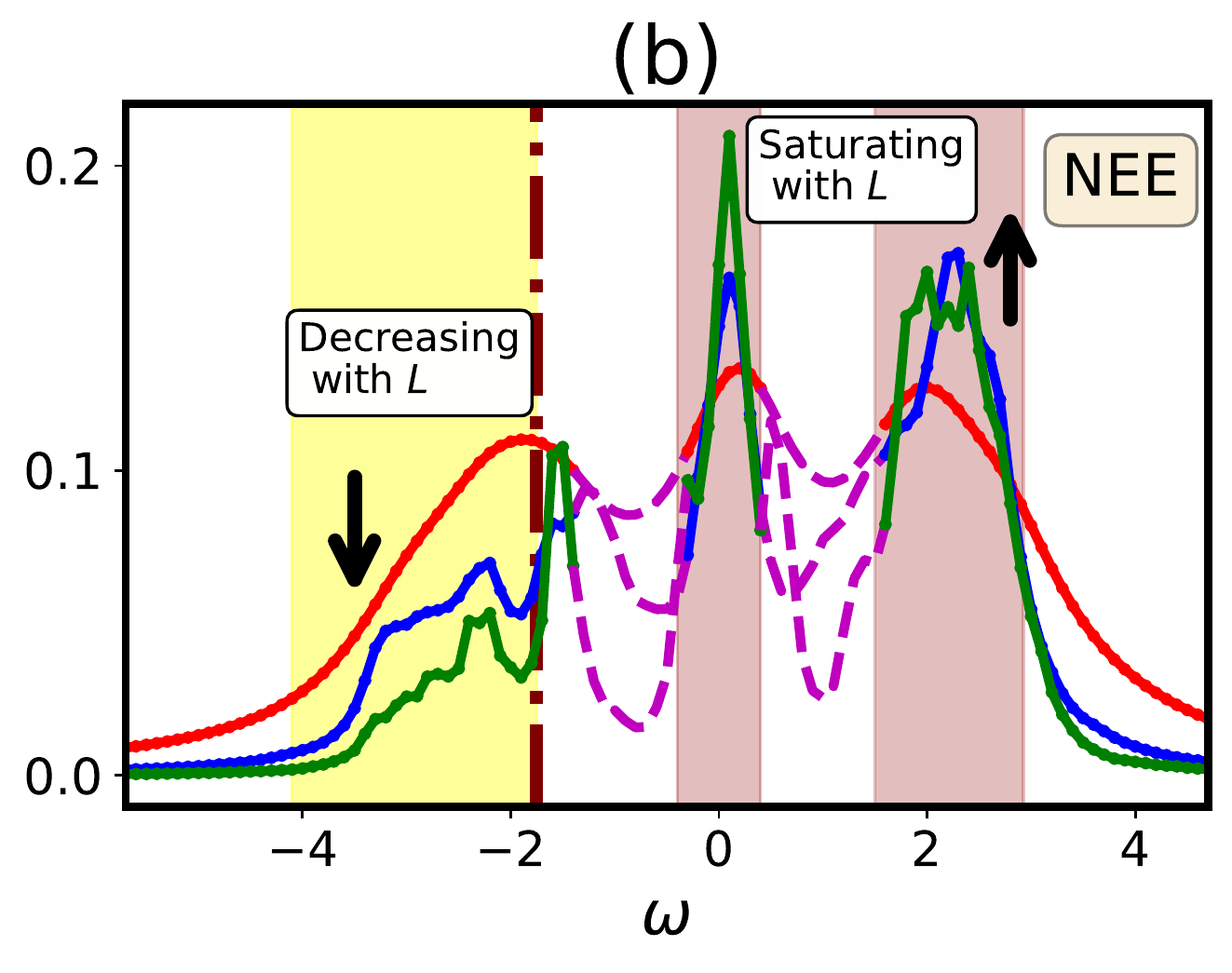}
\includegraphics[width=0.49\columnwidth,height=3.75cm]{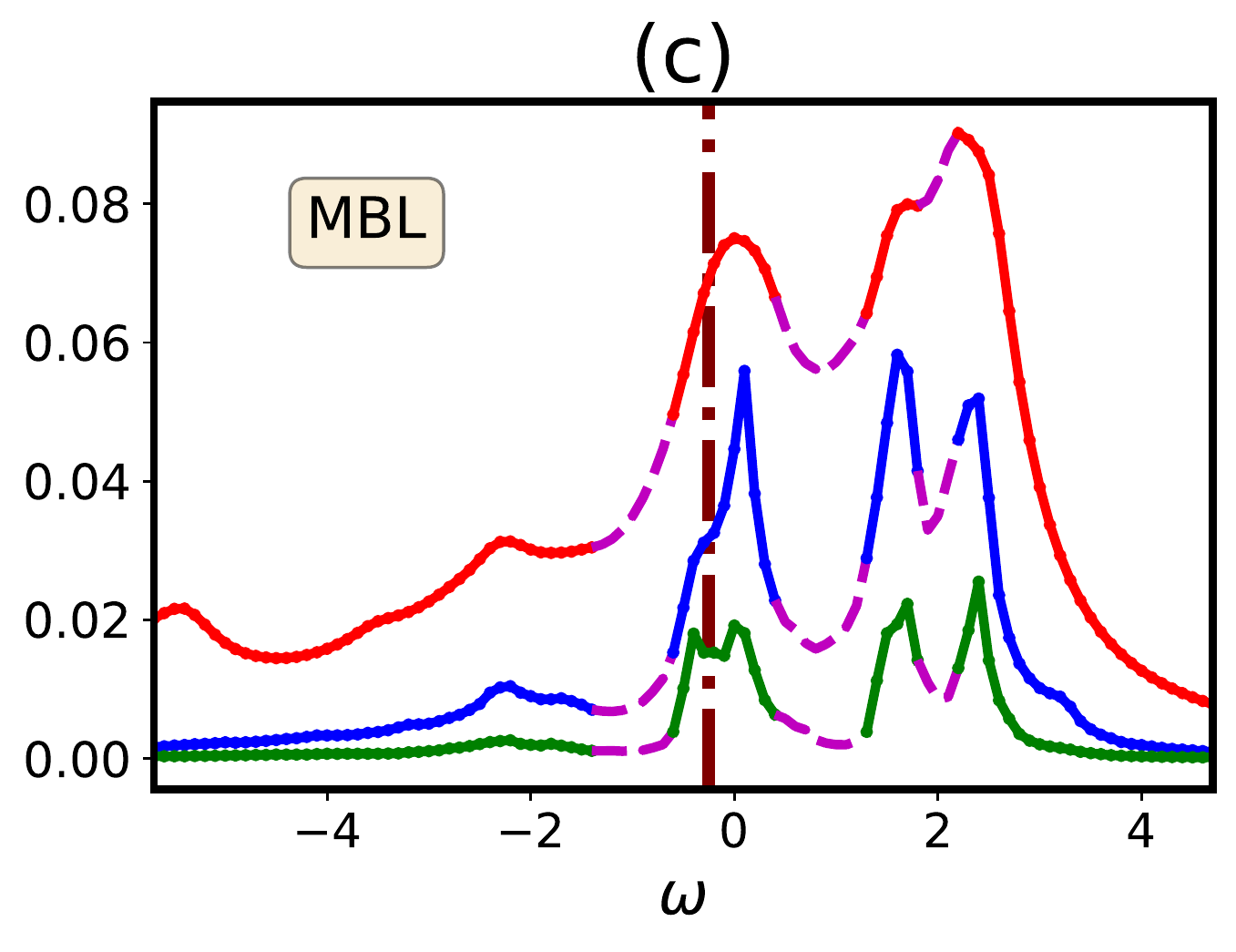}
\includegraphics[width=0.49\columnwidth,height=3.75cm]{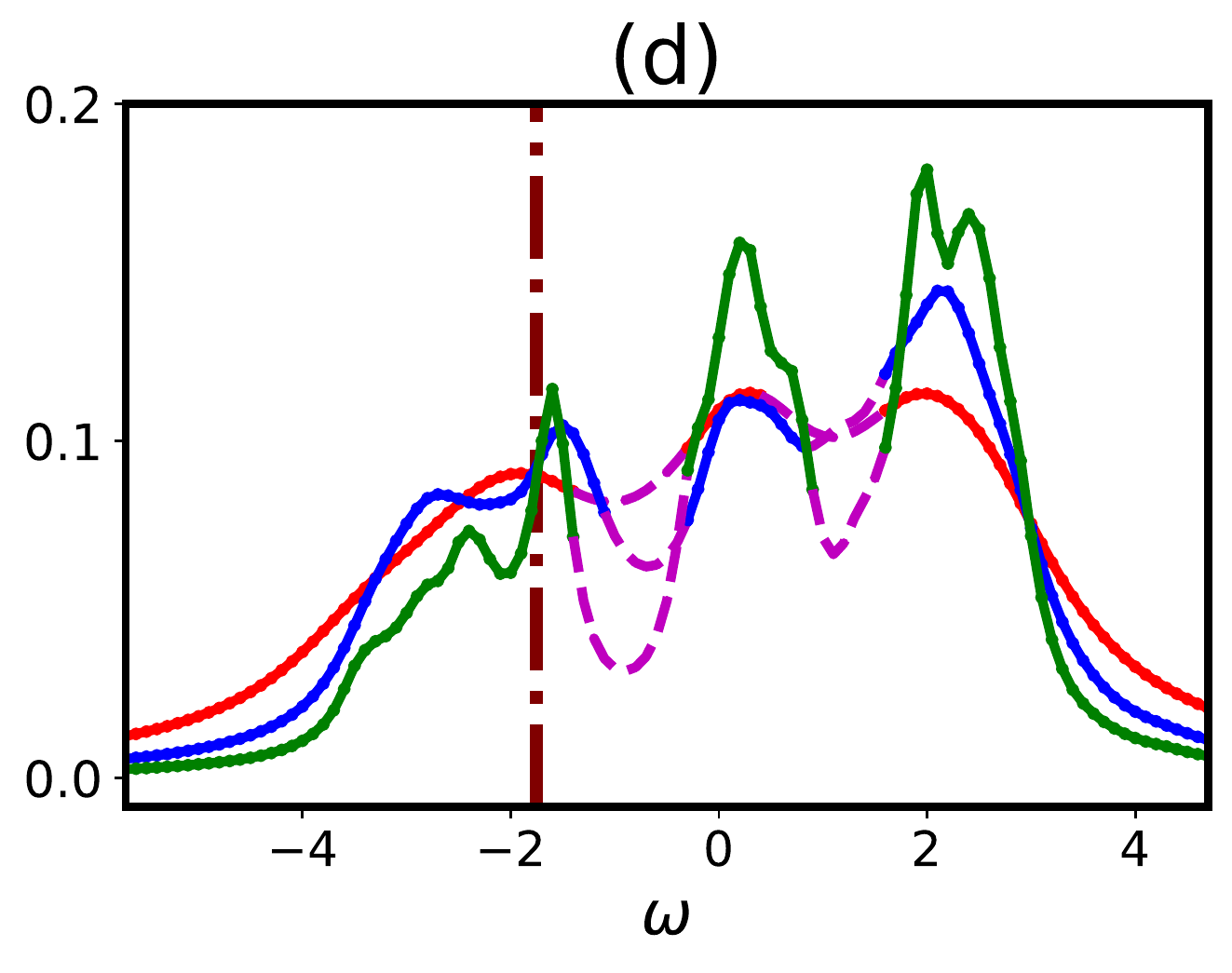}
\caption{{\bf Single-particle excitations in the interacting GAAH model:} (a) $\rho_t$ vs $\omega$ for increasing $L$ in the ergodic phase ($h=0.6$, $\mathcal{E}=0$), (b) NEE phase ($h=0.6$, $\mathcal{E}=-0.49$), (c) MBL phase ($h=1.8$, $\mathcal{E}=-0.49$) and (d) for ($h=0.6$, $\mathcal{E}=-0.66$). The vertical dot-dashed line shows the location of the single particle mobility edge in the non-interacting limit.}
\label{mbl_proximity}
\end{figure*} 

In the non-interacting limit, $(V=0)$, Eq.\eqref{eq:SPGreenFn} can be simplified and the local single-particle density of states (LDOS) can be written as,
\begin{eqnarray}
\rho(i,\omega) =\frac{1}{\pi} \sum_{\nu=1}^{L} |\psi_\nu(i)|^2 \frac{\eta_s}{(\omega-\epsilon_\nu)^2 + \eta_s^2}.
\label{ldos_nonint}
\end{eqnarray}
Here $\psi_\nu(i)$ and $\epsilon_\nu$ are the single particle eigen function and energy, that can be obtained by diagonalizing the non-interacting GAAH model. In this case, the broadening parameter $\eta_s$ is chosen to be the mean single-particle energy level spacing ($\sim 1/L$). We then calculate the typical value $\rho_t(\omega)$ of LDOS, and the average value of LDOS $\rho_a(\omega)$ as discussed earlier. 

The $L$ dependence of $\rho_t(\omega)$ can be used to detect the single-particle mobility edge $\epsilon_c=\text{sgn}(h)(2|t|-|h|)/\alpha$ \cite{ganeshan2015nearest} of the GAAH model in the non-interacting limit ($V=0$). Apart from the mobility edge, the single-particle spectrum of the GAAH model also has gaps, i.e. $\mathcal{O}(1)$ interval of $\omega$ that does not contain any eigenenergy $\epsilon_\nu$. We use $\rho_t(\omega)$ in combination with $\rho_a(\omega)$ to classify for small finite systems -- (a) localized excitation, when $\rho_t(\omega)$ decreases and $\rho_a(\omega)$ approaches a finite value with $L$, (b) delocalized excitation, when  both $\rho_t(\omega)$ and $\rho_a(\omega)$ tend to saturate with $L$, and (c) \emph{gapped} excitation, when both $\rho_t(\omega)$ and $\rho_a(\omega)$ decrease with $L$.
 In Fig.~\ref{nonint}, we show that single-particle excitations are localized for $\omega<\epsilon_c$, where $\rho_t$ decreases with $L$ but $\rho_a$ does not. For $\omega>\epsilon_c$, the excitations are delocalized, and both $\rho_t$ and $\rho_a$ remain finite in the thermodynamic limit. In Fig.~\ref{nonint}, the gapped region are marked by dashed curves where both $\rho_t(\omega)$ and $\rho_a(\omega)$ decrease with $L$. Hence by combining both $\rho_t(\omega)$ and $\rho_a(\omega)$, we are able to detect the mobility edge as well as the gapped region in the single-particle excitation spectrum of the GAAH model. In the following, we employ the same diagnostics to look for localized and delocalized excitations in the interacting system.

\subsection{Interacting system: $V\neq 0$}\label{sec4b}
In the interacting case, we use Eq.\eqref{eq:SPGreenFn} to obtain the LDOS via ED for system sizes $L=8,12,16$. Remarkably, as we show in Fig.~\ref{mbl_proximity}(a-c), different types of excitations, i.e., localized, delocalized, and gapped, as discussed in the preceding section, also exist for $V\neq 0$.
For the ergodic phase ($h=0.6,~\mathcal{E}=0$) [Fig.~\ref{mbl_proximity}(a)] $\rho_t(\omega)$ approaches a finite value over the entire band ($|\omega|\lesssim 4$) except the gapped region (dashed line), implying many-body delocalization of all single-particle excitations due to interaction. 

In contrast, in the MBL phase ($h=1.8,~\mathcal{E}=-0.49$) [Fig.~\ref{mbl_proximity}(c)] all single-particle excitations, below and above the non-interacting mobility edge $\epsilon_c$, are localized, as evinced by the reduction of $\rho_t(\omega)$ for all $\omega$ with $L$. This is a direct signature of the MBL proximity effect~\cite{nandkishore2015many,Hyatt2017,Marino2018}. Through this mechanism, an otherwise delocalized system can become localized when coupled with a localized system. The delocalized system effectively sees an additional disorder through the coupling to the localized system~\cite{nandkishore2015many}. The MBL proximity effect has been studied via perturbative and ED calculations~\cite{nandkishore2015many,Hyatt2017,Marino2018} in two coupled chains of particles or spins. In this ladder-like system, one of the chains is in the delocalized phase and the other in the MBL phase, and the chains are coupled via local \emph{density-density} type interaction. Refs.\onlinecite{nandkishore2015many,Hyatt2017,Marino2018} have shown that the delocalized chain can become localized due to the coupling with the MBL chain.

In previous studies \cite{modak2015many,Li2016,Deng2017,Modak2018,Deng2019,ghosh2020transport}, the MBL proximity effect has been invoked to rationalize the existence of the MBL phase in the GAAH model with single-particle mobility edge. In this case, the Hamiltonian of Eq.\eqref{ham} can be rewritten in the basis of the single-particle eigenstates $\psi_{\nu}(i)$ as
\begin{align}
H &=\sum_{\mu}\epsilon_\mu c_\mu^\dagger c_\mu+\sum_{\mu\nu\delta\gamma} V_{\mu\nu\delta\gamma} c_\mu^\dagger c_\nu^\dagger c_\delta c_\gamma,
\end{align}
where $c_\mu^\dagger=\sum_i \psi_\mu^*(i)c_i^\dagger$ and $V_{\mu\nu\gamma\delta}=V\sum_i \psi_\mu^*(i)\psi_\nu^*(i+1)\psi_\gamma(i)\psi_\delta(i+1)$. Thus, the single-particle states for $\epsilon_\nu>\epsilon_c$ constitute the delocalized system and those for $\epsilon_\nu<\epsilon_c$ form the localized system here. They are coupled via more generic and non-local interaction than the simpler models considered in previous studies~\cite{nandkishore2015many,Hyatt2017,Marino2018} of MBL proximity effect. Nevertheless, we can clearly observe the MBL proximity effect in Fig.~\ref{mbl_proximity}(c), where the delocalized single-particle excitations ($\epsilon>\epsilon_c$) of the non-interacting ($V=0$) system are localized in the presence of interaction $V\neq 0$, presumably due to the coupling with the localized single-particle states ($\epsilon<\epsilon_c$). 

On the contrary, in the NEE phase ($h=0.6,~\mathcal{E}=-0.49$), $\rho_t(\omega)$ decreases with $L$ for $\omega\lesssim \epsilon_c$ and approaches a finite value increasing with $L$ for $\omega\gtrsim \epsilon_c$, as shown in Fig.~\ref{mbl_proximity}(b). This clearly indicates the persistence of \emph{many-body} single-particle mobility edge, that separates localized and delocalized excitation even for $V\neq 0$, in the NEE phase. The mobility edge for single-particle excitations can be deduced more clearly in the semilog plots of Fig.~\ref{logrhotyp}(a-b) in Appendix~\ref{app2}. Thus, in the NEE phase, neither the localized single-particle states are able to localize all the delocalized excitations via the MBL proximity effect, nor the delocalized states are able to act as a bath to delocalize all the localized excitations via interaction. However, it is not possible to determine the mobility edge for single-particle excitations accurately for the interacting case ($V\neq 0$), e.g., from Fig.~\ref{mbl_proximity}(b).

Fig.~\ref{mbl_proximity}(d) shows single-particle excitations for ($h=0.6,\E=-0.66$). In terms of the single-particle excitations, the states at this parameter value, which has been previously characterized as part of the MBL phase \cite{ghosh2020transport}, are hardly distinguishable from the NEE states. This is consistent with the level statistics not converging to the Poisson value in this regime as discussed earlier (Appendix~\ref{app1}). Although the states in this regime show MBL-like behavior through $S_A$~\cite{ghosh2020transport}, i.e. $S_A$ approaches an area-law (constant), for system sizes accessible in ED. Note, however, that the computation of the LDOS requires ED in three particle sectors ($N-1,N,N+1$), as evident from Eq.\eqref{eq:SPGreenFn}, and thus is limited to smaller system sizes ($L\leq 16$) than those employed for the calculations of $S_A$, $\delta^2N_A$, and Fock-space diagnostics, discussed later. As a result, the NEE-like single-particle excitation spectrum [Fig.~\ref{mbl_proximity}(d)] for the MBL states at $(h=0.6,~\E=-0.66)$ might be an artifact of the limited system size, and the energy binning too close to the ground state, as discussed earlier in Sec.\ref{sec3}. The NEE-like level spacing statistics (Appendix~\ref{app1}) at this parameter value might also be due to energy binning. Future studies with larger systems and finer energy binning is required for  $(h=0.6,~\E=-0.66)$ to resolve this issue. 

Overall, we find that qualitative distinctions between MBL, NEE, and ergodic phases in the GAAH model can be made based on single-particle excitations in real space, as captured by typical LDOS. Thus the latter provides a diagnostic complementary to standard diagnostics, like entanglement entropy, subsystem particle number fluctuations, and level spacing statistics, to distinguish the phases (Table \ref{tab:Table}). Due to the many-body nature of an interacting system, yet another complementary perspective ~\cite{Altland2017,Logan,logan2019many,Roy2020,Roy2020a,Ghosh.2019,roy2021fock,mace2019multifractal,de2021rare} of the phases and non-ergodic to ergodic phase transition can be obtained by looking at the localization and ergodicity in the Fock space, as we discuss in the next section. 

\section{Fock-space propagator}\label{sec5}
In the Fock-space the Hamiltonian of Eq.~\eqref{ham} can be rewritten as a tight-binding model in terms of the occupation number basis $\{\ket{I}\}$ as~\cite{logan2019many,welsh2018simple,ghosh2019many}
\begin{eqnarray}
H=\sum\limits_{I,J} T_{IJ} \ket{I}\bra{J} + \sum\limits_{I} E_I \ket{I}\bra{I},
\end{eqnarray}
where $\ket{I}=\ket{n_{I1}n_{I2}...n_{IL}}$ with onsite real-space occupation $n_i\in0\text{ or}1$. Here ``FS hopping" $T_{IJ}=-t$ when $\ket{I}$ and $\ket{J}$ are connected by a single nearest neighbor hop in real space and $T_{IJ}=0$ otherwise. The onsite potential at the FS site $I$, $E_I=\sum_i h_i n_{Ii} + V\sum_i n_{Ii} n_{I,i+1}$, acts like correlated disorder \cite{Altland2017,Logan,logan2019many,Roy2020,Roy2020a,Ghosh.2019}. 
The many-body density of states (MDOS) (per FS site) of the GAAH model, $D(E)=(1/\Nf)\sum_n\delta(E-E_n)$, for large $L$ approaches a Gaussian function of the many-body energy $E$ with the mean energy $\bar{E}\propto L$ and variance $\mu_E^2\propto L$, where the proportionality constants are found from ED (Appendix~\ref{app3}). In order to approach a well-defined thermodynamics limit through our numerical calculations we consider the rescaled Hamiltonian $\tilde{H}=H/\sqrt{L}$, as in the earlier studies~\cite{logan2019many,welsh2018simple,roy2020fock}.  

The FS sites can be organized in slices~\cite{sutradhar2022scaling}, such that any site in a particular slice is connected to the sites of nearest-neighbor slices via a single FS hopping, as shown in Fig.~\ref{linscaling}(a). This locality in the FS lattice allows for an efficient implementation of the standard recursive Green's function method~\cite{lee1981anderson,mackinnon1980conductivity,mackinnon1983scaling,verges1999computational}, which has been recently applied to FS lattice~\cite{sutradhar2022scaling} for a system with the random disorder. The scaled retarded FS propagator at energy $E$ is given by $G(\E)=(\E\sqrt{L}+i\eta-\tilde{H})^{-1}$ with a broadening $\eta=1/[\sqrt{L}\Nf D(E)]$, i.e. the scaled mean many-body level spacing, at the energy density $\E=E/L$. Note that that recursive Green's function method~\cite{lee1981anderson,mackinnon1980conductivity,mackinnon1983scaling,verges1999computational} obtains the $G(\E)$ exactly and there is no approximation involved here. The organization [Fig.~\ref{linscaling}(a)] of the FS lattice into slices facilitates a transparent implementation~\cite{sutradhar2022scaling} of the method in the Fock space. Here we also note that the calculations of FS propagator do not have any energy binning issue, unlike the other diagnostics discussed earlier, since the FS propagator by definition is calculated at given energy density $\E$.

In particular, we compute $G_{IJ}(\E)=\langle I\vert G(\E)\vert J\rangle$ for $I,J$ on the middle slice [Fig.~\ref{linscaling}(a)]. The diagonal element $G_{II}$ provides an order parameter \cite{sutradhar2022scaling} for non-ergodic-to-ergodic transition, namely typical value $\Delta_t=\exp{[\langle \ln\Delta_I\rangle]}$ of the imaginary part of the Feenberg self energy $\Delta_I(\E)=\text{Im}[G_{II}^{-1}(\E)]-\eta$. Here $\langle...\rangle$ denotes the average over disorder realizations and FS sites in the middle slice. The off-diagonal elements $G_{IJ}(\E)$ ($I\neq J$) encode information about the non-local propagation of an FS excitation. A \emph{Fock-space localization length} or decay length $\xi_F$ can be extracted from $G_{IJ}$ in the MBL phase, as we discuss below. For numerical computation in the FS, we average over $2000$ and $1000$ values of $\phi$ for $L=16$ and $20$, respectively. For $L=24$ we average over $400$ and $100$ $\phi$ values for the local and non-local propagators, respectively.


\subsection{Nonergodic-to-ergodic transition in the Fock-space}\label{sec6}
We study the transition as a function of energy density for $h=0.6$. Based on standard diagnostics like transport, entanglement entropy, and variance of local observable, previous studies \cite{ghosh2020transport,li2015many} have detected MBL-to-NEE and NEE-to-ergodic transition around energy density $\mathcal{E}_{1}\approx-0.60$ and $\mathcal{E}_{2}\approx-0.39$, respectively. 
\begin{figure}[h]
\centering
\includegraphics[width=0.44\columnwidth,height=3.35cm]{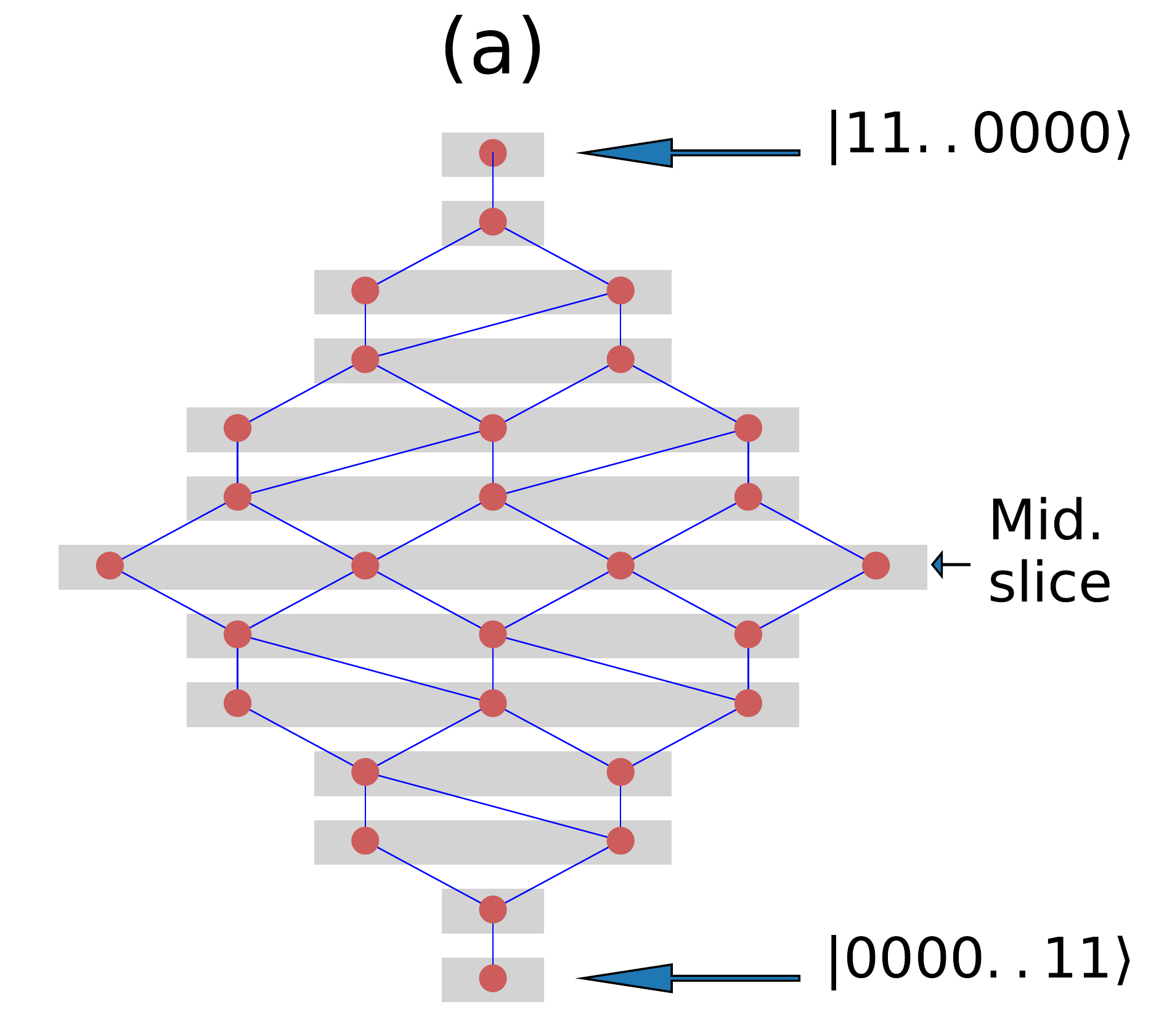}
\includegraphics[width=0.54\columnwidth,height=3.35cm]{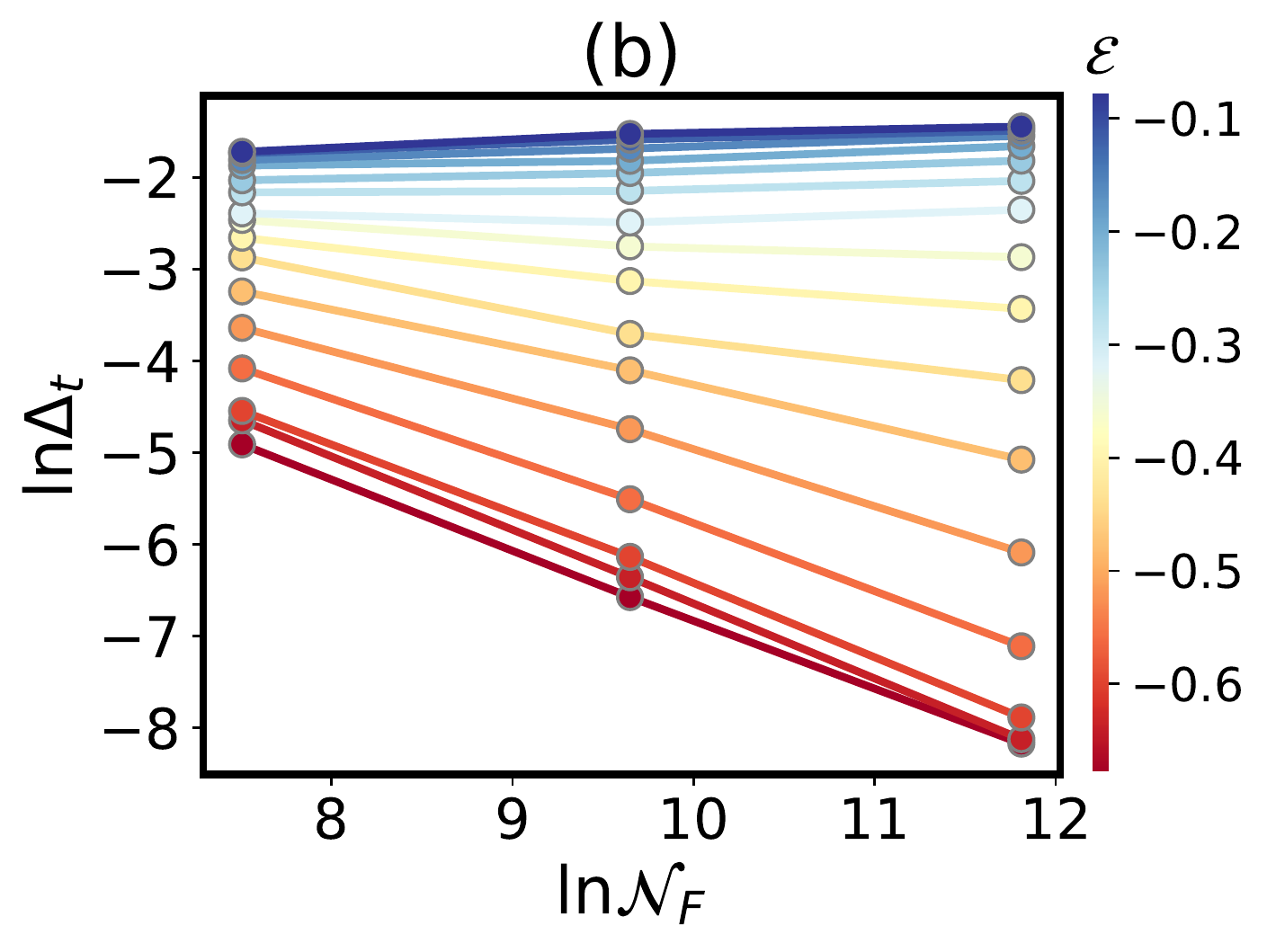}
\includegraphics[width=0.9\columnwidth,height=3.55cm]{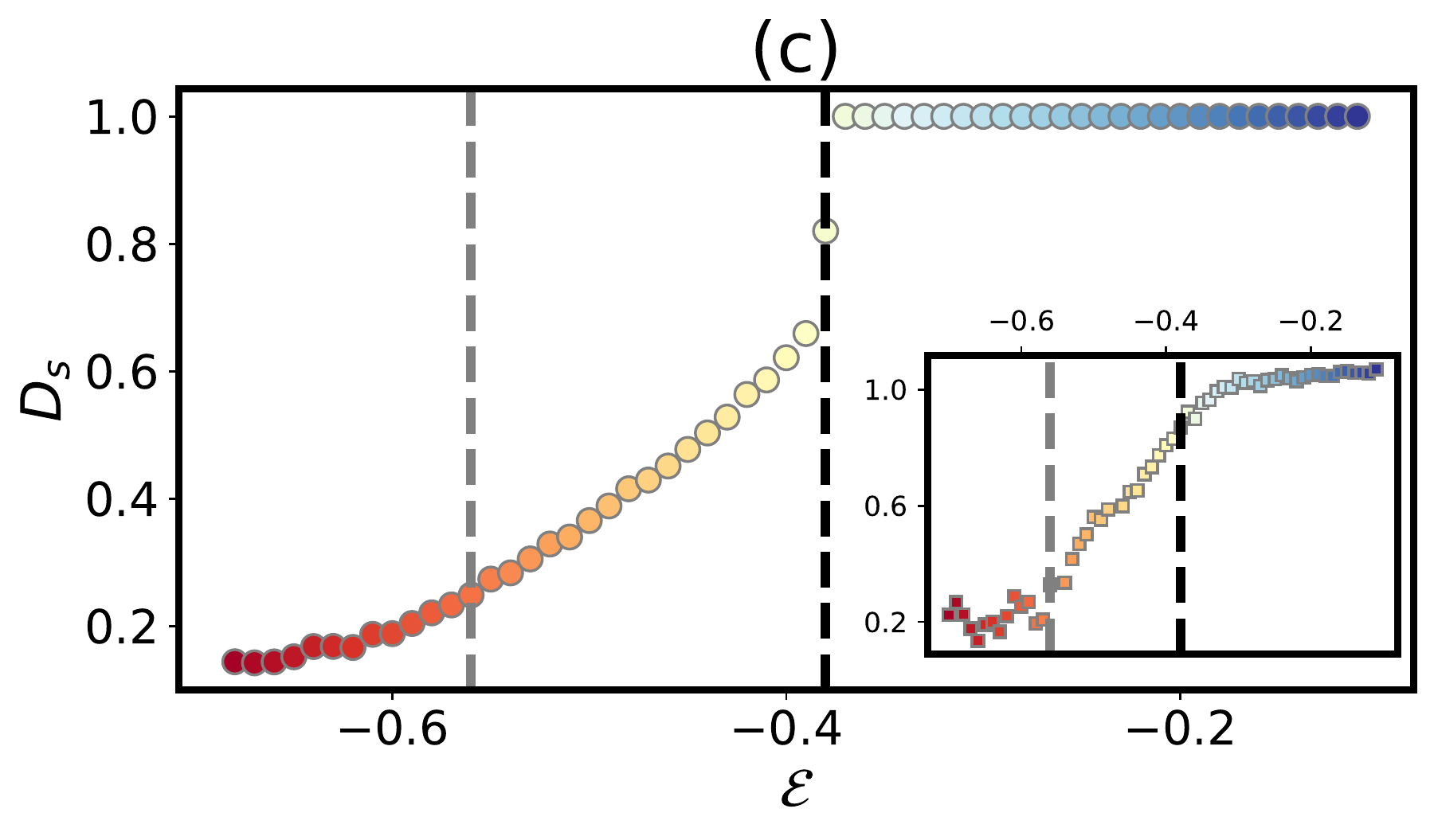}
\caption{{\bf Feenberg self energy and multifractality in the Fock space:} (a) FS lattice constructed out of real-space
occupation-number basis states (orange circles), illustrated
for $L=8$ at quarter filling, starting at the top with $\ket{11..0000}$, i.e. all particles
on the left side, and ending at the bottom with all particles
on the right. The hoppings (blue lines) and the slices (grey
lines) are indicated. (b) $\ln\Delta_t$ as a function of $\ln \Nf$ for increasing $\E$ (color bar). (c) The fractal dimension $D_s$ is found from the finite size scaling theory. $D_s=1$ in the ergodic phase and $D_s<1$ in the nonergodic extended (NEE) and MBL phases. $D_s$ jumps at the nonergodic-ergodic transition point $\E_c=-0.38$ denoted by the dark dashed vertical line. The grey dashed vertical line denotes the MBL-NEE transition at $\E_{nc}=-0.56$, estimated from previous study \cite{ghosh2020transport} and statistics of Feenberg self-energy [Sec.\ref{sec8}]. Inset shows $D_s$ extracted from (a) by directly fitting $\Delta_t\sim \Nf^{-(1-D_s)}$. }
\label{linscaling}
\end{figure}

We compute the imaginary part $\Delta_I(\E)$ of Feenberg self energy for $-0.7\lesssim \E\lesssim -0.1$. $\Delta_I(\E)$ quantifies the inverse lifetime of an excitation created at FS site $I$ with energy $E$~\cite{anderson1958absence}. Hence $\Delta_I$ provides information of ergodicity or its absence. We expect $\Delta_I\sim \mathcal{O}(1)$ in the ergodic phase and $\Delta_I\rightarrow 0$ in the nonergodic phase as $\Nf\rightarrow\infty$ in the thermodynamic limit. In Fig.~\ref{linscaling}(b), we show $\ln\Delta_t$ as a function of $\ln\Nf\propto L$. Deep in the ergodic phase $\Delta_t$ saturates to $\mathcal{O}(1)$ value as $L$ is increased, whereas in the nonergodic phase, which includes both NEE and MBL phases, $\Delta_t$ falls off with a power-law in $\Nf$. 
Also the typical value $D_t(\E)$ of the local many-body density of states $D_I(\E)=(-1/\pi)\mathrm{Im}G_{II}(\E)$ shows similar behavior [Fig.~\ref{energy_ldos} in Appendix.~\ref{app4}].

As discussed in refs.~\onlinecite{sutradhar2022scaling,altshuler2016multifractal}, for non-ergodic phase with multifractal eigenstates, $\Delta_t\sim \eta_c^\theta\sim \Nf^{-(1-D_s)}$ for a broadening parameter $\eta\propto\Nf^{-1}\ll\eta_c$, where $\theta>0$ and $\eta_c\sim\Nf^{-z}$ $(0<z<1)$ is a characteristic energy scale much larger than the mean many-body level spacing. The spectral fractal dimension $D_s=1-z\theta$ lies between $0$ and $1$.
In the inset of Fig.~\ref{linscaling}(c), we show $D_s$ as function of $\E$, extracted from linear fitting of the $\ln\Delta_t$ vs. $\ln\Nf$ plots in Fig.~\ref{linscaling}(b). Deep in the ergodic phase $D_s=1$, whereas in the MBL and NEE phases $0<D_s<1$. This implies that both MBL and NEE states are essentially non-ergodic extended i.e. multifractal.
However, as we will discuss in Sec.~\ref{sec8},  the distinction between NEE and MBL states can be made in terms of the distribution of $\Delta_I$. 

At quarter filling, we only have a few system sizes accessible to our numerics. Thus, following ref.~\onlinecite{sutradhar2022scaling}, we use a finite-size scaling analysis~\cite{garcia2017scaling,mace2019multifractal} across the ergodic-to-non-ergodic transition to estimate $D_s$ more accurately in the thermodynamic limit which we discuss in the next section. 

\begin{figure}[ht!]
\centering
\includegraphics[width=0.95\columnwidth,height=4.5cm]{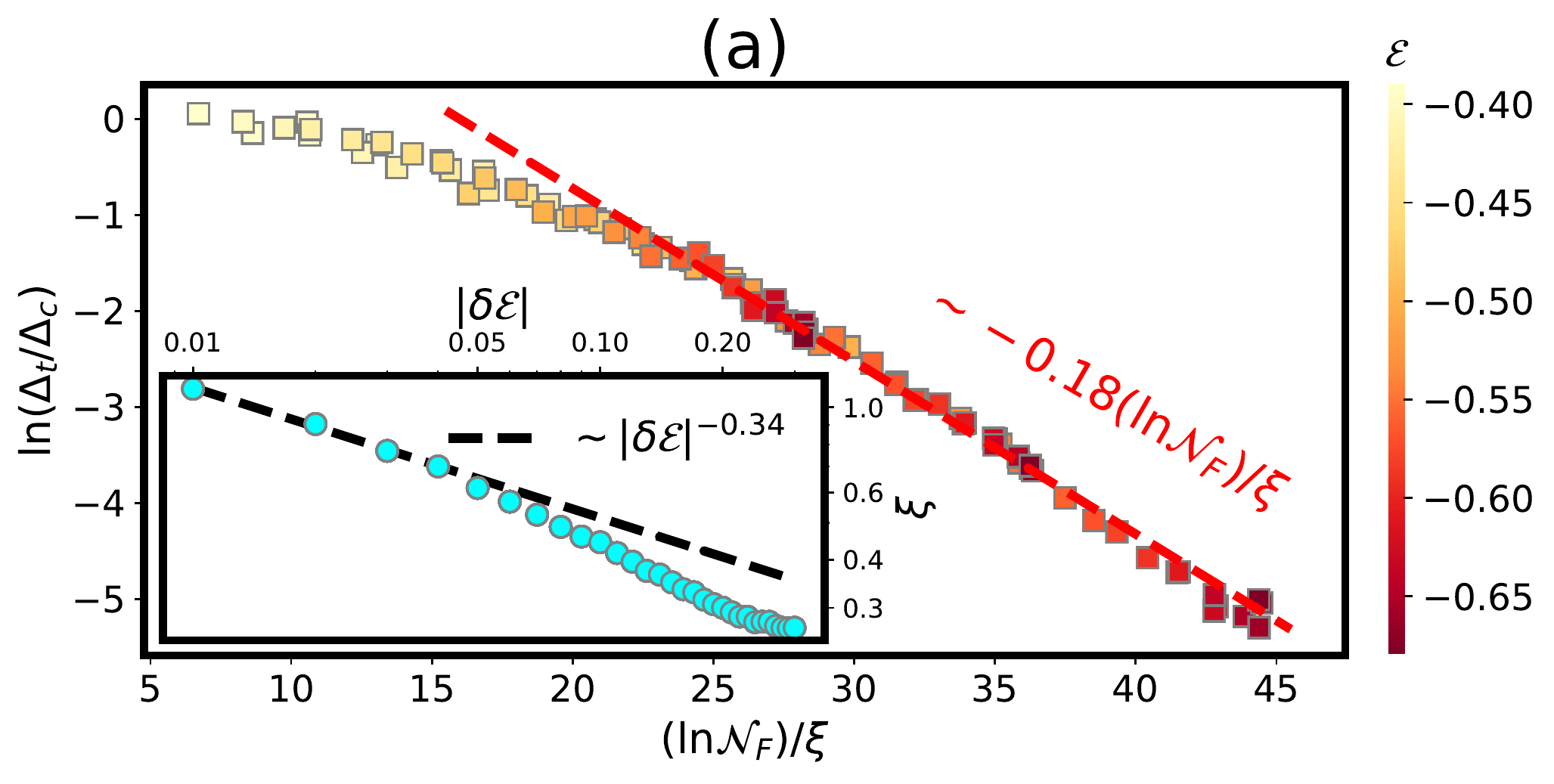}
\includegraphics[width=0.95\columnwidth,height=4.5cm]{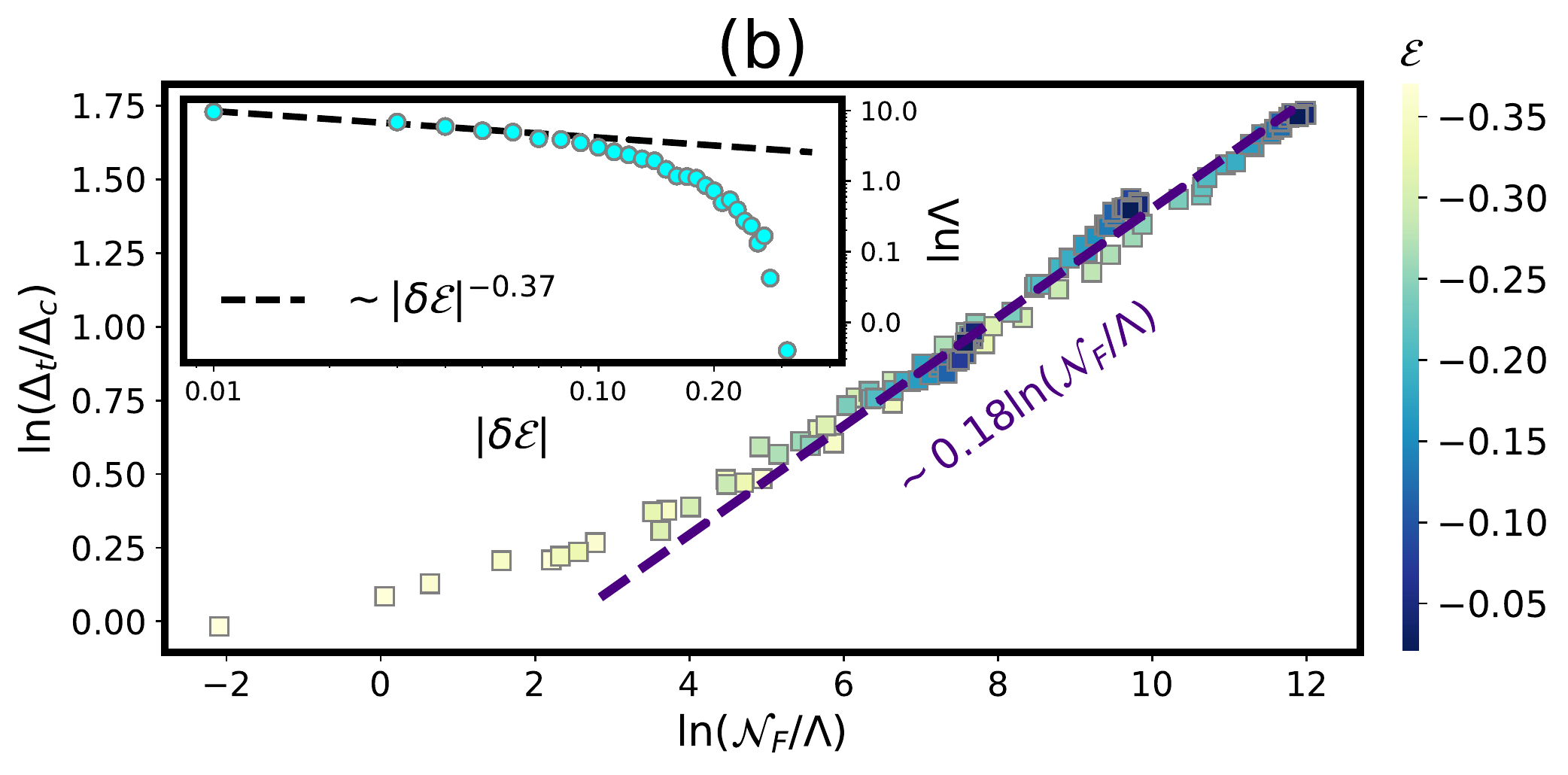}
\caption{{\bf Finite-size scaling collapse across nonergodic-ergodic transition:} (a) Finite-size scaling collapse of $\ln(\Delta_t/\Delta_c)$ in the nonergodic phase using linear scaling. The asymptotic scaling form is given by $\ln(\Delta_t/\Delta_c)=-0.18 (\ln \Nf)/\xi$ where $\ln \Nf\propto L$. Inset shows the the power-law divergence of the correlation length $\xi\sim |\delta\E|^{-\beta}$ with $\beta\simeq 0.34$ and $\delta\E=(\E-\E_c)$.
(b) Finite size scaling of $\ln(\Delta_t/\Delta_c)$ in the ergodic phase with a volumic scaling form $\mathcal{F}_{vol}(\mathcal{N}_F/\Lambda)$ where nonergodic-ergodic transition point is at $\mathcal{E}_c=-0.38$ . In the asymptotic limit $\ln(\Delta_t/\Delta_c)\sim 0.18 \ln(\mathcal{N}_F/\Lambda)$ deep in the ergodic phase implies $(1-D_c)\approx0.18$ at $\mathcal{E}=\mathcal{E}_c$ where $\Delta_c\sim \mathcal{N}_F^{ -(1-D_c)}$. Inset shows KT-like essential singularity of the nonergodic volume $\Lambda\sim \exp[b/(\delta\mathcal{E})^\gamma]$ with $\gamma\approx 0.4$,  $b\sim\mathcal{O}(1)$ and $\delta\mathcal{E}=\mathcal{E}-\mathcal{E}_c$ near $\mathcal{E}_c$. } 
\label{volscaling}
\end{figure}

\subsection{Finite-size scaling for Nonergodic-ergodic transition }\label{sec7}
To analyze the nonergodic-ergodic transition and obtain a more accurate estimate of $D_s$, we perform scaling collapse of our data in Fig.~\ref{linscaling}(b) using the following finite-size scaling form ~\cite{sutradhar2022scaling},
\begin{eqnarray}
\ln\frac{\Delta_t}{\Delta_c}=
\begin{cases}
      \mathcal{F}_{vol}\big(\frac{\mathcal{N}_F}{\Lambda}\big)  & ~~:~ \mathcal{E}>\mathcal{E}_c\\
      \mathcal{F}_{lin}\big(\frac{\ln\mathcal{N}_F}{\xi}\big)  & ~~:~ \mathcal{E}<\mathcal{E}_c,
    \end{cases} 
\label{scaling_ansatz}    
\end{eqnarray}
with $\Delta_c=\Delta_t(\E=\E_c)\sim \Nf^{-(1-D_c)}$.
In the entire nonergodic phase, which includes MBL and NEE phases, we are able to obtain a data collapse using the linear scaling for $\E<\E_c$, where $\xi$ plays the role of correlation length in the FS~\cite{mace2019multifractal,roy2021fock,sutradhar2022scaling}. In the asymptotic limit, $x=(\ln\Nf)/\xi\gg1$, the scaling function is given by $\Fl(x)\sim-(1-D_c)x$ with  $\xi=(1-D_c)/(D_c-D_s)$~\cite{sutradhar2022scaling}. Fig.~\ref{volscaling}(a) shows the scaling collapse of the data in the nonergodic phase for $\E_c=-0.38$. The fit to asymptotic scaling leads to $\ln(\Delta_t/\Delta_c)=-0.18(\ln\Nf)/\xi$ with $\xi\sim |\delta\E|^{-\beta}$ where $\delta\E=\E-\E_c$ and $\beta\simeq0.34$, as shown in the inset of Fig.~\ref{volscaling}(a). The asymptotic scaling form implies the critical spectral fractal dimension $D_c=0.82$. The $D_s$ extracted from $\xi$ is shown in Fig.~\ref{linscaling}(c).

In the ergodic phase, we use volumic scaling for $\E>\E_c$ where $\Lambda$ represents the nonergodic volume in FS~\cite{garcia2017scaling,mace2019multifractal,sutradhar2022scaling}. The scaling collapse is shown in Fig.~\ref{volscaling}(b). For $x=\Nf/\Lambda\gg1$, the asymptotic scaling form is $\Fv(x)\sim (1-D_c)\ln x$ \cite{sutradhar2022scaling}. From the scaling collapse, we find $\ln(\Delta_t/\Delta_c)=0.18\ln(\Nf/\Lambda)$ with $D_c=0.82$, which is consistent with $D_c$ extracted from the asymptotic scaling in the non-ergodic phase, as discussed in the preceding paragraph. The extracted $D_c$ is also consistent with $D_c\approx0.8$ directly found [Fig.~\ref{linscaling}(c)(inset)] by fitting the data of Fig.~\ref{linscaling}(b) with $\Delta_t\sim \Nf^{-(1-D_s)}$. The nonergodic volume $\Lambda$ shows a KT-like essential singularity such that $\Lambda\sim \exp[b/(\delta\E)^\gamma]$ where $b\sim \mathcal{O}(1)$ and $\gamma\approx0.4$, as shown in Fig.~\ref{volscaling}(b) (inset).
In the ergodic phase $\Delta_t\sim\Lambda^{-(1-D_c)}$~\cite{sutradhar2022scaling} such that it continuously vanishes as $\Lambda$ diverges on approaching the critical point. We find similar kind of volumic and linear scaling collapses for a choice of $\E_c$ within $-0.38 \pm 0.02$ although the optimized scaling collapse is obtained when $\E_c=-0.38$.

Hence, this analysis reveals that $D_s=1$ throughout the ergodic phase and it discontinuously jumps to $D_s<1$ in the non-ergodic phase with a critical $D_s=D_c\simeq 0.8$. In the next section, we compare the fractal dimension directly extracted from many-body wavefunction with the spectral fractal dimension $D_s$ computed from the typical Feenberg self-energy.

\begin{figure}[ht!]
\centering
\stackon{\includegraphics[width=0.4925\columnwidth,height=3.6cm]{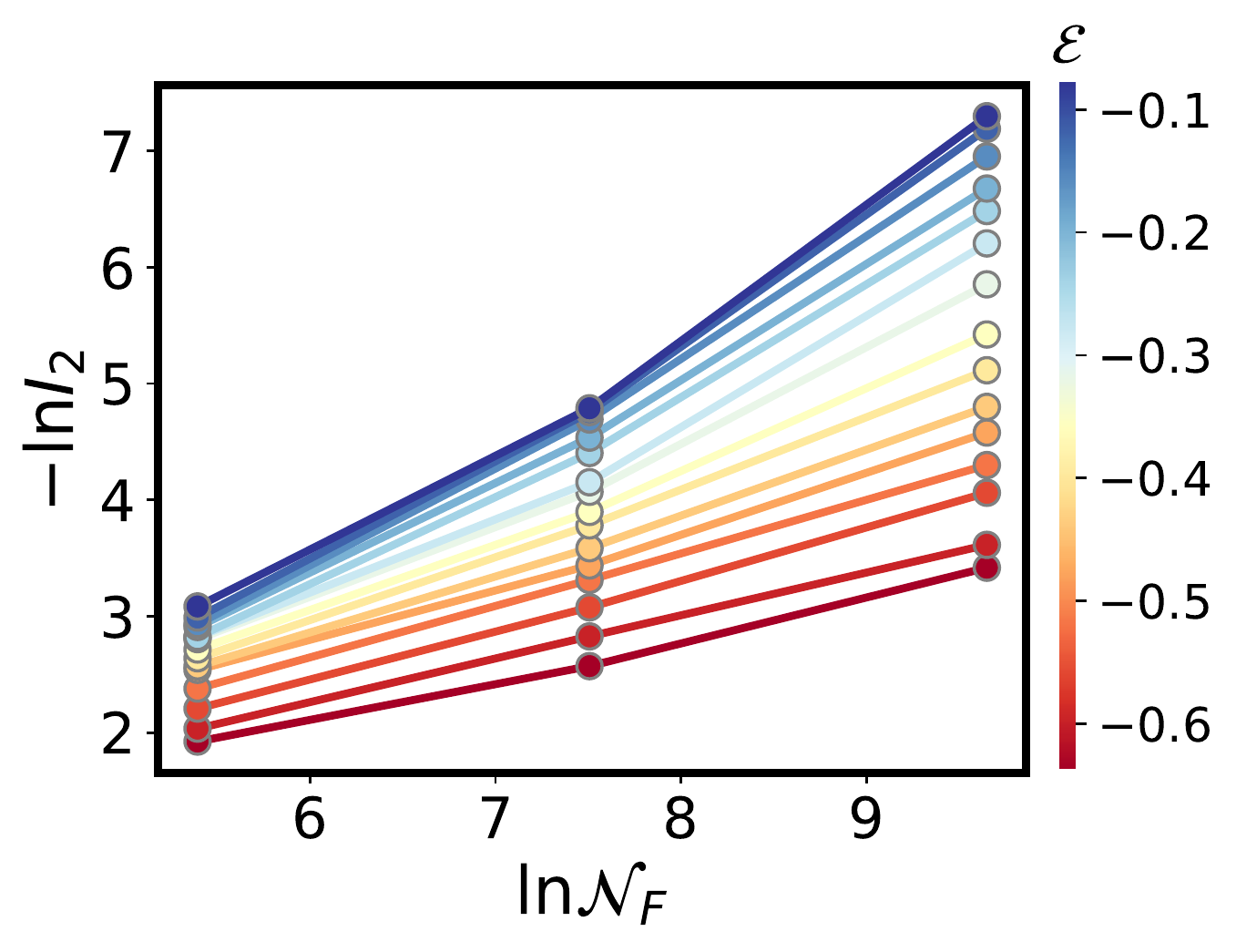}}{(a)}
\stackon{\includegraphics[width=0.4925\columnwidth,height=3.6cm]{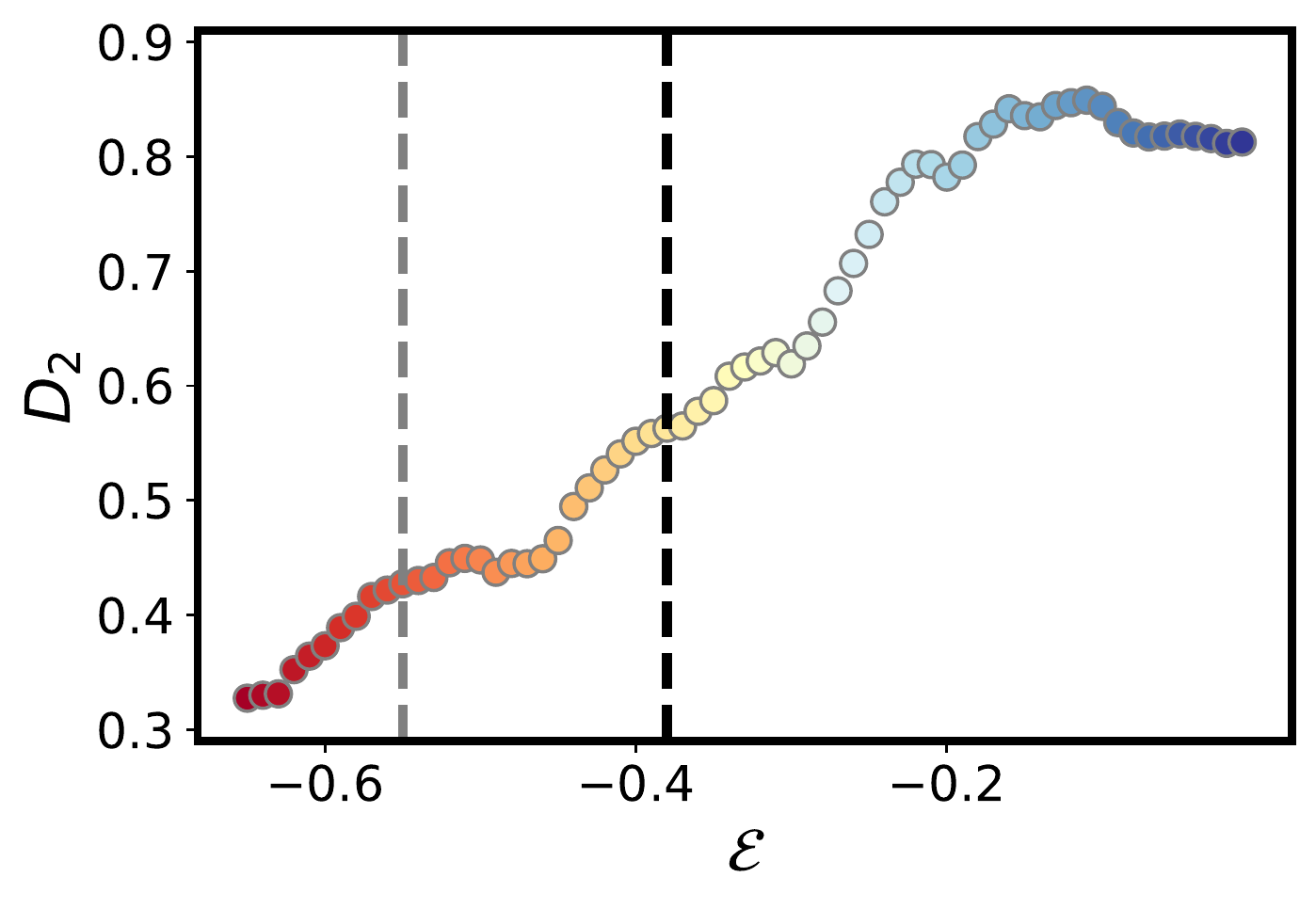}}{(b)}
\stackon{\includegraphics[width=0.4925\columnwidth,height=3.6cm]{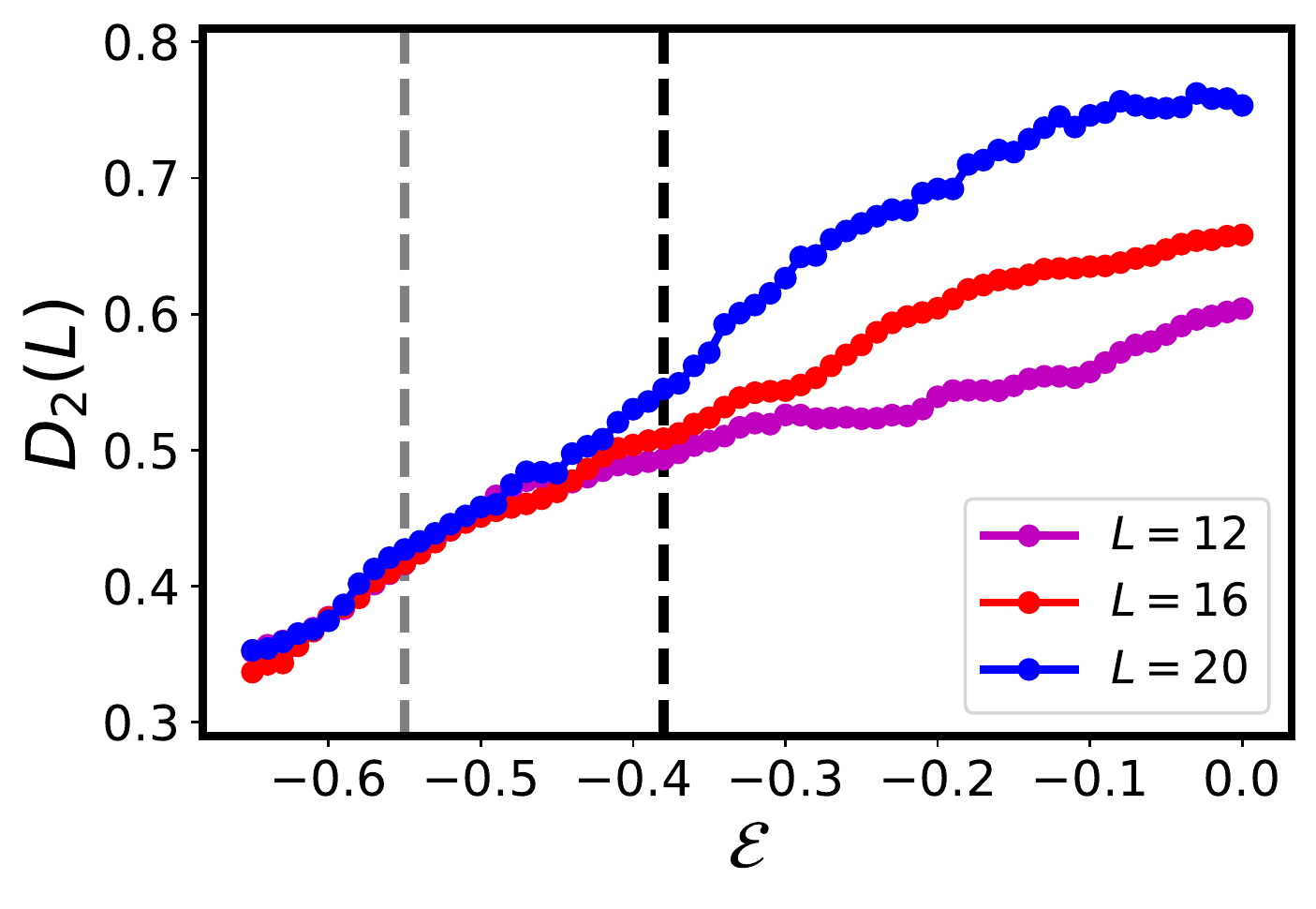}}{(c)}
\stackon{\includegraphics[width=0.4925\columnwidth,height=3.6cm]{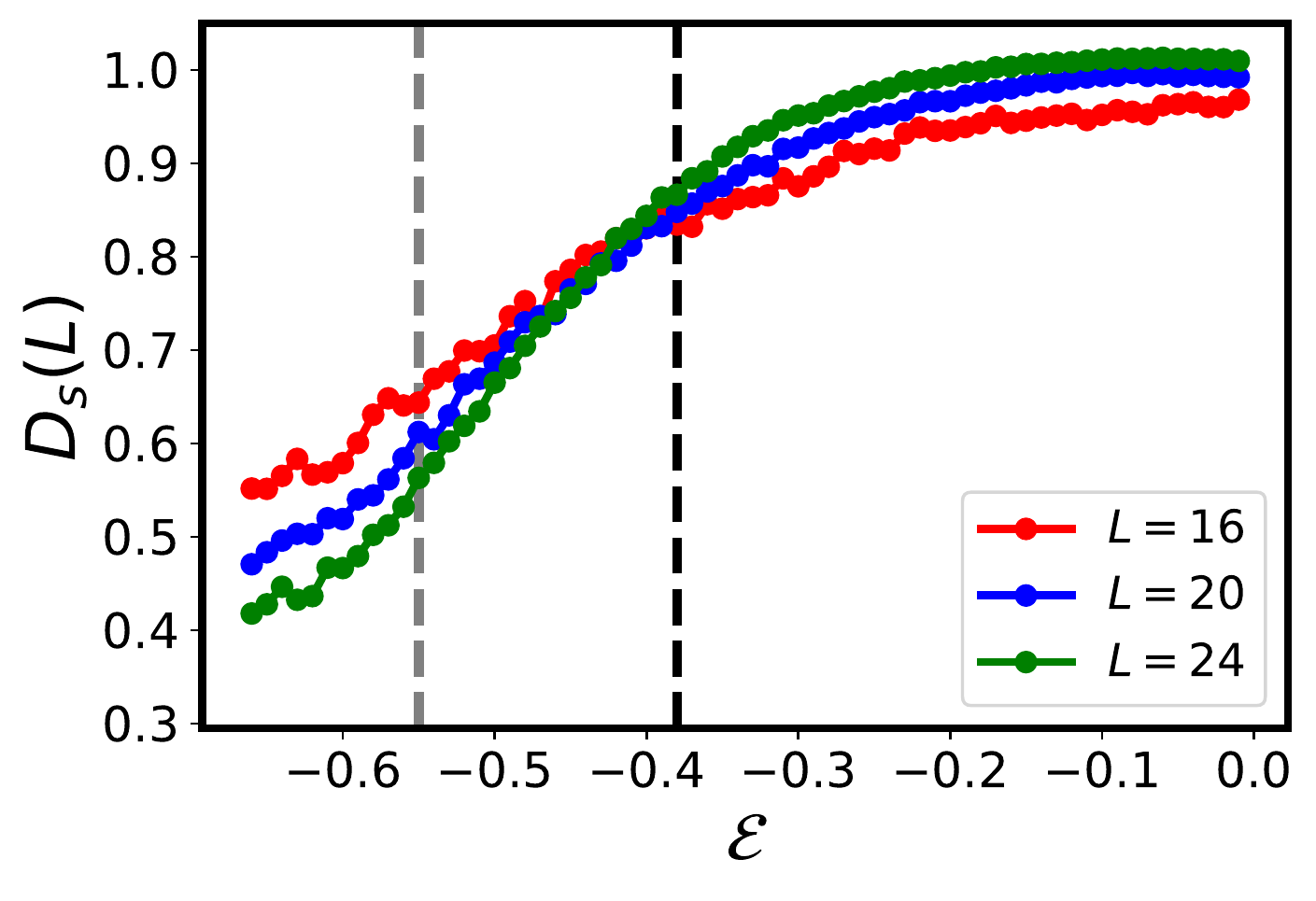}}{(d)}
\caption{{\bf Comparison of $D_2$ and $D_s$:} (a) Plots of $(-\ln I_2)$ as a function of $\ln \mathcal{N}_F$ for increasing $\mathcal{E}$. (b) The slopes $D_2$ extracted by linear fitting as a function of $\mathcal{E}$. 
(c) Variation of $D_2(L)$ with $\mathcal{E}$ for increasing system size $L$. (d) Variation of $D_s(L)$ with $\mathcal{E}$ for increasing system size $L$. The vertical light and dark dashed lines indicate the MBL-NEE and NEE-ergodic transitions, respectively.}
\label{ipr}
\end{figure}

\subsection{Comparison between fractal dimensions from inverse participation ratio and FS self-energy}\label{sec9}
The inverse participation ratio (IPR) is one of the most important quantities in the context of localization transition. The IPR can be obtained as $I_2=\sum_{J=1}^{\mathcal{N}_F} |\Psi_J|^4$ for a normalized eigenstate $\ket{\Psi}=\sum_{J}\Psi_J\ket{J}$. $I_2$ scales with $\mathcal{N}_F$ as $I_2\sim \mathcal{N}_F^{-D_2}$, where $D_2$ is the fractal dimension. $D_2=0$ and 1 imply localized and ergodic states, respectively, whereas $0<D_2<1$ corresponds to a (multi) fractal state. 
The fractal dimension $D_2$ and spectral dimension $D_s$ have been shown to match with each other exactly for a non-interacting particle in the presence of uncorrelated disorder on Bethe lattice and Rosenzweig-Poter random matrices~\cite{altshuler2016multifractal}. No explicit results showing the comparison of $D_2$ and $D_s$ are available in the literature for the MBL phase, which occurs in the presence of correlated disorder on the FS lattice. Hence, it is worth making an attempt to compare them in our system. 

In Fig.~\ref{ipr}(a) we show $(-\ln I_2)$ as a function of $\ln \mathcal{N}_F$. The slope of the curves gives fractal dimension $D_2$. The values of $D_2$ extracted from the linear fitting of these plots are shown as a function of $\mathcal{E}$ in Fig.~\ref{ipr}(b). The behavior of $D_2$ with $\mathcal{E}$ matches with that of $D_s$ [Fig.~\ref{linscaling}(c)], albeit only qualitatively. This may be due to smaller system sizes accessed in ED ($L\leq 20$) to compute $D_2$ compared to those ($L\leq 24$) for extracting $D_s$ from the recursive Green's function method. The linear fitting to extract $D_2$ clearly seems to be an underestimation, especially in the ergodic phase, as the slopes of the $(-\ln I_2)$ vs. $\ln \mathcal{N}_F$ curves clearly increase with $\ln \mathcal{N}_F$. To illustrate this more clearly we show plots of $D_2(L)=-\frac{\ln I_2}{\ln \mathcal{N}_F}$ as a function of $\mathcal{E}$ for increasing system sizes $L$ in Fig.~\ref{ipr}(c). On the other hand, in Fig.~\ref{ipr}(d) we show plots of $D_s(L)=1+\frac{\ln \Delta_t}{\ln \mathcal{N}_F}$  as a function of $\mathcal{E}$ for increasing $L$. It is evident that the plots of $D_2(L)$ suffer more from a finite-size effect than that of $D_s(L)$. We note that the relation between $D_2$ and $D_s$ needs more careful investigation, which is beyond the scope of this work. 

\subsection{The distribution of Feenberg self-energy in the MBL, NEE, and ergodic phases}\label{sec8}
\begin{figure}[h!]
\centering
\stackon{\includegraphics[width=0.49\columnwidth,height=3.3cm]{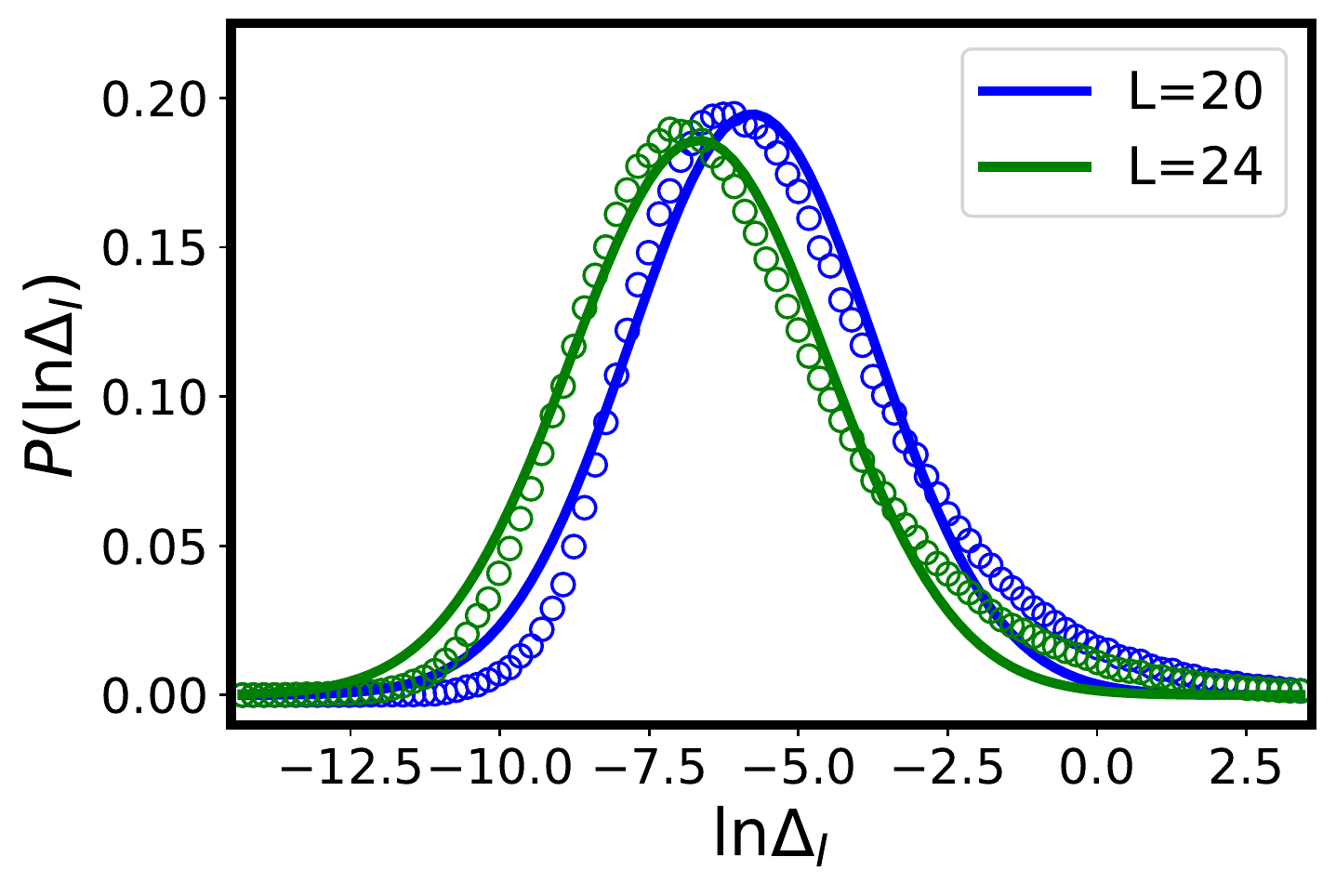}}{(a)}
\stackon{\includegraphics[width=0.49\columnwidth,height=3.3cm]{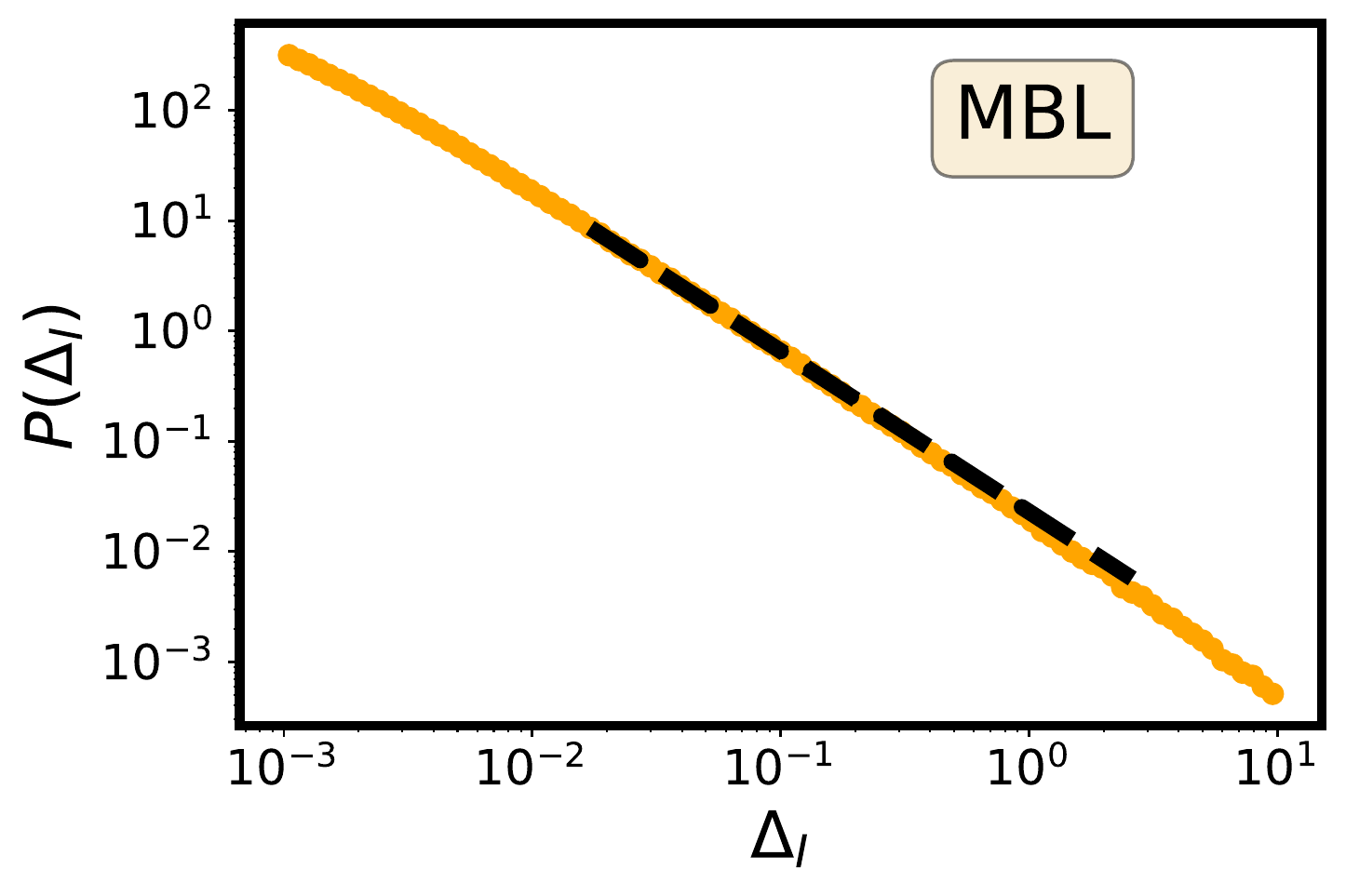}}{(f)}
\stackon{\includegraphics[width=0.49\columnwidth,height=3.3cm]{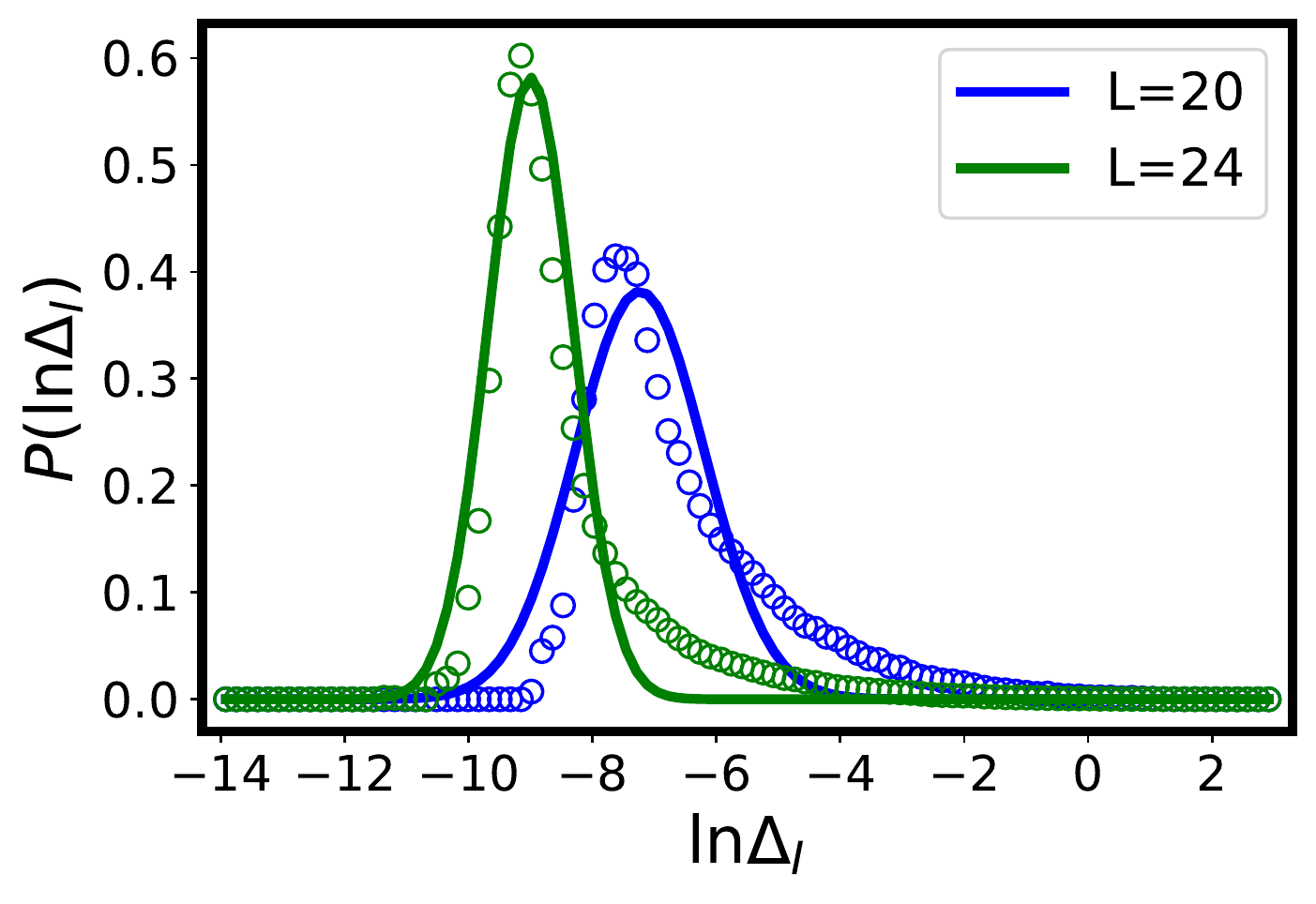}}{(b)}
\stackon{\includegraphics[width=0.49\columnwidth,height=3.3cm]{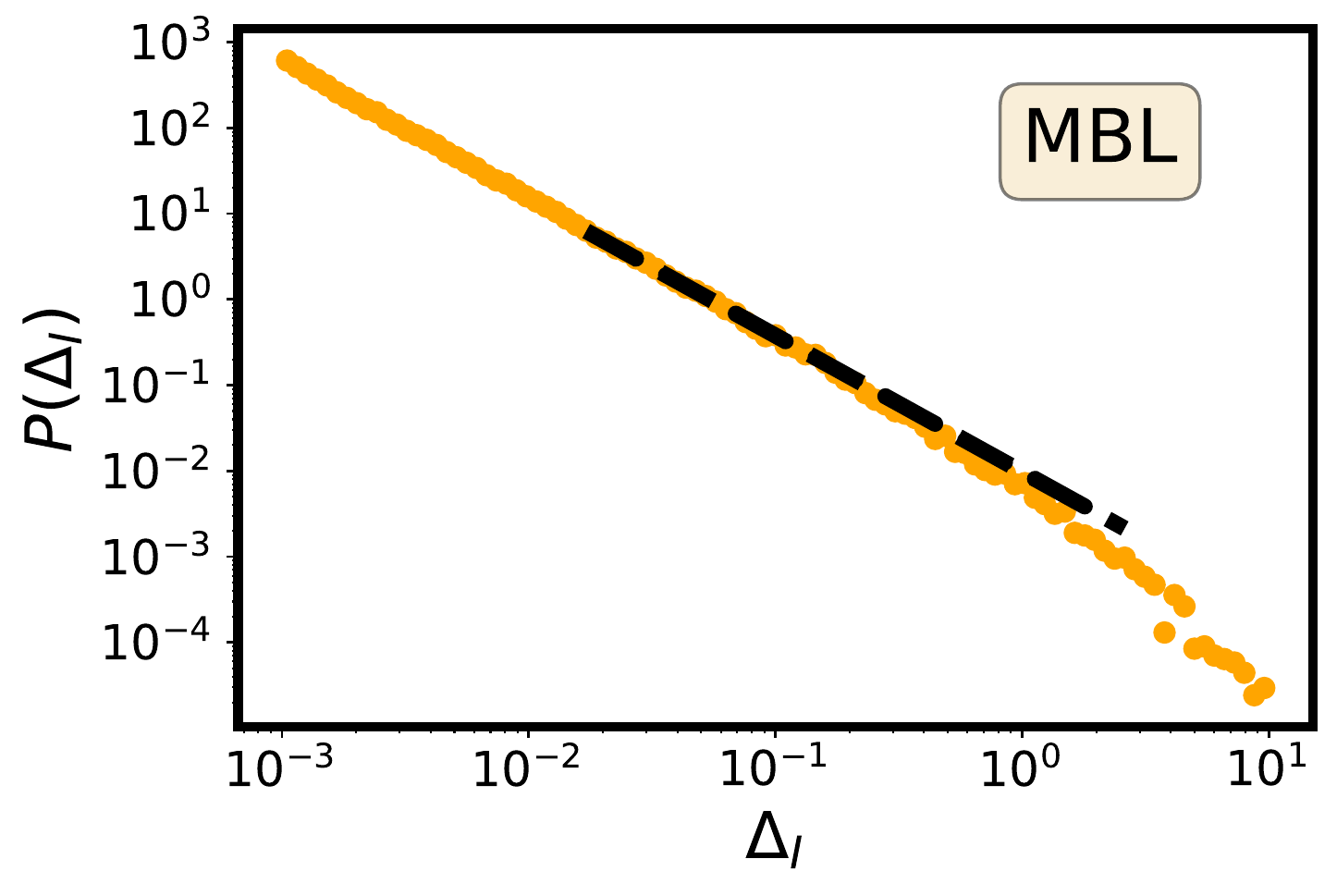}}{(g)}
\stackon{\includegraphics[width=0.49\columnwidth,height=3.3cm]{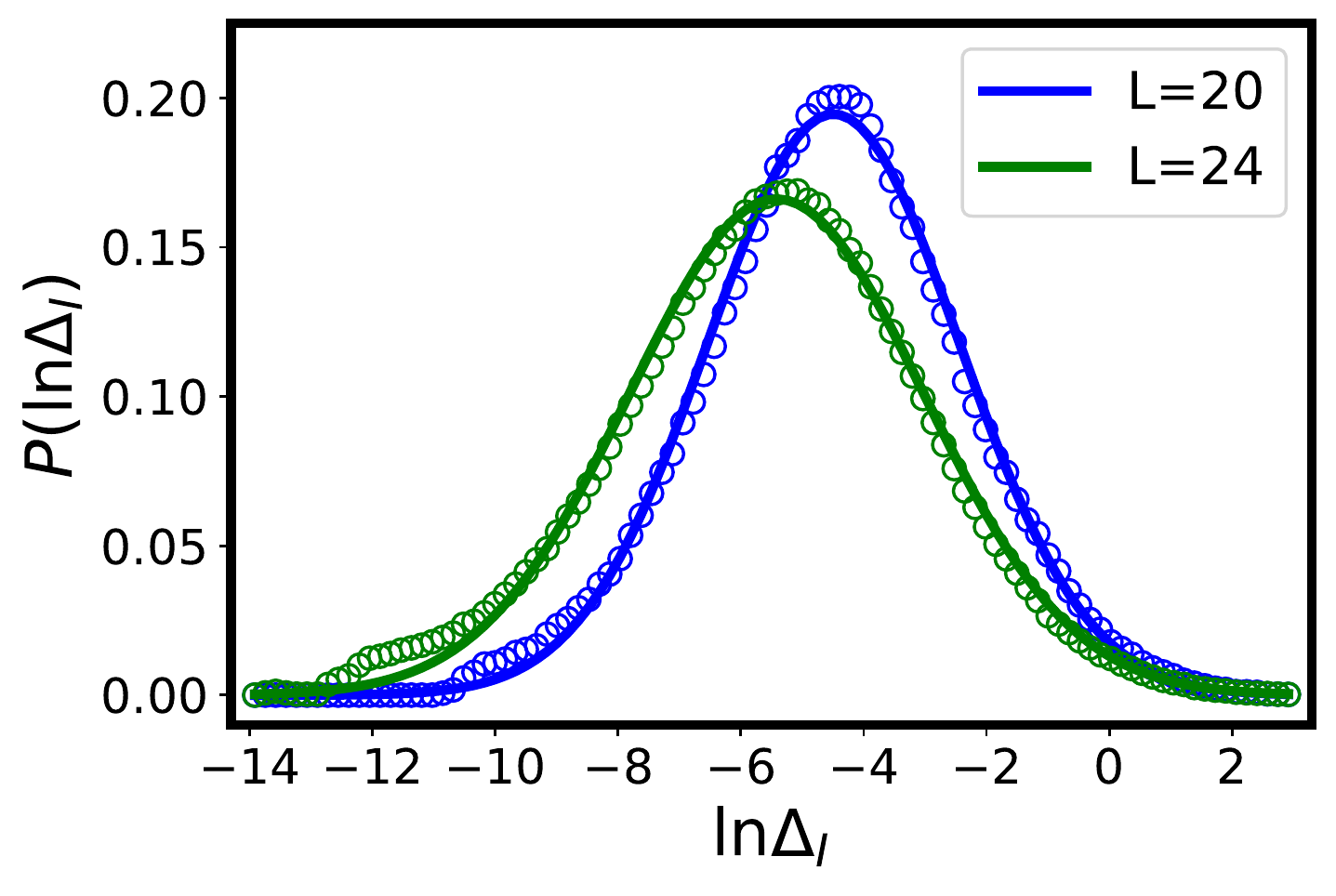}}{(c)}
\stackon{\includegraphics[width=0.49\columnwidth,height=3.3cm]{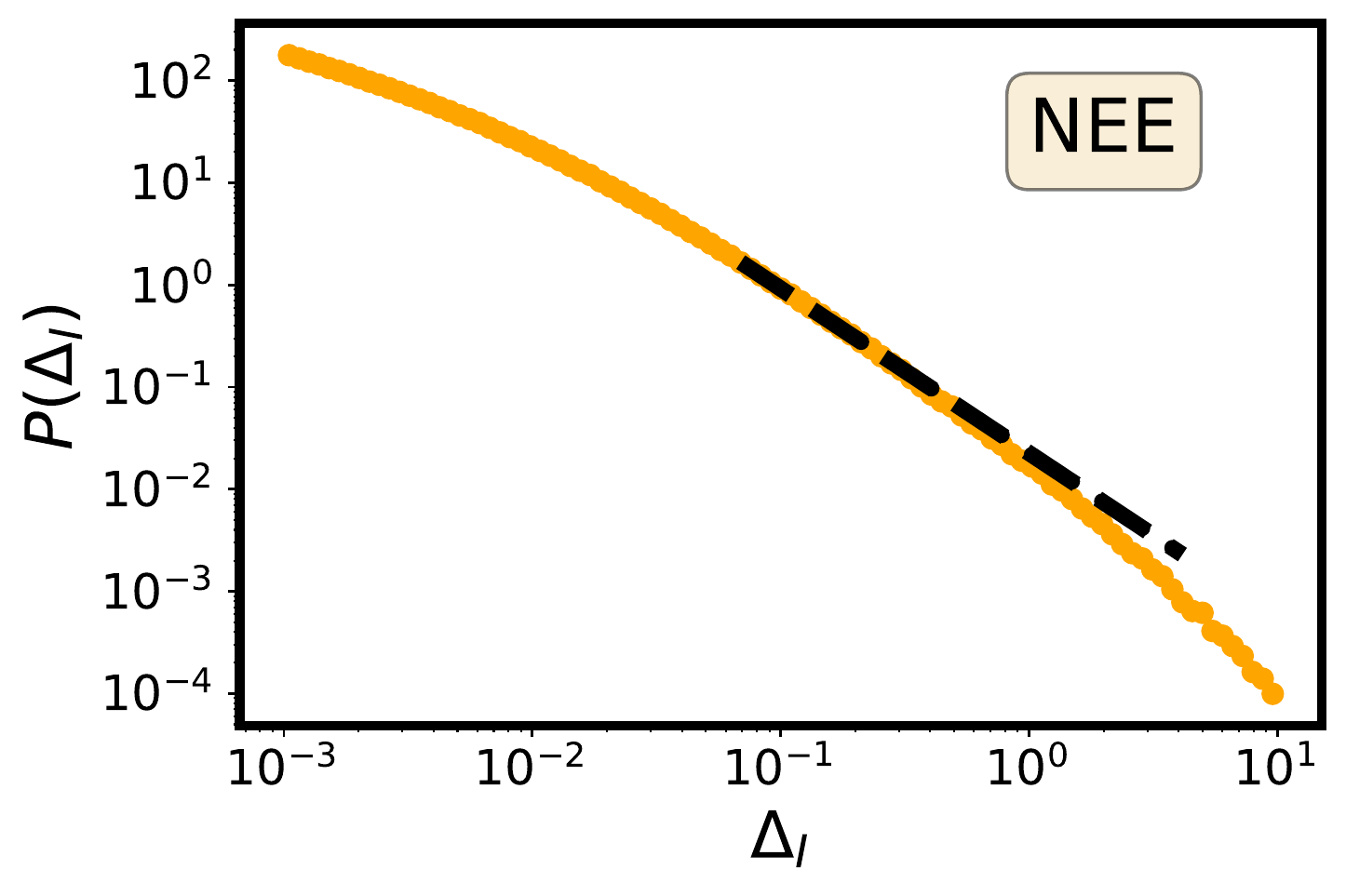}}{(h)}
\stackon{\includegraphics[width=0.49\columnwidth,height=3.3cm]{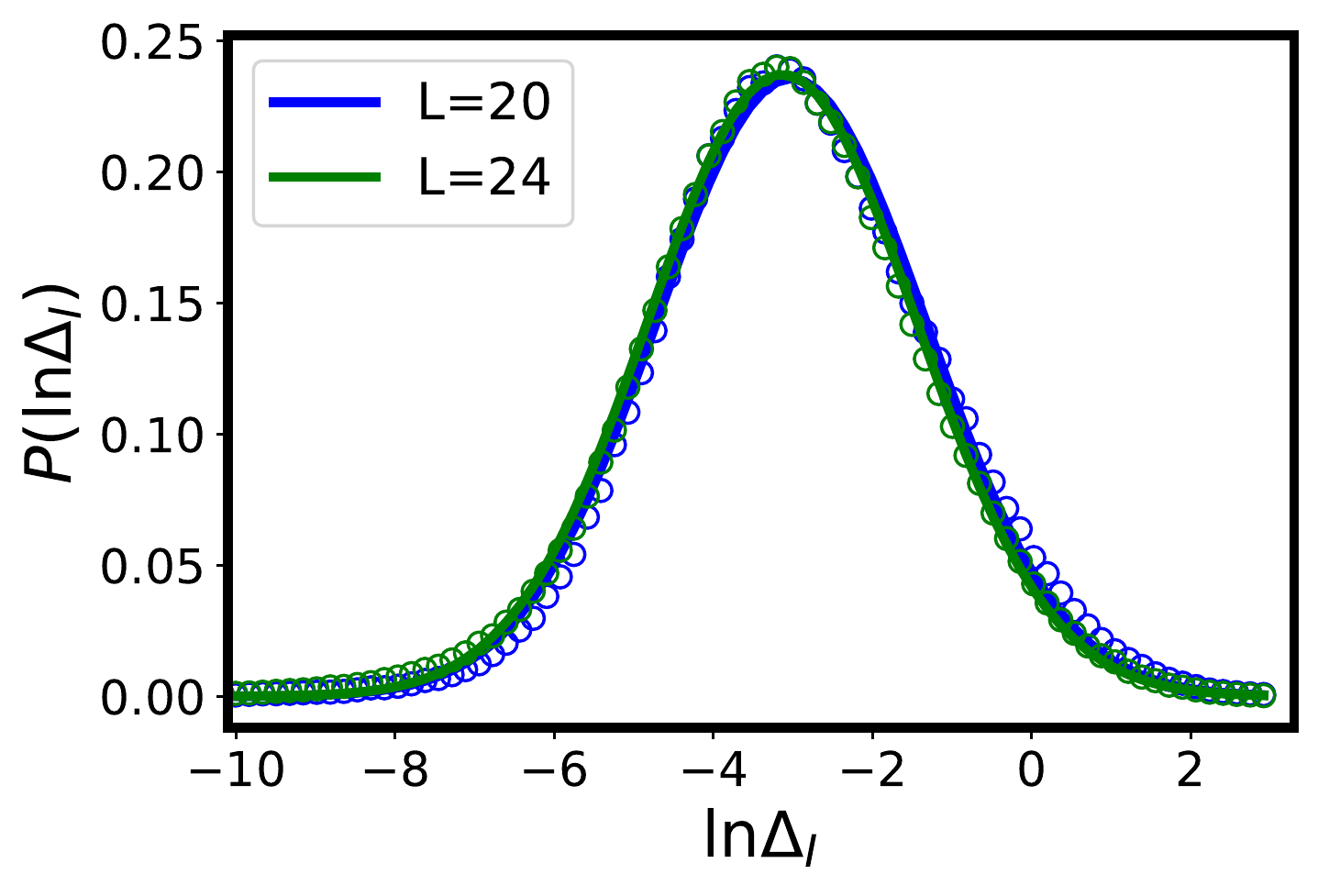}}{(d)}
\stackon{\includegraphics[width=0.49\columnwidth,height=3.3cm]{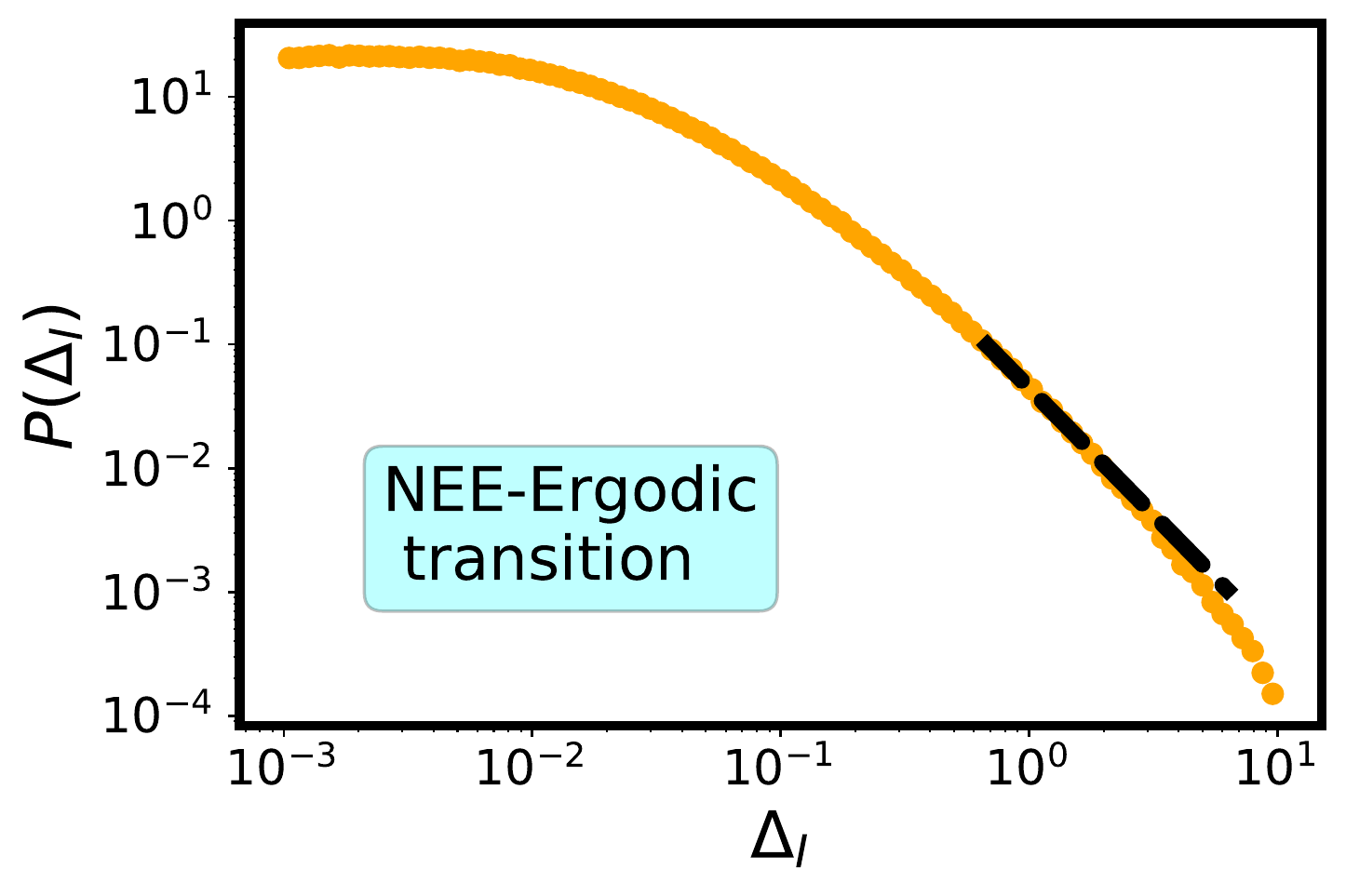}}{(i)}
\stackon{\includegraphics[width=0.49\columnwidth,height=3.3cm]{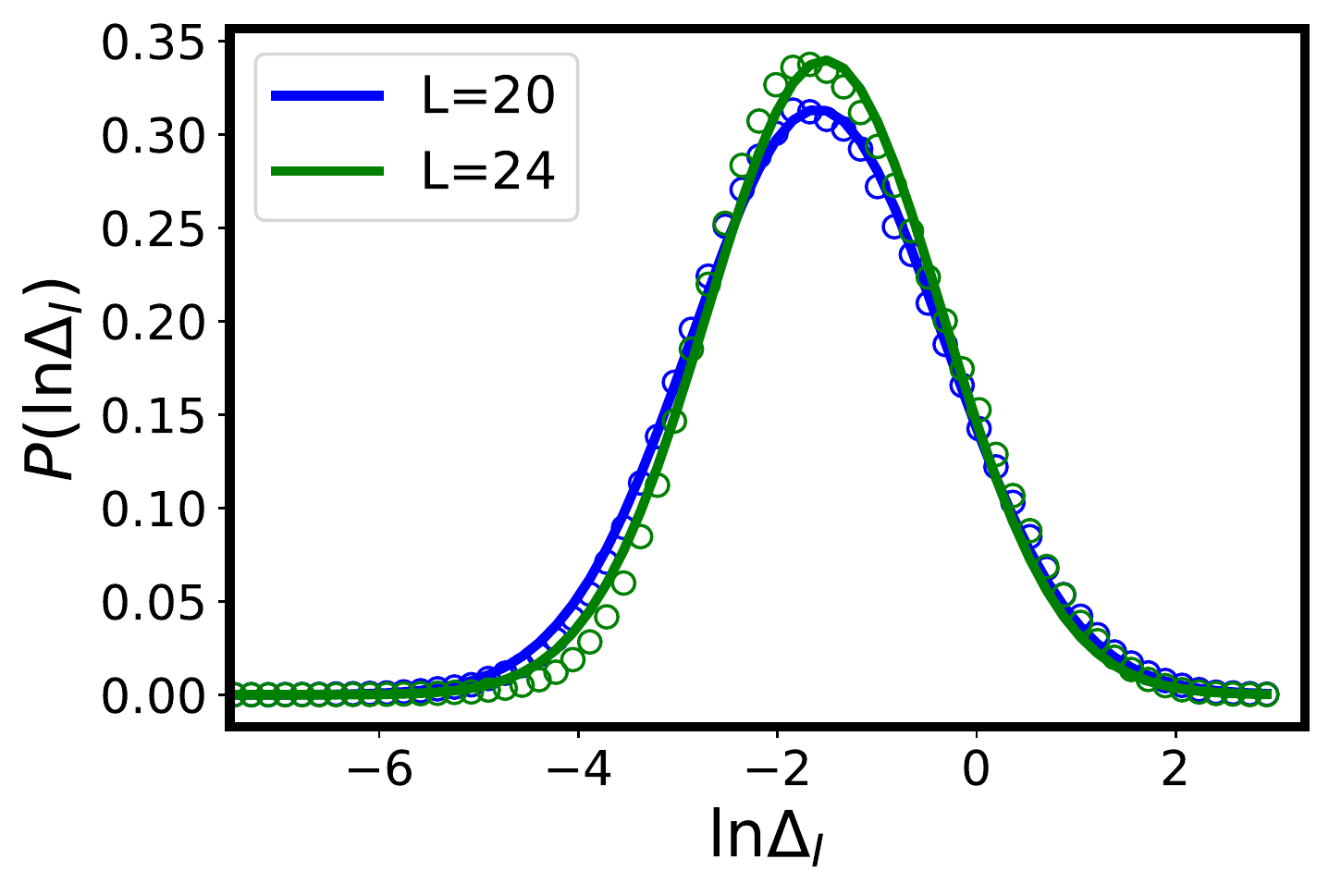}}{(e)}
\stackon{\includegraphics[width=0.49\columnwidth,height=3.3cm]{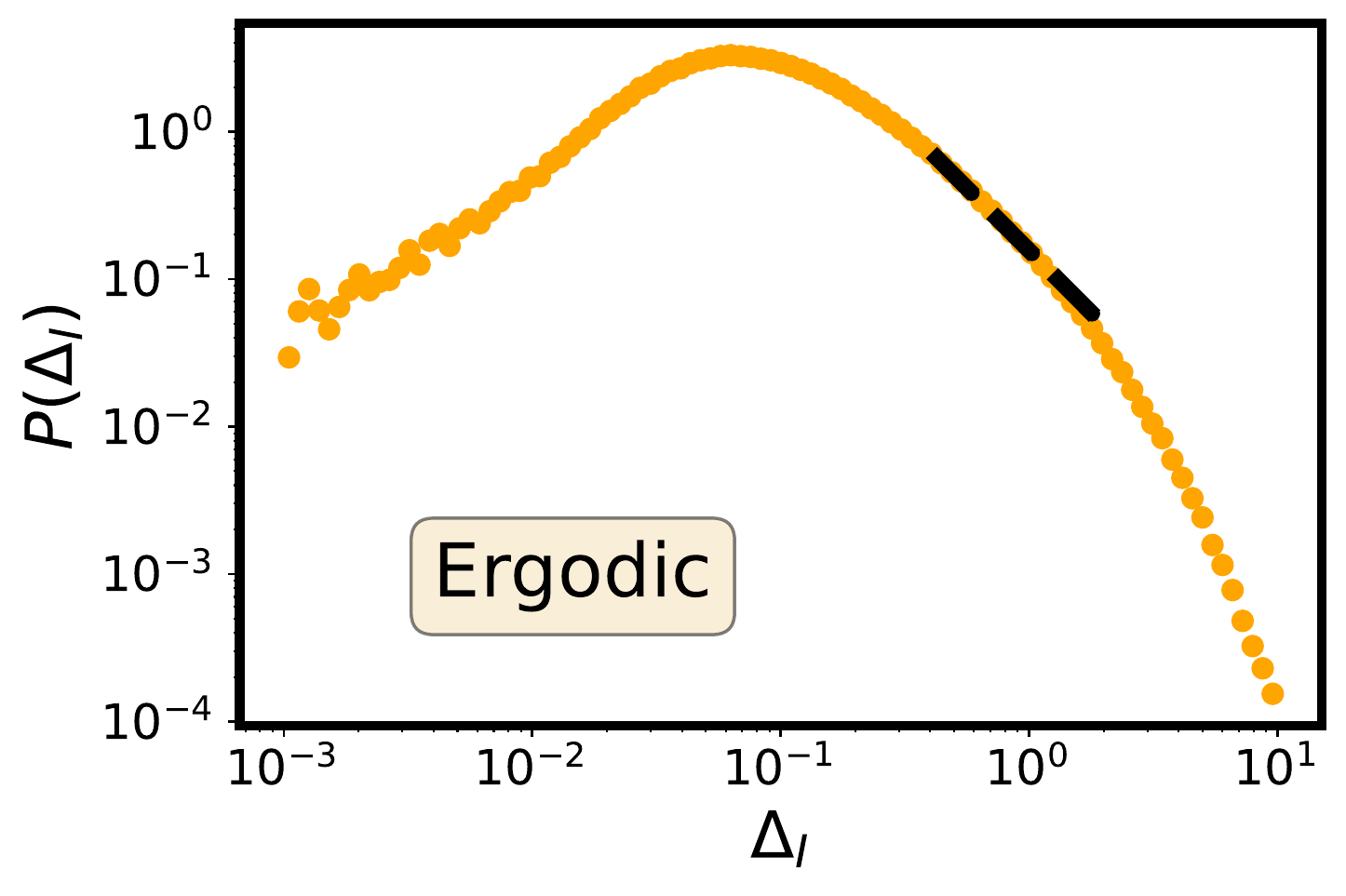}}{(j)}
\caption{{\bf Distributions of $\ln\Delta_I$ and $\Delta_I$:} (a-e) Probability distribution $P(\ln\Delta_I)$ in the MBL phase for $(h=1.8,\mathcal{E}=-0.49)$ and $(h=0.6,\mathcal{E}=-0.66)$, NEE phase $(h=0.6,\mathcal{E}=-0.50)$, at the nonergodic-ergodic transition $(h=0.6,\mathcal{E}=-0.38)$ and in the ergodic phase $(h=0.6,\mathcal{E}=-0.10)$, respectively.
(f-j) Probability distribution $P(\Delta_I)$ in the same phases as mentioned for Figs.~(a-e), respectively. The dashed dark lines indicating the power-law dependence in the distributions are drawn as a guide to eye.}
\label{dist_logself}
\end{figure}

To gain a better understanding of the statistical properties of $\Delta_I$, we also study its distribution. In Fig.~\ref{dist_logself}, we plot the distribution $P(\ln\Delta_I)$ of $\ln\Delta_I$ in the MBL [Fig.~\ref{dist_logself}(a,b)], NEE and the ergodic phases [Fig.~\ref{dist_logself}(c,e)], and at the NEE-ergodic transition [Fig.~\ref{dist_logself}(d)]. In an earlier study~\cite{logan2019many} on a model with the random disorder, $P(\ln\Delta_I)$ was found to be close to a Gaussian, i.e. $P(\Delta_I)$ is a log-normal distribution, deep in the delocalized phase. We also find the log-normal (LN) distribution to be a good description of our data, deep in the ergodic phase ($h=0.6,\E=-0.10$) as shown by the Gaussian fit in Fig~\ref{dist_logself}(e). In contrast, in the non-ergodic phases, especially in the MBL phase, $P(\ln\Delta_I)$ significantly deviates from the Gaussian fit as shown in Fig.~\ref{dist_logself}(a,b) for the MBL ($h=0.6,\E=-0.66$) and in Fig.~\ref{dist_logself}(c) for NEE ($h=0.6,\E=-0.50$) phases. At the NEE-ergodic transition ($h=0.6,\E=-0.38$), $P(\ln\Delta_I)$ looks scale-invariant, i.e., $P(\ln\Delta_I)$ becomes independent of system size, and the distribution is close to a Gaussian, as shown in Fig.~\ref{dist_logself}(d). However, a closer inspection through various non-Gaussian measures reveals deviations from the scale invariance and Gaussian distribution, as we discuss below.

The LN distribution is ubiquitous
for non-interacting systems on regular lattices with an uncorrelated disorder, both in the localized and delocalized phases \cite{Schubert2010}, and naturally arises in the non-linear $\sigma$ model description of such systems \cite{Mirlin1996,Mirlin2000}. On the FS lattice, however, the disorder is highly correlated \cite{logan2019many,roy2020fock,ghosh2019many,Altland2017} 
and $P(\Delta_I)$ is found to have a long L\'evy-like power-law tail in the thermodynamic limit in a self-consistent theory of localization on the FS lattice~\cite{logan2019many}. We also find power-law tails in $P(\Delta_I)$ as shown in Fig.~\ref{dist_logself}(f-j). The power-law tail of $P(\Delta_I)$ is also an indicator of non-Gaussianity and is quite evident in both the nonergodic phases [see Fig.~\ref{dist_logself}(f,g) and Fig.~\ref{dist_logself}(h)]. Less prominent power-law tails in $P(\Delta_I)$ can be inferred even in the ergodic phase and at the NEE-ergodic transition [see Fig.~\ref{dist_logself}(i,j)].
We find that the nonergodic-ergodic transition corresponds to a shift of the most probable value of $\Delta_I$ in $P(\Delta_I)$ from zero to non-zero values across the transition point. This is of course consistent with the behavior of $\Delta_t$ in Fig.\ref{linscaling} and Fig.\ref{energy_ldos}(a) (Appendix \ref{app4}). At the transition point $P(\Delta_I)$ for small $\Delta_I$ [Fig.~\ref{dist_logself}(i)] shows a plateau.

\begin{figure}[ht!]
\centering
\stackon{\includegraphics[width=0.4925\columnwidth,height=3.4cm]{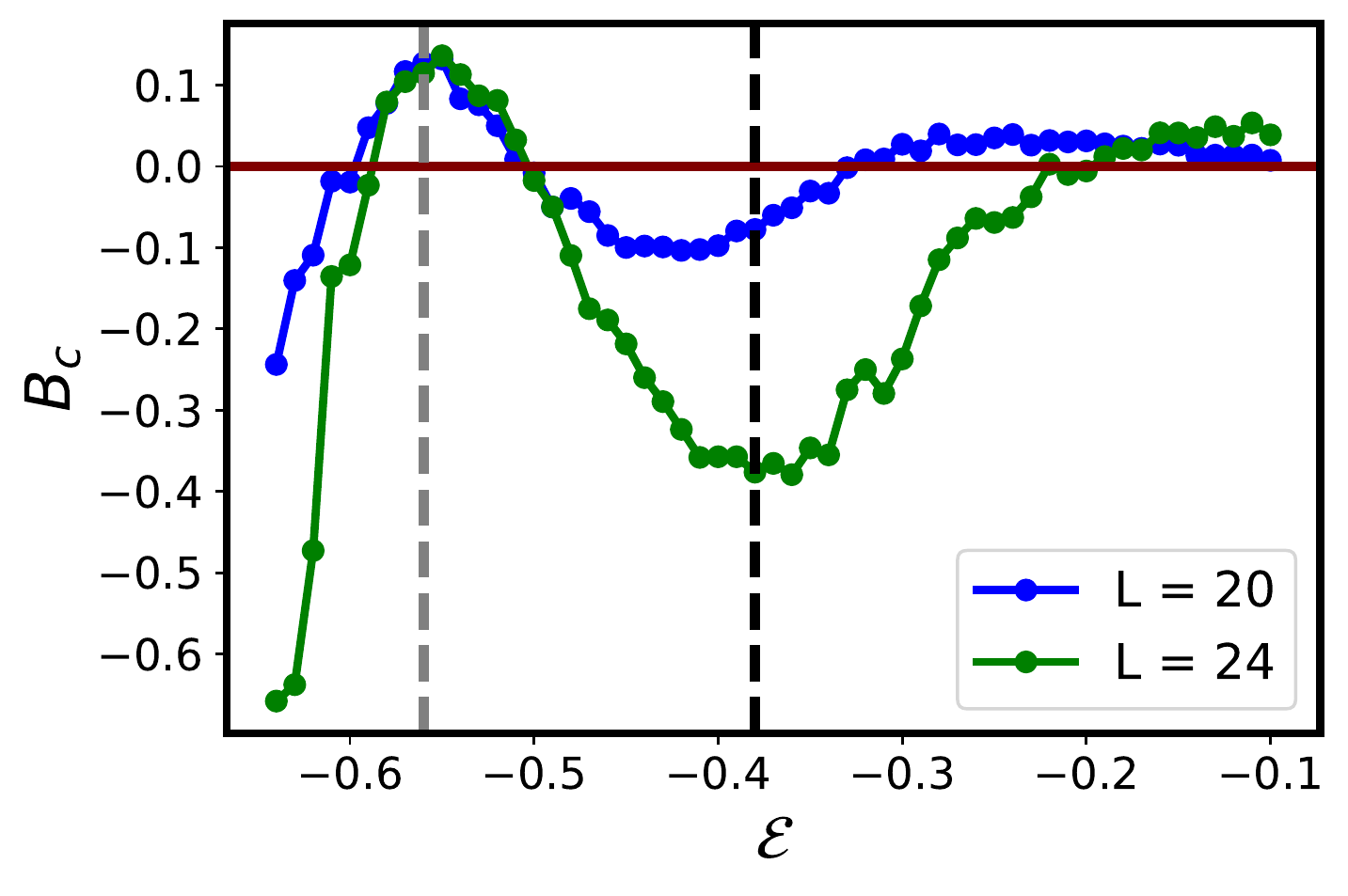}}{(a)}
\stackon{\includegraphics[width=0.4925\columnwidth,height=3.4cm]{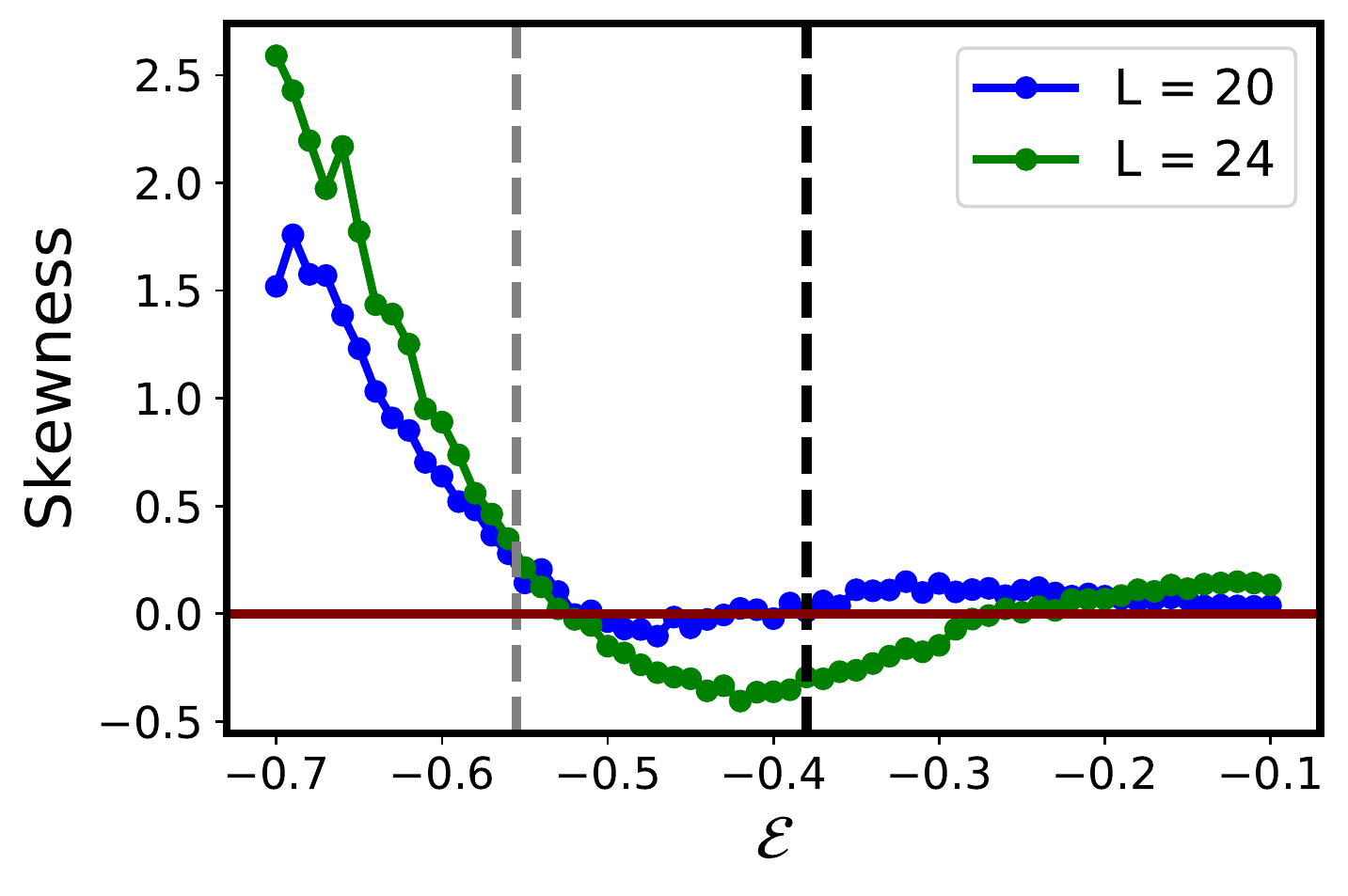}}{(b)}\\
\stackon{\includegraphics[width=0.85\columnwidth,height=3.8cm]{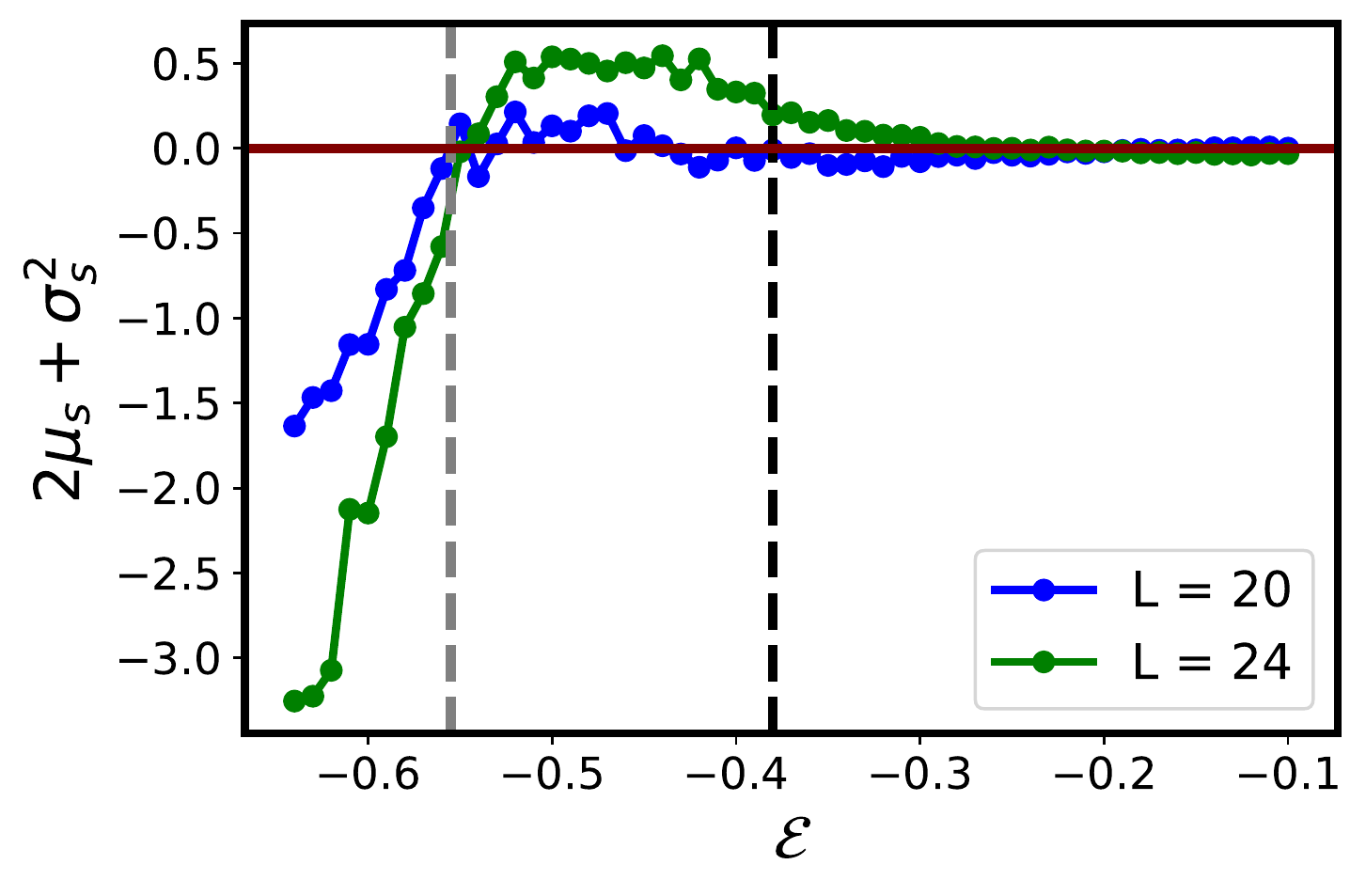}}{(c)}
\caption{{\bf Measures of non-Gaussianity of the distributions of $\Delta_I$:} (a) Binder cumulant $B_c$, and (b) skewness of $P(\ln\Delta_I)$ as a function of energy density $\E$ in the MBL, NEE and ergodic phases. (c) $(2\mu_s + \sigma^2_s)$ calculated for the distribution $P(\tilde{\Delta}_I)$ plotted as a function of energy density $\E$. The grey and dark vertical lines in all the plots denote the MBL-NEE and nonergodic-ergodic phase transitions, respectively.}
\label{dist_nongauss}
\end{figure}

We quantitatively characterize the quality of the Gaussian fits in Figs.\ref{dist_logself}(a-d) via the Binder cumulant, $B_c=1-\braket{x^4}/3\braket{x^2}^2$ for $x=\ln\Delta_I$, and the skewness. The non-zero value of $B_c$ quantifies the deviation from the Gaussian distribution, and the non-zero skewness measures the asymmetry of $P(\ln\Delta_I)$ around the mean. As shown in Fig.~\ref{dist_nongauss}(a), we find that $B_c\approx 0$ deep in the ergodic phase, whereas it is negative deep in the non-ergodic phase. Interestingly, with increasing $\E$ starting from the lowest energy density ($\simeq -0.7$), $B_c$ changes sign and then seems to reach a system size-independent positive value at $\E=\E_{nc}\approx-0.56$, which is consistent with the previous estimate~\cite{ghosh2020transport} of critical energy density for the MBL-NEE transition. We thus take $\E_{nc}$ as an estimate for the putative MBL-NEE transition \cite{ghosh2020transport}. With further increase of $\E$, $B_c$ remains negative more or less throughout the NEE phase and across the NEE-ergodic transition at $\E_c\approx-0.38$, where $|B_c|$ shows a broad peak. The magnitude of the peak increases with $L$. The skewness of $P(\ln\Delta_I)$, shown in Fig.~\ref{dist_nongauss}(b), is also $\sim 0$ deep in the ergodic phase and deviates from zero in the most part of the non-ergodic phases and close to the NEE-ergodic transition. The skewness tends to reach a system-size independent value at the putative MBL-NEE transition and changes its sign thereafter with increasing $\E$. 

Another way to verify the applicability, or lack thereof, of the LN distribution is to look at $\tilde{\Delta}_I$, namely $\Delta_I$ normalized by its (arithmetic) mean. In this case, since the distribution $P(\tilde{\Delta}_I)$ is normalized and has a unit mean, the LN distribution $P(\tilde{\Delta}_I)=\exp[-(\ln\tilde{\Delta}_I-\mu_s)^2/2\sigma^2_s]/(\sqrt{2\pi\sigma^2_s}\tilde{\Delta}_I)$ implies $2\mu_s=-\sigma^2_s$~\cite{Schubert2010}. As shown in Fig.~\ref{dist_nongauss}(c), we find this relation to be satisfied quite well deep in the ergodic phase. Approaching the ergodic-NEE transition $\E_c$ from the ergodic side, $(2\mu_s+\sigma^2_s)$ starts deviating from zero. $(2\mu_s+\sigma^2_s)$ also changes sign at the putative MBL-NEE transition $\E_{nc}\approx -0.56$ and becomes $L$ independent just like the Binder cumulant and skewness [Fig.~\ref{dist_nongauss}(a,b)]. Here it is worth mentioning that the full distribution $P(\ln\Delta_I)$ in MBL phase for two choices of parameters, as shown in  Fig.~\ref{dist_logself}(a,b), are quite different. However, the non-Gaussian measures like the Binder cumulant, skewness and $(2\mu_s + \sigma_s^2)$ have very similar value and system-size dependence.

In summary, the behaviors of the Binder cumulant, skewness, and $(2\mu_s+\sigma^2_s)$ for the distribution $P(\ln\Delta_I)$ imply the general non-Gaussian nature of the distribution, except deep in the ergodic phase. Moreover, the non-Gaussianity measures of $P(\ln\Delta_I)$ also tend to become system size independent at the MBL-NEE transition $\E_{nc}$. Hence, unlike the typical value $\Delta_t$ discussed in Secs.\ref{sec6},\ref{sec7}, the distribution of Feenberg self-energy in the FS seems to show a signature of a transition as a function of energy density $\E$ at $\E_{nc}$ for $h=0.6$. This is in spite of the fact that the states for $h=0.6$ and $\E<\E_{nc}$ show anomalous behaviors, in between MBL and NEE, when all the other diagnostics, like entanglement entropy, subsystem particle number fluctuations~\cite{ghosh2020transport}, level spacing statistics (Appendix~\ref{app1}) and real-space single-particle excitations [Fig.\ref{mbl_proximity}((d)], are combined. Here we would like to note that, currently, we do not have any theoretical understanding of behavior of the distribution of the FS Feenberg self-energy and its various Gaussian/non-Gaussian characteristics in the MBL, NEE, and ergodic phases and across the transitions. We report here the apparent scale invariance of the non-Gaussian characteristics at the putative MBL-NEE transition as interesting observations. In the future, it will be worthwhile to get a better theoretical understanding of the distribution of the local self-energy and find out suitable finite-size scaling ansatzes for the non-Gaussianity parameters across the MBL-NEE and NEE-ergodic transition.

In the next section, we show that another distinction between the MBL and NEE states can be obtained from the system-size dependence of an FS localization length.



\section{Fock-space localization length}\label{sec10}
\begin{figure}[h]
\centering
\includegraphics[width=0.495\columnwidth,height=3.6cm]{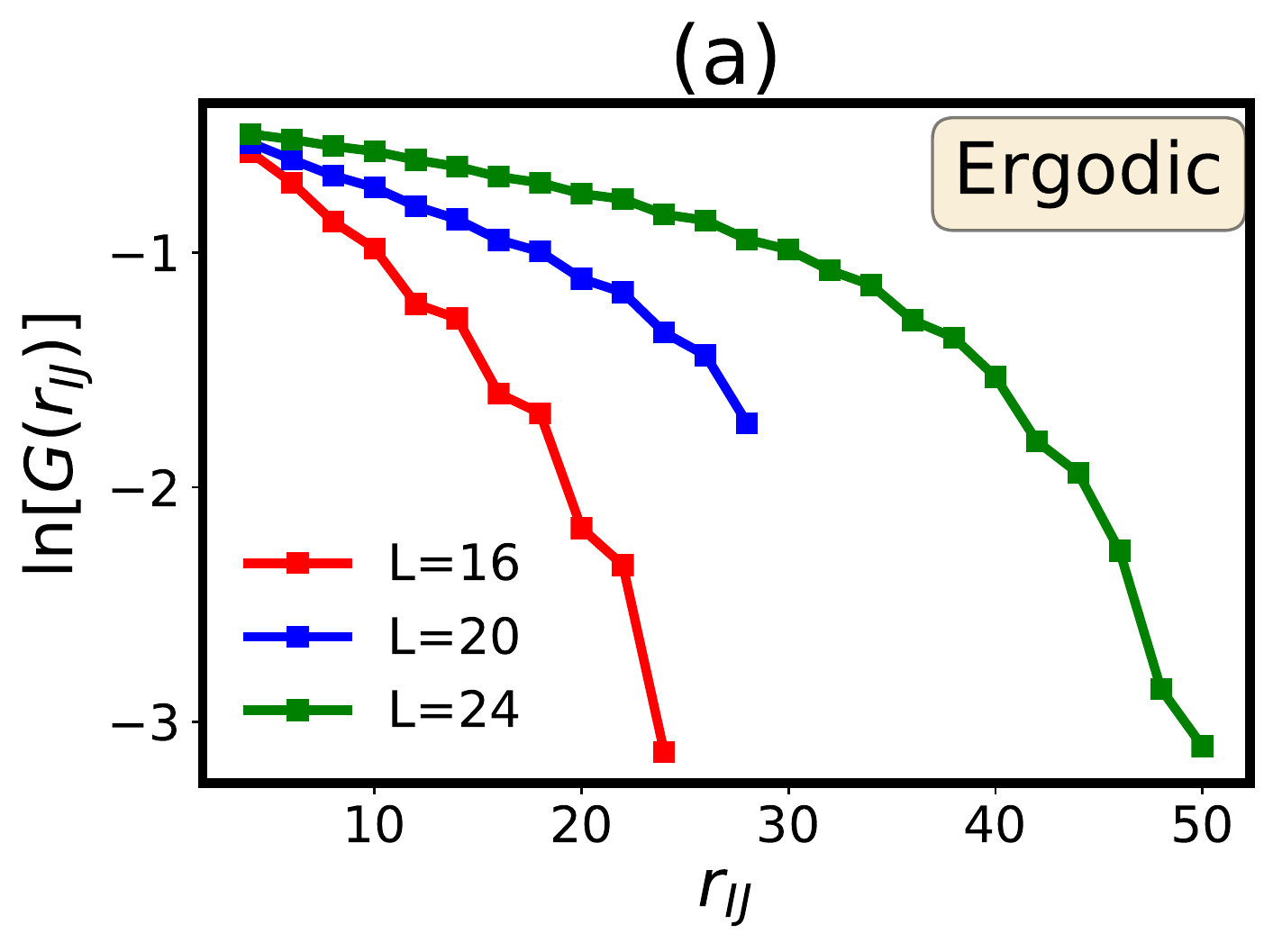}
\includegraphics[width=0.49\columnwidth,height=3.6cm]{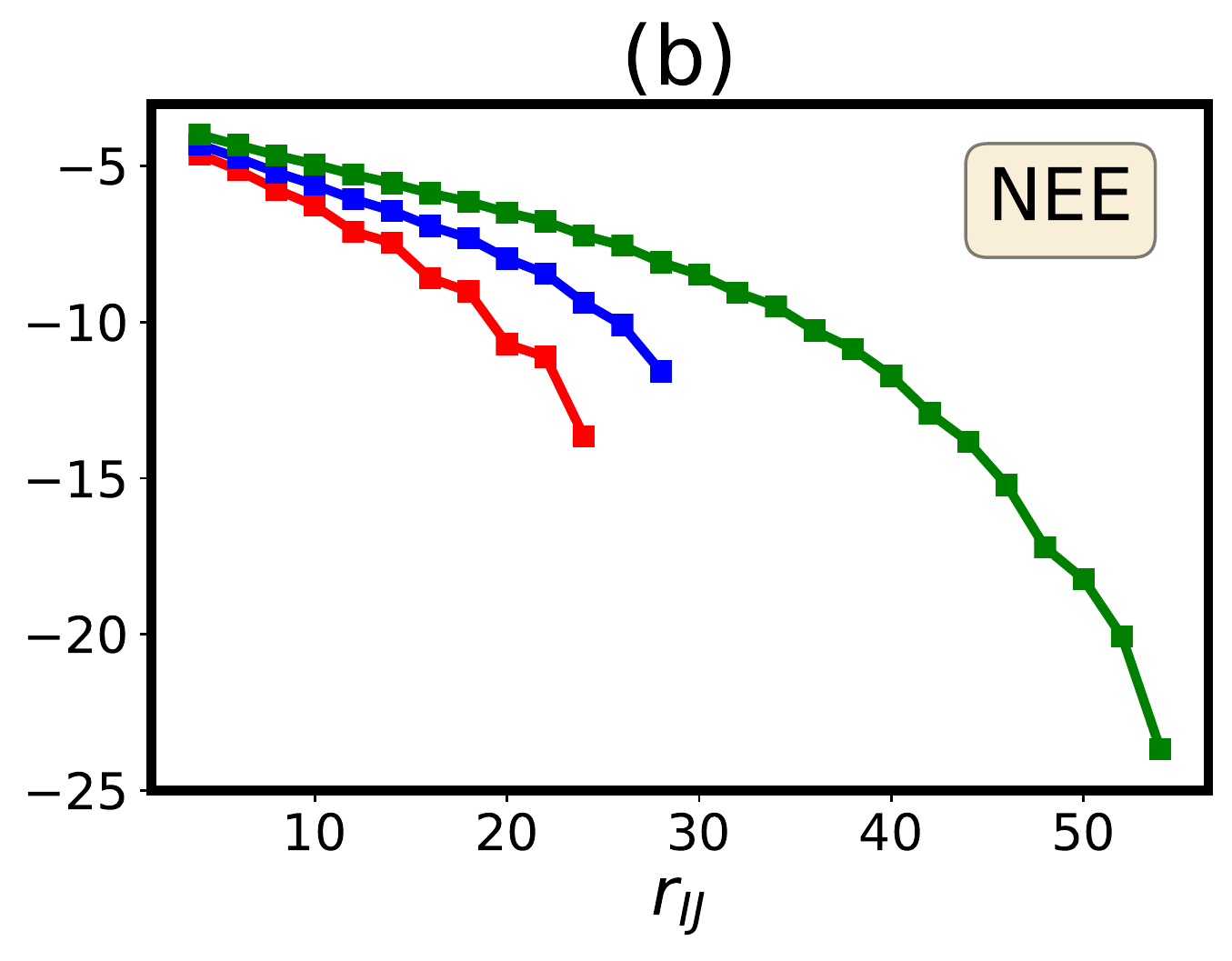}
\includegraphics[width=0.495\columnwidth,height=3.6cm]{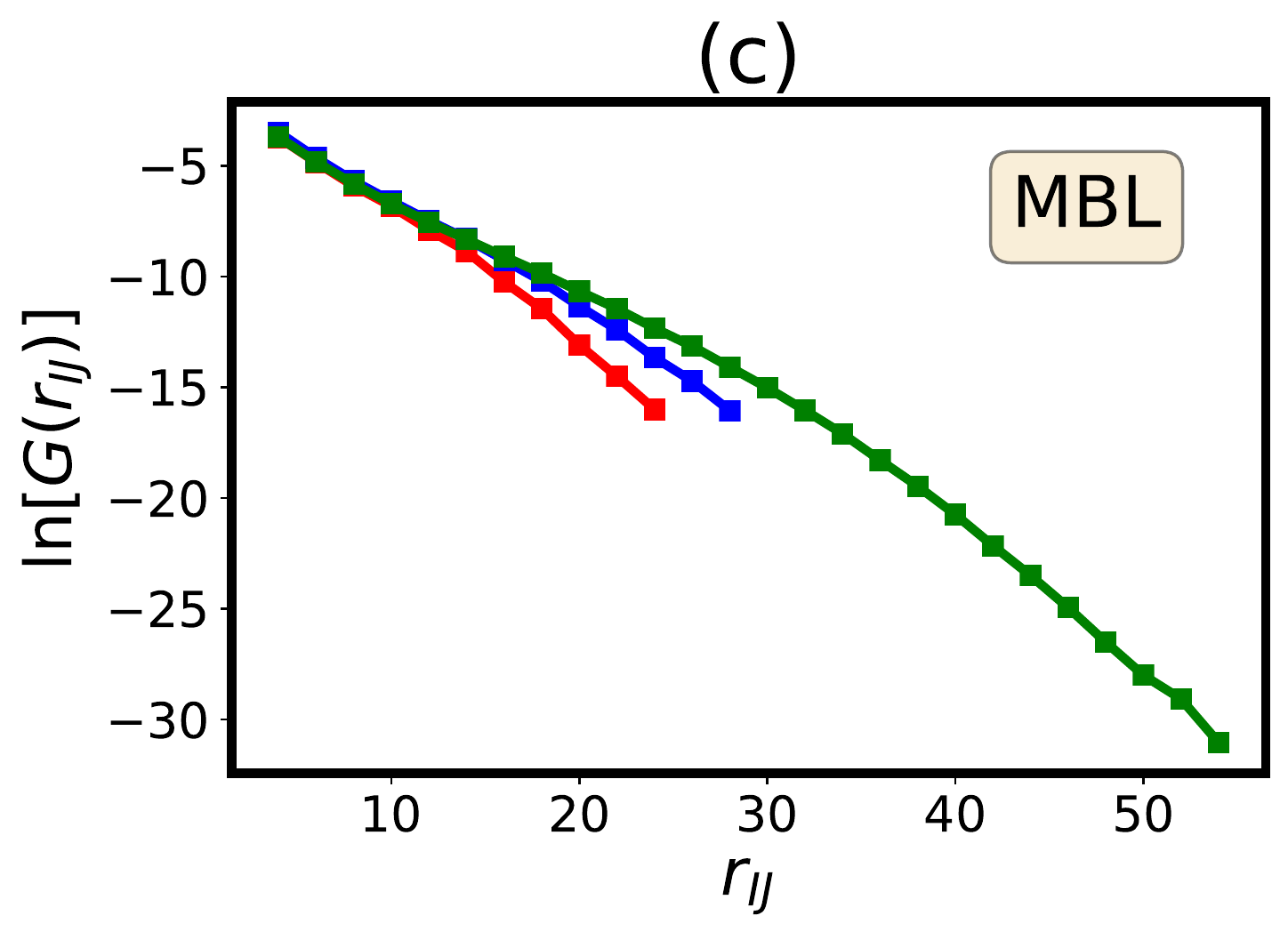}
\includegraphics[width=0.49\columnwidth,height=3.6cm]{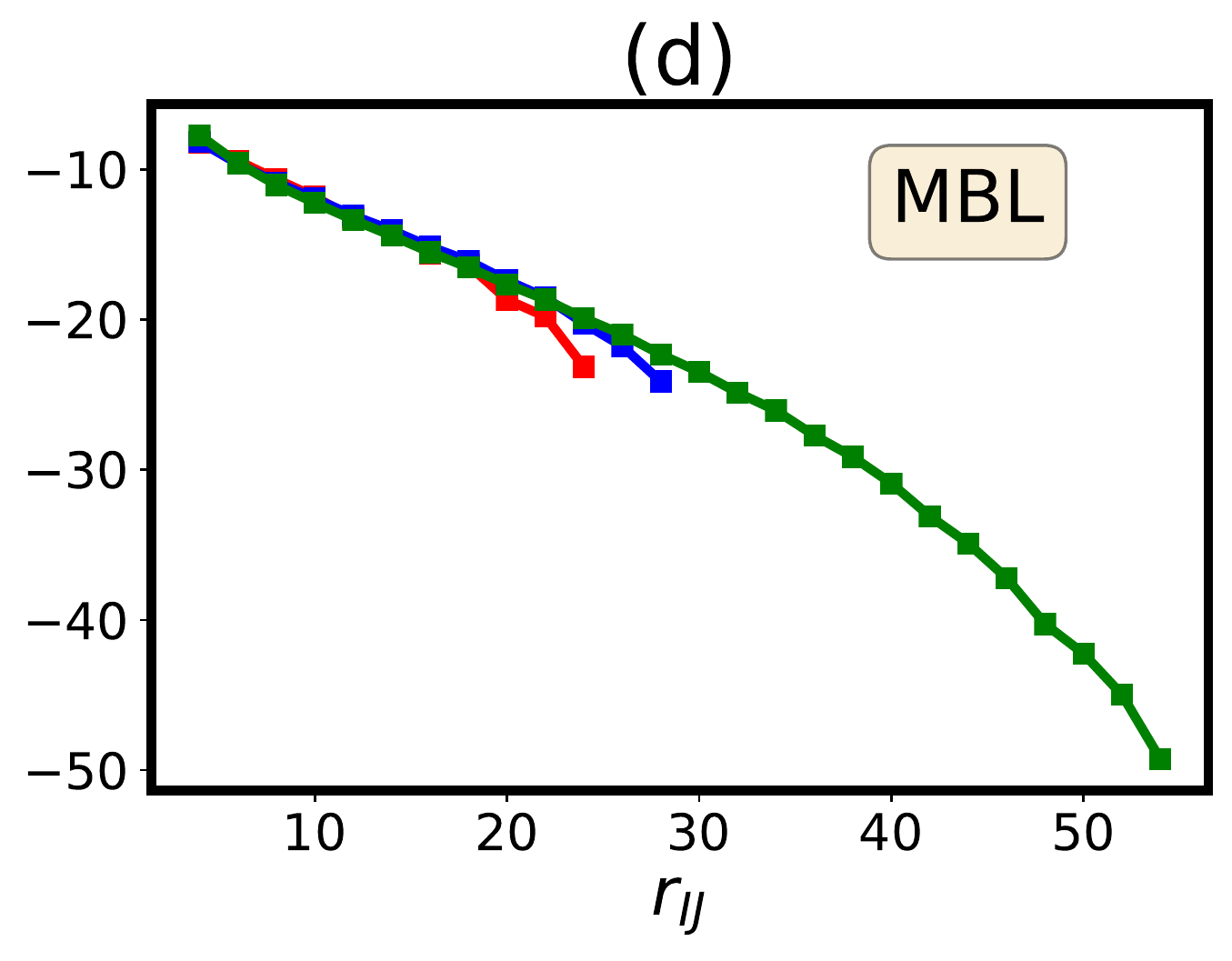}
\caption{{\bf Fock-space localization length:} (a-d) Variation of $\ln[G(r_{IJ})]$ with FS-hopping distance $r_{IJ}$ in the ergodic $(h=0.6, \E=-0.18)$, NEE $(h=0.6, \E=-0.46)$ and MBL phases $[(h=1.8, \E=-0.49), (h=0.6, \E=-0.66)]$, respectively. 
$G(r_{IJ})$ for the smallest $r_{IJ}$ has been scaled such that  plots for different $L$ all start from the same point.
}
\label{localization_length}
\end{figure}

We now show that, unlike $\Delta_t(L)$, the typical non-local propagator $G(r_{IJ})=\exp{[\langle \ln G_{IJ}(\E)\rangle]}$ can tell NEE and MBL states apart. Here $\langle \dots\rangle$ denotes average over $\phi$ and all the off-diagonal elements $G_{IJ}$ for pair of FS sites connected by hopping distance $r_{IJ}$, i.e. the minimum number of nearest-neighbor hops to reach from $I$ to $J$ on the middle slice. Fig.~\ref{localization_length}(a-d) show $\ln[G(r_{IJ})]$ as a function of $r_{IJ}$ for increasing $L$ in the ergodic, NEE and MBL [Fig.~\ref{localization_length}(c-d)] phases, respectively, for $(h,\E)$ values same as in Fig.~\ref{mbl_proximity}(a-d). In all the phases, the plots show a linear regime, $\ln G(r_{IJ})\propto r_{IJ}$, before deviating from the linearity at larger $r_{IJ}$ depending on $L$. The deviation of linearity of $\ln G(r_{IJ})$ vs. $r_{IJ}$ for larger $r_{IJ}$'s in Fig.~\ref{localization_length}(a-c) presumably corresponds to rare hopping paths and associated multiple length scales in the FS. The linear regime implies existence of an FS decay length $\xi_F(L)$ through the relation $G(r_{IJ})\sim \exp{[-r_{IJ}/\xi_F(L)]}$. The curves for different $L$ approximately overlap for a certain range of $r_{IJ}$ in the MBL phase ($h=1.8,~\E=-0.49$)[Fig.~\ref{localization_length}(c)] indicating a localization length $\xi_F$ almost independent of $L$ or weakly dependent on $L$, unlike the decay length in the ergodic ($h=0.6,~\E=-0.18$) and NEE phases ($h=0.6,~\E=-0.46$) in Figs.~\ref{localization_length}(a,b), where $\xi_F$ evidently has quite strong dependence on $L$.

In Fig.~\ref{localization_length}(d), we show $\ln[G(r_{IJ})]$ as a function of FS-hopping distance $r_{IJ}$ for $h=0.6$ and $\mathcal{E}=-0.66$. Similar to Fig.~\ref{localization_length}(c), which corresponds to the MBL phase, in Fig.~\ref{localization_length}(d), we also find that all the curves in the linear regime, $\ln G(r_{IJ})\propto r_{IJ}$, for different $L$  almost overlap, implying $G(r_{IJ})\sim exp(-r_{IJ}/\xi_F)$ with $L$-independent $\xi_F$. Hence, the states for ($h=0.6,~\E=-0.66$) [Fig.~\ref{localization_length}(d)] show MBL behavior, consistent with that of $S_A$~\cite{ghosh2020transport}, unlike the NEE-like behavior seen in single-particle excitations spectrum [Fig.~\ref{mbl_proximity}(d)] and in level-spacing statistics [Fig.~\ref{levelstat}(d)].  

\begin{figure}
\includegraphics[width=0.8\columnwidth,height=4.7cm]{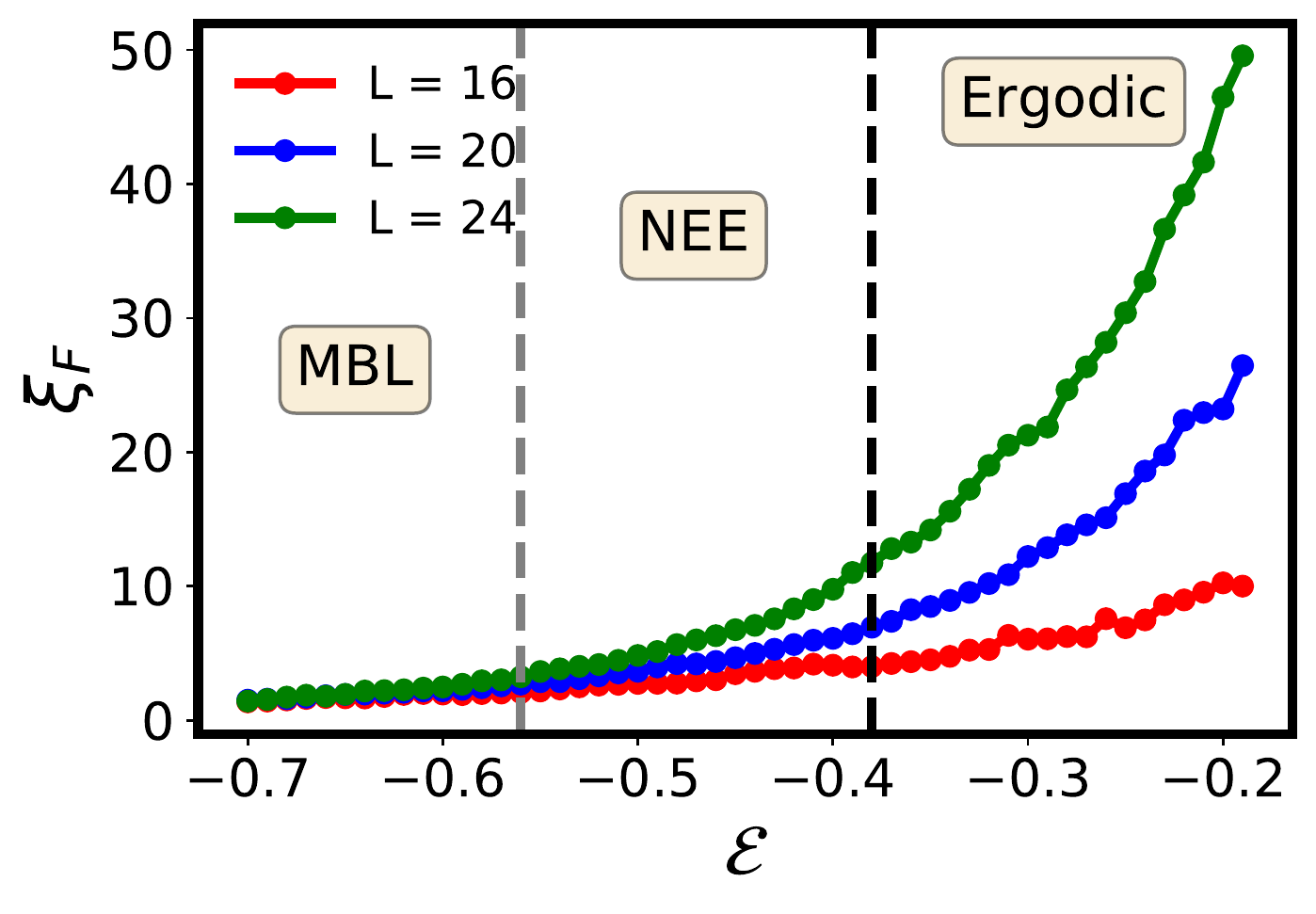}
\caption{{\bf Fock-space transitions from localization length:} System-size $L$ dependence of FS localization length $\xi_F$ extracted from $G_{IJ}$ (see main text) as a function of $\E$ for $h=0.6$. The grey and dark dashed vertical lines denote the MBL-NEE and NEE-ergodic transitions at $\E_{nc}\approx-0.56$ and $\E_c\approx-0.38$, respectively.}
\label{llength}
\end{figure}
We show the $L$-dependence of $\xi_F$ as a function of $\E$ in Fig.~\ref{llength} for $h=0.6$ across the MBL, NEE, and ergodic phases. It is evident that $\xi_F$ is almost independent of $L$ in the MBL phase, whereas $\xi_F$ increases with $L$ in the NEE and ergodic phases. Based on this, and the non-Gaussianity of the distribution of $\ln\Delta_I$ [Sec.~\ref{sec8}], we deduce an MBL-to-NEE transition around $\E\approx -0.56$ consistent with previous estimates \cite{ghosh2020transport}. We leave a detailed finite-size scaling analysis of the FS localization length across the MBL-NEE and NEE-ergodic phase transitions for future study. 


\section{Discussion and Conclusions}\label{sec11}
In summary, by combining the diagnostics of real-space single-particle excitation, and multifractality, statistics of local self-energy and decay length in Fock space, we provide a direct signature of MBL proximity effect and classification of the MBL, NEE, and ergodic phases in the GAAH model. 

Like all \emph{first-principle} numerical studies, e.g. via ED, of MBL phenomena and MBL-ergodic transition, our computations of standard diagnostics, like entanglement entropy, and single-particle excitation spectrum are limited to relatively small systems. Our exact recursive Greens' function calculations for FS propagators access slightly larger systems, almost comparable \cite{sutradhar2022scaling} to those achieved by state-of-the-art shift-invert \cite{luitz2015many} and polynomially filtered \cite{Sierant2020} ED studies. The latter can access up to $L\lesssim 24$ for random Heisenberg or XXZ models. However, ref.\onlinecite{Sierant2020} estimates a system size $L\gtrsim 50$ to conclusively access the MBL-ergodic transition. But, even this estimation is only based on the extrapolation of the data for $L\leq 24$. The same issues of strong finite-size effects, of course, exist for extrapolating the classification of the MBL, NEE and ergodic phases and the characterization of the phases transitions in the GAAH model to the thermodynamic limit $L\to \infty$ in all previous studies \cite{li2015many,modak2015many,Li2016,Deng2017,Modak2018,Nag2017,ghosh2020transport,Deng2019,Pomata2020}, and in our current work.

The finite-size artifacts in small-system exact numerical studies have become more of a concern recently since the region of stability of the MBL phase accessed in the ED studies of models with the random disorder has been questioned \cite{Polkovnikov,Huse,Chandran,Chandran2,Sels2022}. It has been argued that the MBL transition captured via ED is a finite-size crossover due to the possibility of long-range resonances and avalanche instability from rare weak-disorder regions in larger system sizes, not accessible via ED. In the same vein, the region of stability of the non-ergodic phases in the GAAH model could also be affected by long-range resonances in larger system sizes. 
However, unlike the systems with the random disorder, the quasiperiodic systems are not susceptible to the usual avalanche instability \cite{deRoeck} due to the absence of rare weak-disorder regions \cite{agarwal}. Hence the non-ergodic phases, like the MBL phase, in the GAAH model could be more robust. However, the NEE states may be more fragile than the MBL states and become unstable in the thermodynamics limit \cite{Pomata2020}. For example, the stability of the NEE phase has been highly debated and led to huge controversies~\cite{Biroli2012,DeLuca2014,Altshuler2016,altshuler2016multifractal,Tikhonov2016,Tikhonov2016a,Sonner2017,garcia2017scaling,Tikhonov2019,Tikhonov2021} for non-interacting fermions on locally tree-like graph with uncorrelated random disorder.  

Nevertheless, our results of single-particle excitations, Fock-space multifractality, self-energy statistics and localization length suggest curious distinctions between MBL, NEE, and ergodic phases in the GAAH model for finite systems, and put forward the NEE phase an intriguing possibility if it persists in the thermodynamic limit. However, we note that
the distinction between MBL and NEE phases as a function of many-body energy density in the GAAH model is not easy to
capture within the system sizes accessed in our current work, and would require larger systems and finer energy binning in future.

Moreover, recently, the GAAH model has been realized experimentally in cold-atomic systems of bosons with mean-field-like interactions~\cite{an2021interactions}. The single-particle mobility edge, weakly renormalized by the mean-field interactions, has also been seen in this experiment. In the future, it will be interesting to realize the interacting GAAH model [Eq.\eqref{ham}] in a similar setup. While the FS diagnostics, of course, cannot be probed in the experiment, our results on the single-particle excitations in Sec.\ref{sec4b} suggest a feasible way to distinguish MBL, NEE, and ergodic phases based on the existence, or lack thereof, of the single-particle mobility edge in the interacting system. In Ref.\onlinecite{an2021interactions}, the mobility edge has been detected through the participation ratio. In future, with development and improvement of energy-resolved spectroscopies~\cite{Kollath2007,Stewart2008,Gaebler2010,Perali2011,Jiang2011} for the cold-atomic systems, it might be possible to probe LDOS. Theoretically, it will be an interesting future direction to study short- and long-range resonances~\cite{Garratt2022,Garratt2021,Huse,Chandran,Chandran2}, as well as more subtle Fock-space correlations and their connections to entanglement \cite{Tomasi2020,Roy2022} in the non-ergodic phases of the GAAH model.

\section*{Acknowlegement}
We acknowledge useful discussions with Soumi Ghosh, Sthitadhi Roy, David Logan, and Subroto Mukerjee during collaboration on related topics. NR thanks Indian Institute of Science for support through IoE-IISc fellowship program. SB acknowledges support from SERB (ECR/2018/001742), DST, India. JS and SB acknowledge support from QuST, DST, India.



\appendix
\renewcommand{\theequation}{\thesection.\arabic{equation}}
\renewcommand{\thefigure}{A\arabic{figure}}

\setcounter{figure}{0}


\begin{figure}[ht!]
\centering
\stackon{\includegraphics[width=0.495\columnwidth,height=3.7cm]{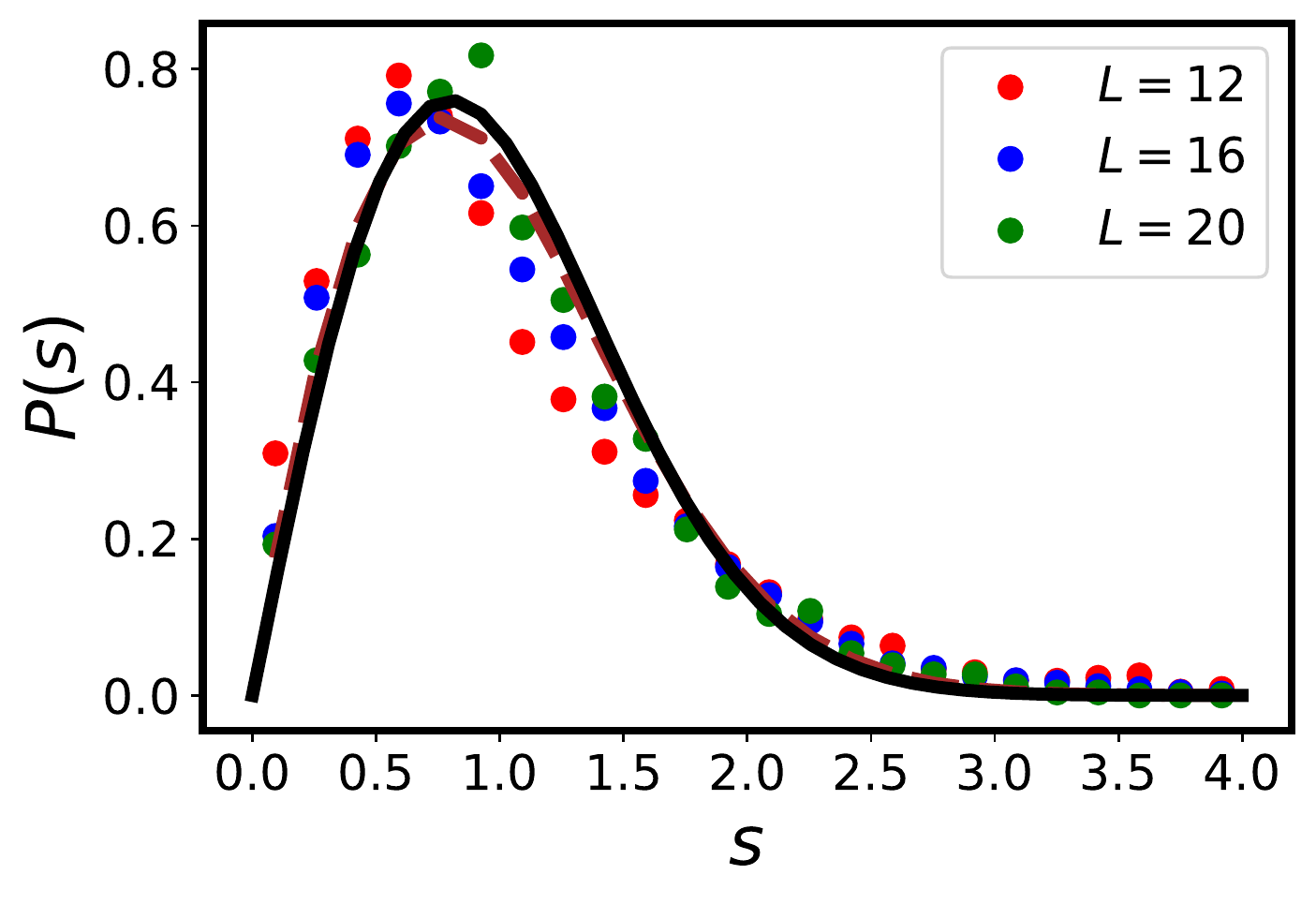}}{(a)}
\stackon{\includegraphics[width=0.49\columnwidth,height=3.7cm]{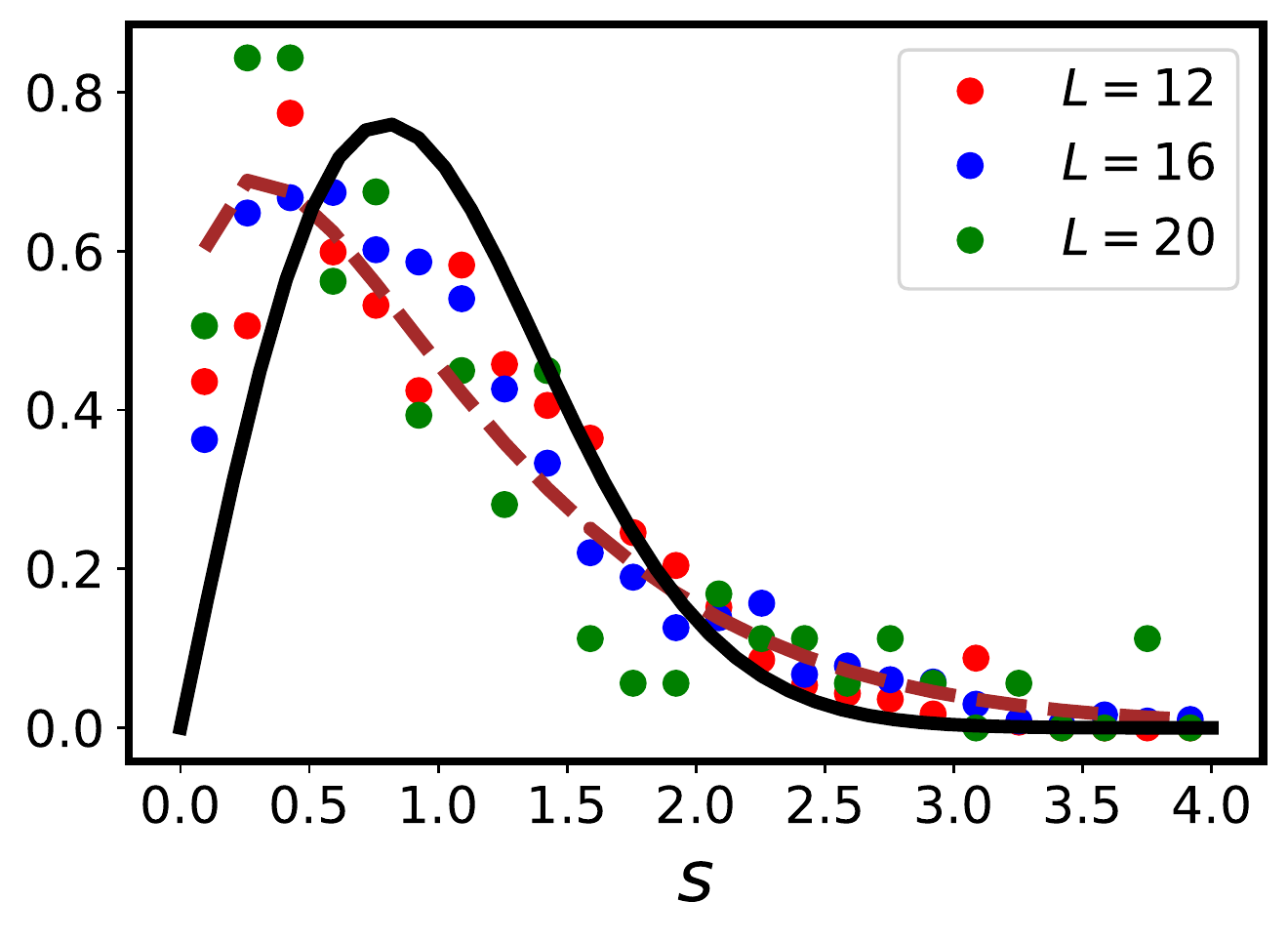}}{(b)}
\stackon{\includegraphics[width=0.495\columnwidth,height=3.7cm]{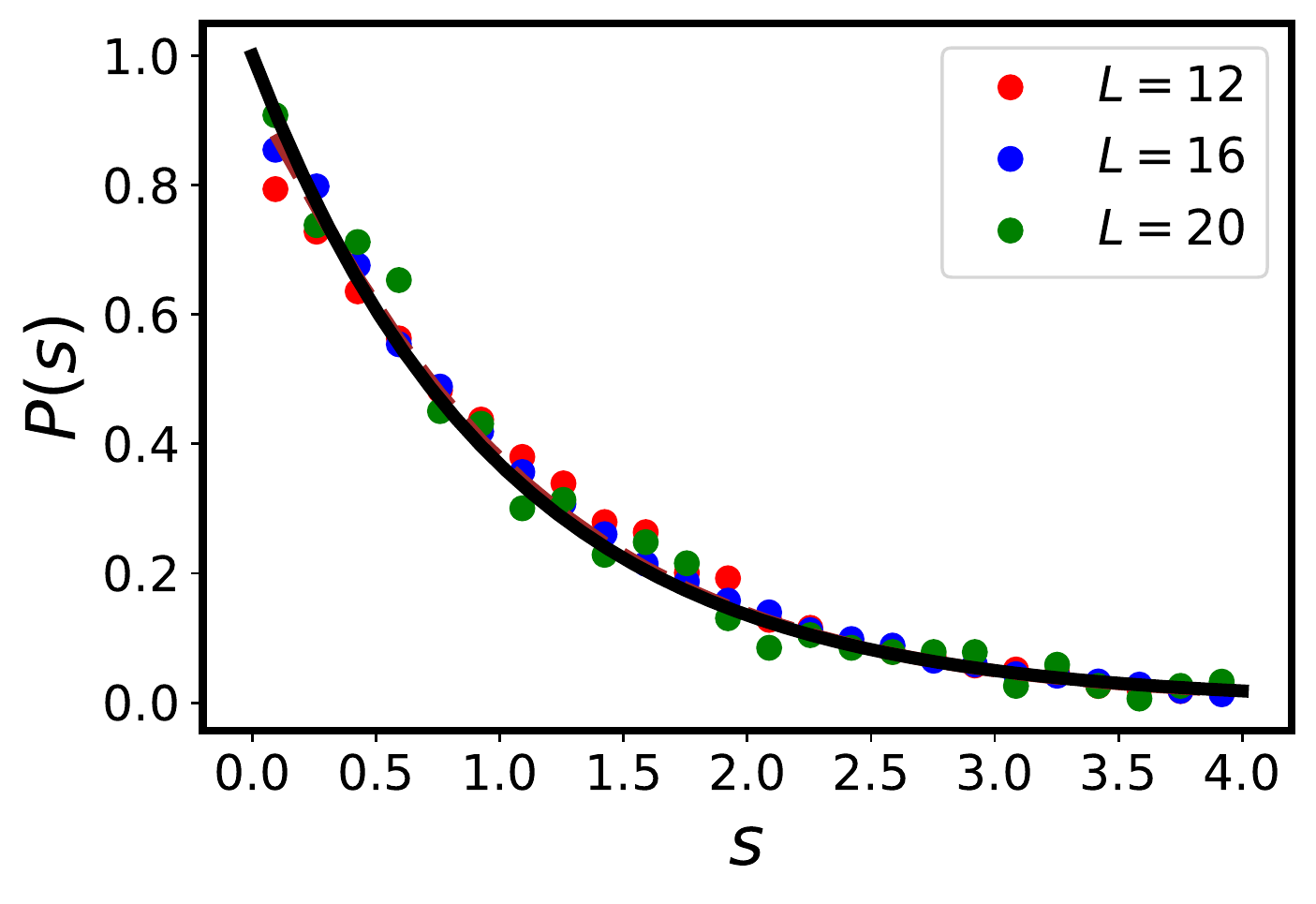}}{(c)}
\stackon{\includegraphics[width=0.49\columnwidth,height=3.7cm]{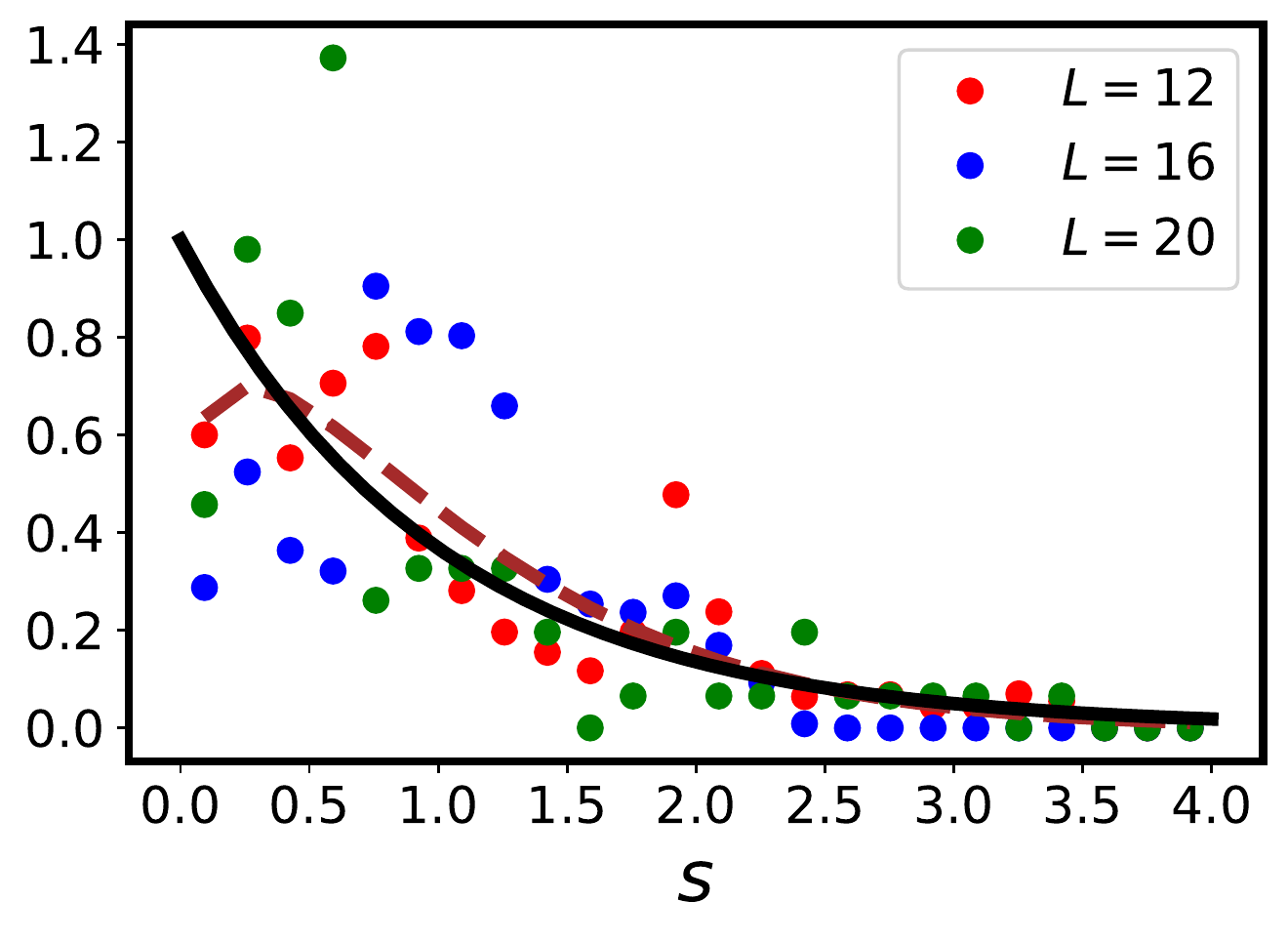}}{(d)}
\caption{{\bf Levelstatistics:} (a-d) Probability distribution $P(s)$ of consecutive energy-level spacing for increasing system sizes $L=12,16,20$ and parameters $(h,\E)$ used in Fig.~\ref{mbl_proximity}(a-d), respectively. The solid dark lines represent GOE distribution whereas the dashed lines represent the curves with data fitted to the Brody distribution for $L=20$.}
\label{levelstat}
\end{figure}

\section{Energy level statistics in different phases}\label{app1}
As already mentioned, it is hard to use the level spacing ratio $r$ for any controlled finite-size scaling analyses as $r$ strongly fluctuates in the non-ergodic phases of the GAAH model [Fig.~\ref{old_measures}(d)]~\cite{ghosh2020transport}. In Fig.~\ref{levelstat}(a-d) we show the distributions $P(s)$ of consecutive energy-level spacing $s$, normalized by the (arithmetic) mean level spacing, for the same values of $(h,\E)$ as in Fig.~\ref{mbl_proximity}(a-d). For system size $L=20$ we fit the data with the Brody distribution given by $P(s)=As^a\exp(-As^{a+1}/(a+1))$ where $A=(a+1)\Gamma(\frac{a+2}{a+1})^{(a+1)}$. $a=1$ and $a=0$ correspond to GOE and Poisson distributions, respectively. From Fig.~\ref{levelstat}(a) for ergodic phase ($h=0.6$,~$\E=0$) we find $a\approx0.91$ approaching the GOE distribution. From Fig.~\ref{levelstat}(b) we find that the data in the NEE phase ($h=0.6$,~$\E=-0.49$) do not really fit well to Brody's distribution and substantially deviate from either of GOE and Poisson distributions. 
Fig.~\ref{levelstat}(c) shows Poisson distribution with $a\approx0.02$ in the MBL ($h=1.8$,~$\E=-0.49$) phase. On the contrary, in Fig.~\ref{levelstat}(d), for ($h=0.6$,~$\E=-0.66$), which is expected to be in the MBL phase \cite{ghosh2020transport}, the distribution does not conform to the Poisson distribution, as discussed earlier.

\section{Semilog plots of $\rho_t$}
\label{app2}
\begin{figure}[h!]
\centering
\stackon{\includegraphics[width=0.495\columnwidth,height=3.7cm]{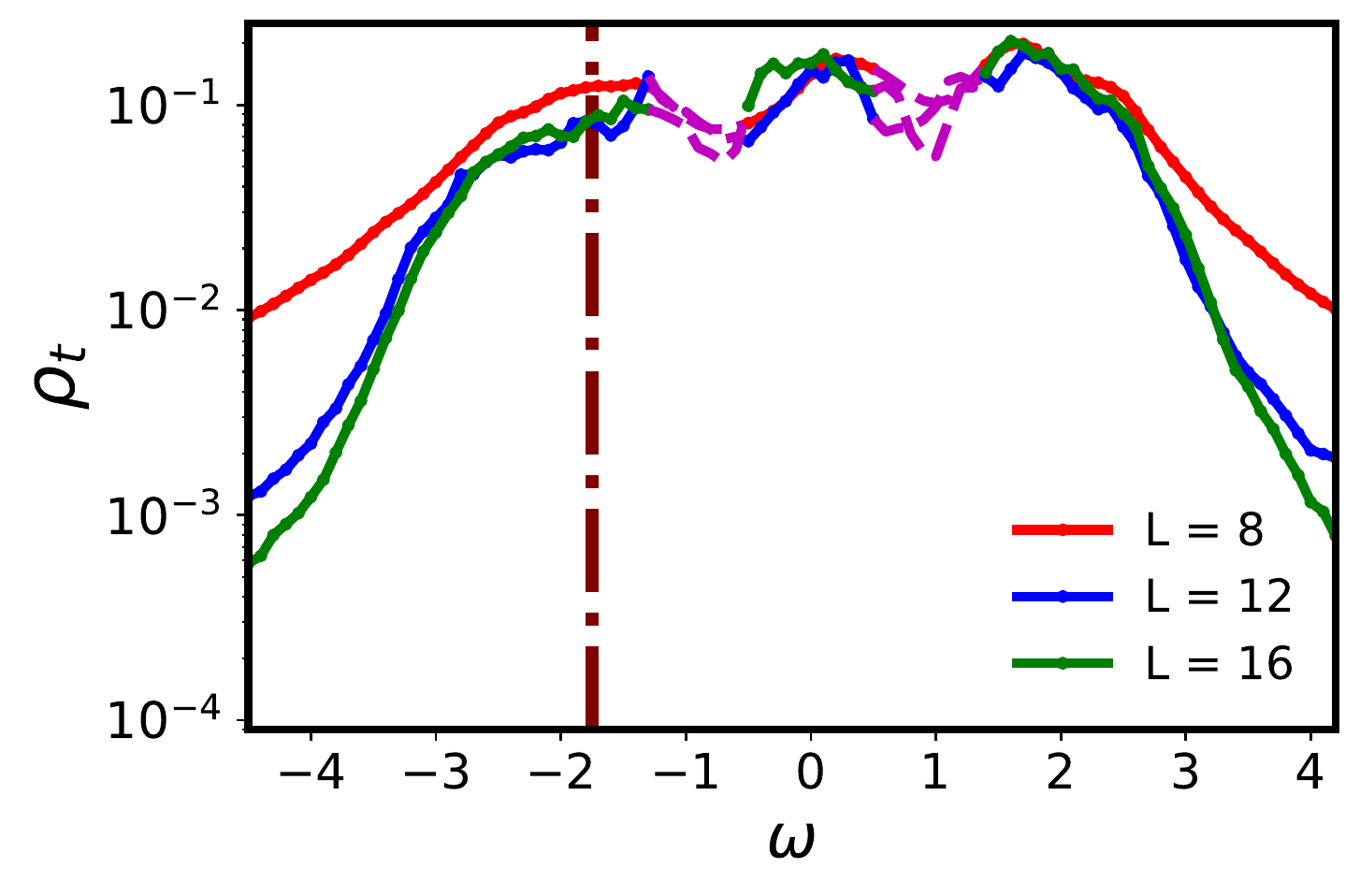}}{(a)}
\stackon{\includegraphics[width=0.49\columnwidth,height=3.7cm]{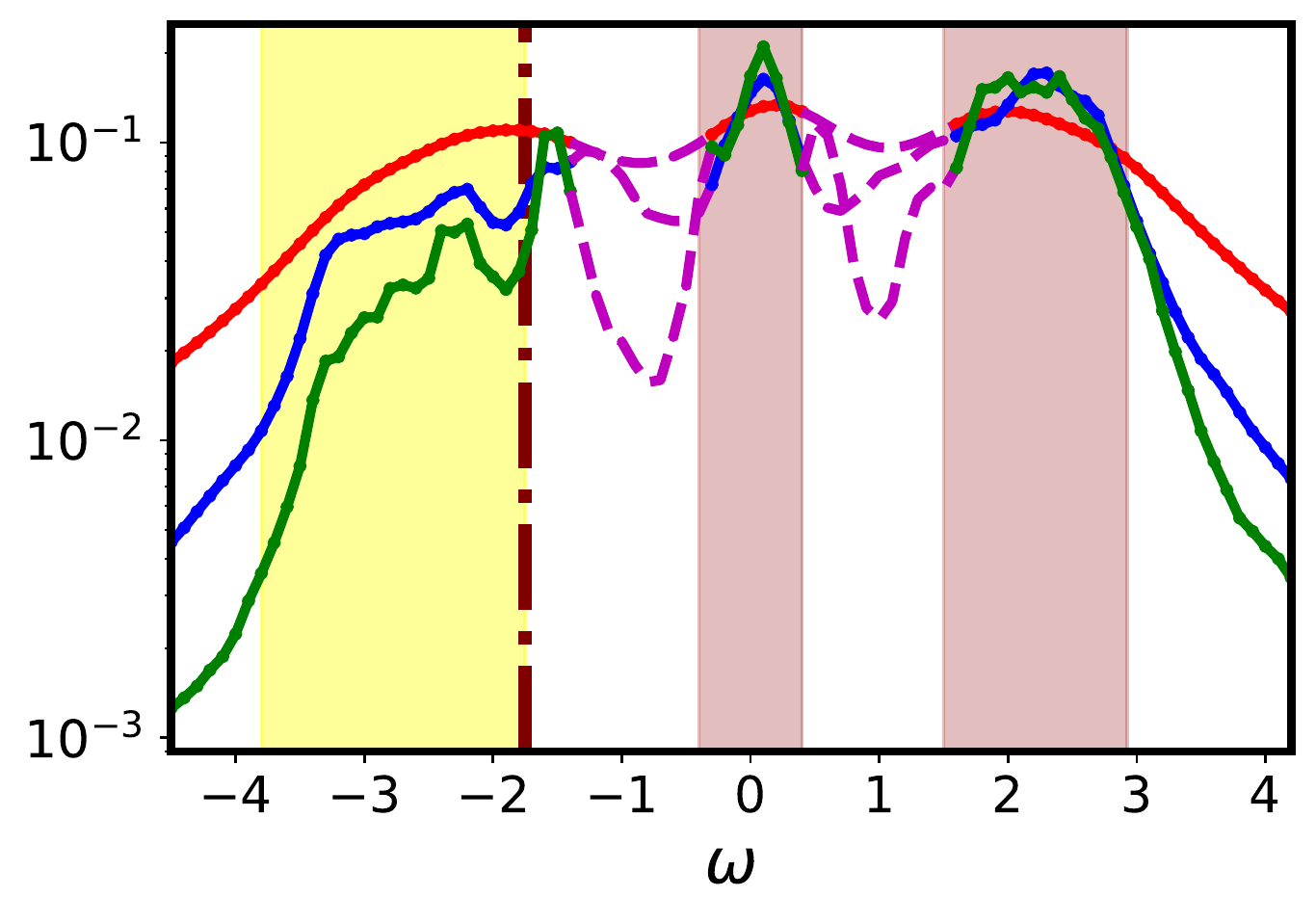}}{(b)}
\caption{{\bf Fig.~\ref{mbl_proximity}(a-b) in semilogscale:} (a-b) $\rho_t(\omega)$ vs. $\omega$ for (a) ergodic phase ($h=0.6,~\E=0$) and (b) NEE phase ($h=0.6,~\E=-0.49$) with increasing system size. For the NEE phase, $\rho_t$ decreases in the yellow shaded region, unlike the ergodic state where it saturates everywhere. The vertical dot-dashed lines represent the non-interacting single-particle mobility edge $\epsilon_c$.}
\label{logrhotyp}
\end{figure}

In Figs.~\ref{logrhotyp}(a-b) we replot Figs.\ref{mbl_proximity} with the typical LDOS $\rho_t$ in logscale as a function of $\omega$ for the ergodic [Fig.~\ref{logrhotyp}(a)] and NEE [Fig.~\ref{logrhotyp}(a)] states, respectively, to clearly bring out the existence of a mobility edge in single-particle excitations in the NEE phase. The plots make it evident that for $\omega$ within the bandwidth ($|\omega|\lesssim 4$) of the system, in the NEE phase, $\rho_t$ decreases with $L$ for single-particle excitations below $\omega\approx \epsilon_c$ and saturates for single-particle excitations above $\omega\approx \epsilon_c$. On the other hand, for ergodic state $\rho_t(\omega)$ tends to saturate with $L$ for all the single-particle excitations.  

\section{Many-body density of states and the choice of the broadening parameter $\eta$}\label{app3}
Here we show that the disorder-averaged many-body density of states (MDOS) in the quasiperiodic GAAH model approaches a Gaussian function in the thermodynamic limit, similar to that in the models with random disorder~\cite{welsh2018simple,logan2019many}.
The Gaussian MDOS is given by~\cite{logan2019many}
\begin{eqnarray}
D(E)=\frac{1}{\sqrt{2\pi}\mu_E} \exp\bigg[{-\frac{(E-\bar{E})^2}{2\mu_E^2}}\bigg] ,
\label{gauss_MDOS}
\end{eqnarray} 
where $E$ is the many-body energy. $\bar{E}$ and $\mu_E$ are the mean energy and the standard deviation, respectively. An analytical estimate ($\mu_E^{Th}$) for the standard deviation in energy for our one-dimensional model can be obtained following method similar to that described in ref.~\onlinecite{welsh2018simple}. 
\begin{figure}[h!]
\centering
\includegraphics[width=0.995\columnwidth,height=3.9cm]{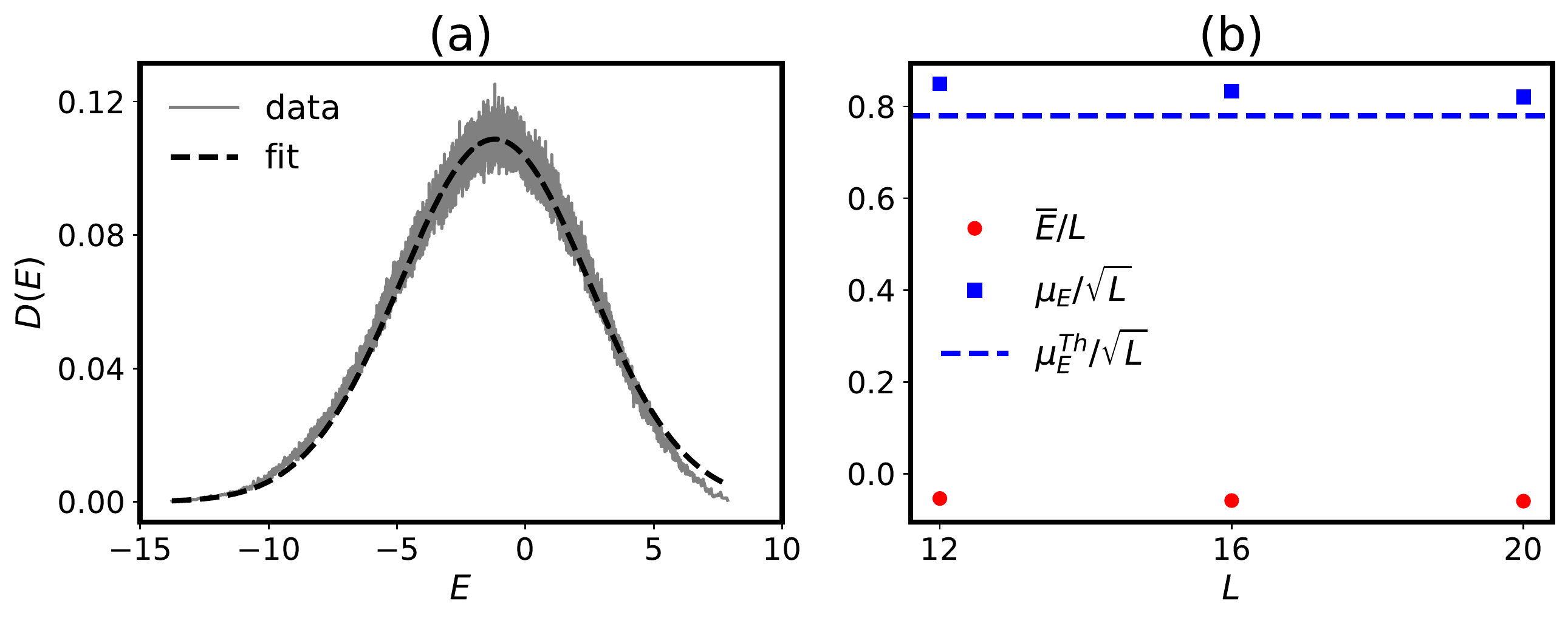}
\includegraphics[width=0.995\columnwidth,height=3.9cm]{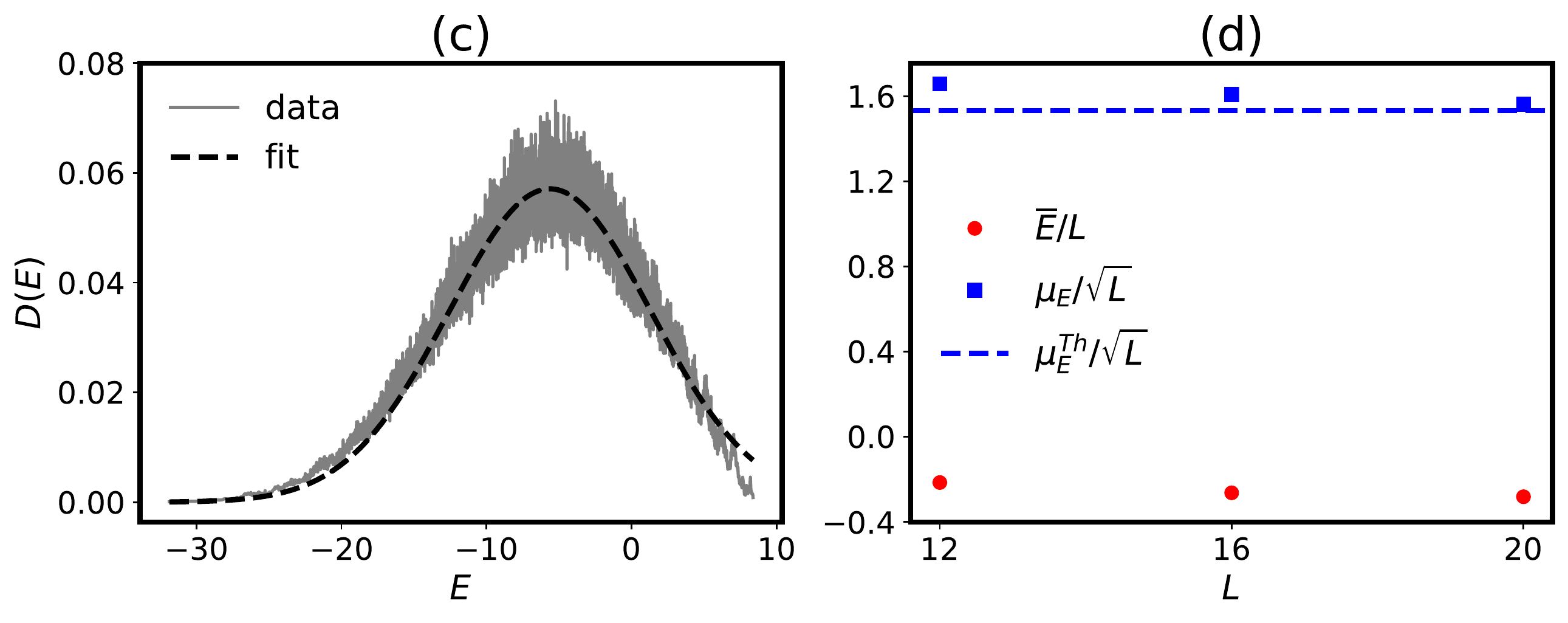}
\caption{{\bf Gaussian many-body density of states:} (a) The average many-body density of states $D(E)$ vs many-body energy $E$ for $h=0.6$ and $L=20$. The dashed line is a Gaussian fit to the data. (b) The values of $\bar{E}/L$, $\mu_E/\sqrt{L}$ extracted from Gaussian fit and theoretical value of $\mu_E^{Th}/\sqrt{L}$ as a function of $L$ for $h=0.6$ . (c) $D(E)$ vs $E$ and a Gaussian fit for $h=1.8$ and $L=20$. (d) $\bar{E}/L$, $\mu_E/\sqrt{L}$ and $\mu_E^{Th}/\sqrt{L}$ as a function of $L$ for $h=1.8$ . The disorder realizations over $\phi$ are $16000,4000$ and $120$ for $L=12,16$ and $20$ respectively.}
\label{MDOS}
\end{figure}
The expression is given by   
\begin{eqnarray}
{\mu_E^{Th}}
= \sqrt{L}&&\bigg[2t^2f(1-f) + f(1-f)  \langle\epsilon^2\rangle + V^2 f^2(1-f)^2 \nonumber\\
&&+ 4V\langle\epsilon\rangle f^2(1 - f)\bigg]^{1/2} ,
\label{mu_theory}
\end{eqnarray}
with filling fraction $f$ and disorder-averaged onsite potential $\langle\epsilon\rangle$. In case of random disorder and quasiperiodic AAH model $\langle\epsilon\rangle=0$. For the GAAH model, $f=1/4$ and $\langle\epsilon\rangle=-5h/6$. Fig.~\ref{MDOS}(a) shows MDOS numerically calculated for $h=0.6$ and $L=20$ via exact diagonalization. We then extract $\mu_E$ and $\bar{E}$ via Gaussian curve fitting. In Fig.~\ref{MDOS}(b), we show $\bar{E}/L$ and $\mu_E/\sqrt{L}$, which we compare with $\mu_E^{Th}/\sqrt{L}$, as a function of $L$. We find that $\bar{E}/L$ is essentially a constant implying extensivity of $\bar{E}$. $\mu_E/\sqrt{L}$ is almost a constant approaching the analytically estimated value in Eq.~\ref{mu_theory} as $L$ is increased.
Fig.~\ref{MDOS}(c) shows the MDOS for $h=1.8$ and $L=20$. The MDOS in this case shows more fluctuations than that for $h=0.6$. Fig.~\ref{MDOS}(d) shows similar behavior of $\bar{E}/L$ and $\mu_E/\sqrt{L}$ as in Fig.~\ref{MDOS}(b).

For numerical calculations in FS, we assume $\mu_E\propto\sqrt{L}$, $\bar{E}\propto L$ and we use the proportionality constants obtained from numerical fitting for $L=20$, which are closest to the analytical values. We then obtain $D(E)$ by replacing $\mu_E(L)$ and $\bar{E}(L)$ in Eq.~\ref{gauss_MDOS}. The broadening parameter $\eta(E)$ is then given by $\eta(E)=1/[\mathcal{N}_F D(E)]$. For the energy density $\mathcal{E}$, the broadening parameter $\eta(\mathcal{E})=1/[L\mathcal{N}_F D(E)]$.
   
\begin{figure}[ht!]
\centering
\includegraphics[width=0.995\columnwidth,height=4.0cm]{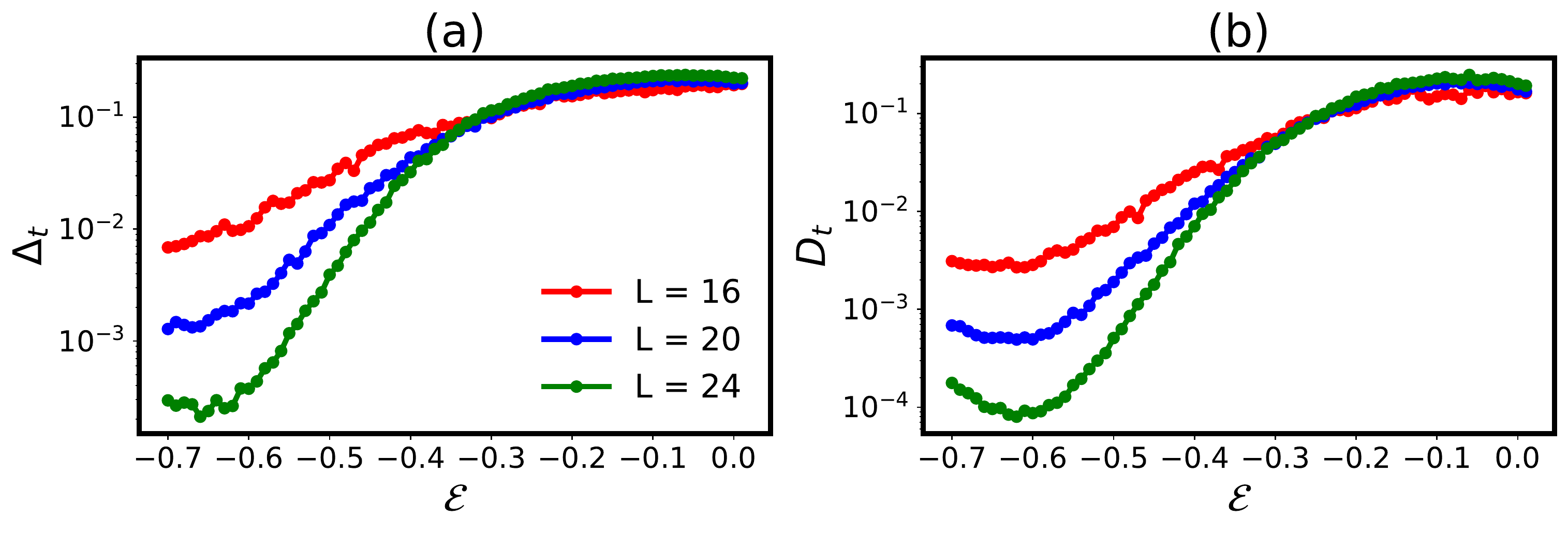}
\caption{{\bf Feenberg self-energy and LDOS in FS:} (a) $\Delta_t$ as a function of the many-body energy density $\mathcal{E}$ for increasing values of $L$. (b) $D_t$ vs $\mathcal{E}$ for increasing values of $L$. Disorder realizations over $\phi$ used for the plots are $2000,1000$ and $400$ for $L=16,20$ and $24$ respectively.}
\label{energy_ldos}
\end{figure}

\section{Nonergodic-ergodic transition: self-energy and LDOS in Fockspace}\label{app4}
Here we discuss the nonergodic-to-ergodic transition in terms of the typical value of the imaginary part of the Feenberg self-energy $\Delta_t$, which acts as an order parameter for the transition. The typical value $D_t(\E)$ of the local many-body density of states $D_I(\E)=(-1/\pi)\mathrm{Im}G_{II}(\E)$ also shows similar behavior. 

Previous work~\cite{ghosh2020transport} has shown the existence of MBL, NEE, and ergodic phases for $h=0.6$ as the many-body energy density $\mathcal{E}$ is increased from lower to higher values~\cite{ghosh2020transport}. As discussed in the main text, $\Delta_t$ and $D_t$ are calculated from the geometric mean of $\Delta_I$ and $D_I$, respectively, by averaging over disorder realizations and sites $I$ on the middle slice.
In Fig.~\ref{energy_ldos}(a-b), we show the variation of $\Delta_t$ and $D_t$ with $\mathcal{E}$ for increasing system sizes $L$. In the ergodic phase ($\mathcal{E}>\mathcal{E}_c$ with $\mathcal{E}_c\approx -0.4$) both $\Delta_t$ and $D_t$ approaches $\mathcal{O}(1)$ value in the thermodynamic limit, i.e. with increasing $L$. On the other hand, in the nonergodic phases ($\mathcal{E}<\mathcal{E}_c$) both the quantities decrease exponentially with $L$ and vanish in the thermodynamic limit. Thus, though $\Delta_t$ and $D_t$ can be used to distinguish nonergodic and ergodic phases, these quantities show qualitatively similar behavior in the MBL and NEE phases. This is evident in terms of the fractal dimension $D_s$, extracted from $\Delta_t\sim \mathcal{N}_F^{ -(1-D_s)}$ behavior in Fig.~\ref{linscaling}(b), where $0<D_s<1$ indicating multifractal nature of both the MBL and NEE states. Analysis of $D_t$ also leads to similar conclusions.\\

\begin{figure}[h]
\centering
\includegraphics[width=0.75\columnwidth,height=4.6cm]{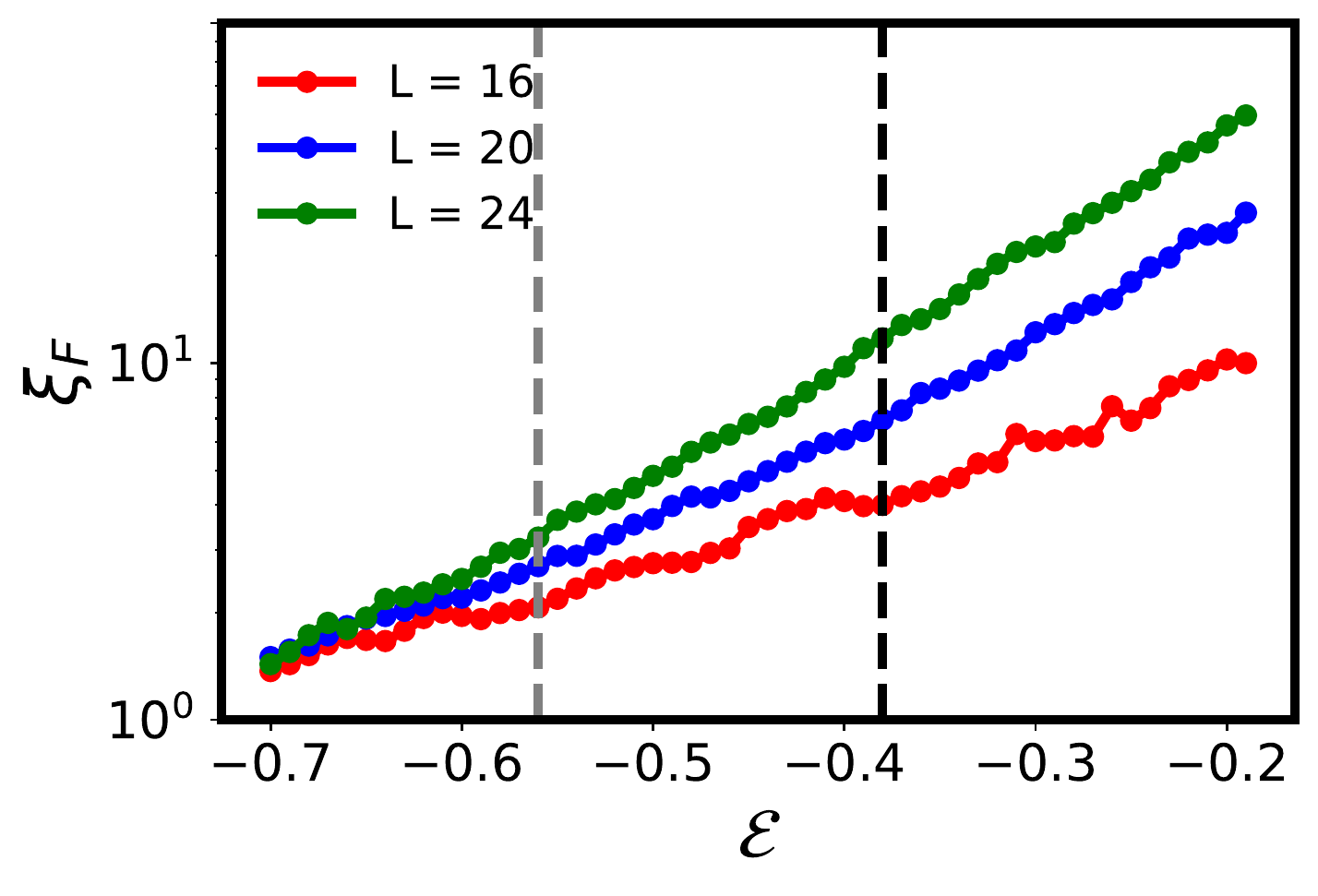}
\caption{{\bf Fig.~\ref{llength} in logscale:} FS localization length $\xi_F$ in logscale (of Fig.~\ref{llength}) as a function of $\mathcal{E}$ for increasing system sizes $L=12,16,20$. The vertical light and dark dashed lines indicate the MBL-NEE and NEE-ergodic transitions, respectively, based on previous study~\cite{ghosh2020transport} and analyses of Feenberg self-energy in Sec.\ref{sec5}.}
\label{FS_length}
\end{figure}

\section{Semilog plots of localization length}
\label{app5}
Fig.~\ref{FS_length} shows FS localization length in logscale as a function of $\E$.
There is no sharp crossing point in the plots, but evidently $\xi_F$ becomes very weakly $L$-dependent in the MBL phase, and it acquires clear $L$ dependence in the NEE phase ($-0.56\lesssim \E\lesssim 0.4$) and in the ergodic phase ($\E\gtrsim -0.4$).


\bibliography{refs}

\end{document}